\def \VersionLong {}
\documentclass[a4paper,10pt]{llncs}
\makeatletter
\AtBeginDocument{%
  \@ifpackageloaded{hyperref}
  {\def\@doi#1{\href{https://doi.org/#1}
      {\ttfamily https://doi.org/#1}\egroup}}
  {\def\@doi#1{\ttfamily https://doi.org/#1\egroup}}
  \def\doi{\bgroup\catcode`\_=12\relax\@doi}}
\makeatother
\makeatletter
\def\@biblabel#1{[#1]}

\makeatother
\makeatletter
\def\orcidID#1{\smash{\href{https://orcid.org/#1}{\protect\raisebox{-1.25pt}{\protect\includegraphics{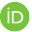}}}}}
\makeatother

\usepackage[utf8]{inputenc}
\usepackage[english]{babel}
\usepackage[T1]{fontenc}
\usepackage{csquotes}

\newcommand{\ourTitle}{Exemplifying parametric timed specifications over signals with bounded behavior}
\newcommand{\ourKeywords}{specification, timed automata, hybrid automata, signals}
\newcommand{\ourAbstract}{%
	Specifying properties can be challenging work.
	In this paper, we propose an automated approach to exemplify properties given in the form of automata extended with timing constraints and timing parameters, and that can also encode constraints over real-valued signals.
	That is, given such a specification and given an admissible automaton for each signal, we output concrete runs exemplifying real (or impossible) runs for this specification.
	Specifically, our method takes as input a specification, and a set of admissible behaviors, all given as a subclass of rectangular hybrid automata, namely timed automata extended with arbitrary clock rates, signal constraints, and timing parameters.
	Our method then generates concrete runs exemplifying the specification.
}

\usepackage[svgnames,table]{xcolor}

\definecolor{algocommentscolor}{rgb}{.4, .4, .7}
\usepackage[ruled,vlined,linesnumbered]{algorithm2e}
	\SetKwInOut{Input}{input}
	\SetKwInOut{Output}{output}
	\SetKw{kwTo}{to}
	\SetKw{kwDownTo}{down to}
	
	\SetCommentSty{mycommfont}

\usepackage{subcaption}

\usepackage{float} %

\usepackage{paralist} %

\newenvironment{ienumerate}
	{\ifdefined\VersionLong\begin{enumerate}\else\begin{inparaenum}[\itshape i\upshape)]\fi}
	{\ifdefined\VersionLong\end{enumerate}\else\end{inparaenum}\fi}

\newenvironment{oneenumerate}
	{\ifdefined\VersionLong\begin{enumerate}\else\begin{inparaenum}[1)]\fi}
	{\ifdefined\VersionLong\end{enumerate}\else\end{inparaenum}\fi}

\usepackage{amsmath} %
\usepackage{amssymb} %

\usepackage[misc,geometry]{ifsym} %

\usepackage{graphicx}
\graphicspath{{figures}}

\ifdefined \VersionLong
	\newcommand{\LongVersion}[1]{#1}
	\newcommand{\ShortVersion}[1]{}
\else
	\newcommand{\LongVersion}[1]{}
	\newcommand{\ShortVersion}[1]{#1}
\fi

\ifdefined\VersionLong
	\usepackage[backend=biber,backref=true,style=alphabetic,url=false,doi=true,defernumbers=true,sorting=anyt,maxnames=99]{biblatex} %
	\renewbibmacro*{doi+eprint+url}{%
		\iftoggle{bbx:doi}
			{\color{black!40}\footnotesize\printfield{doi}}
			{}%
		\newunit\newblock
		\iftoggle{bbx:eprint}
			{\usebibmacro{eprint}}
			{}%
		\newunit\newblock
		\iftoggle{bbx:url}
			{\usebibmacro{url+urldate}}
			{}%
	}

\fi
\ifdefined\VersionWithComments
	\usepackage{draftwatermark}
	\SetWatermarkText{draft}
	\SetWatermarkScale{9}
	\SetWatermarkColor[gray]{0.9}
\fi
\definecolor{darkblue}{rgb}{0, 0, 0.7}

\usepackage[
		pdfauthor={Etienne Andre, Masaki Waga, Natsuki Urabe, Ichiro Hasuo},%
		pdftitle={\ourTitle{}},
		breaklinks  = true,
		colorlinks  = true,
	\ifdefined \VersionWithComments
	\fi
		citecolor   = blue!50!blue,
		linkcolor   = darkblue,
		urlcolor    = blue!50!green,
	]{hyperref}

\usepackage[capitalise,english,nameinlink]{cleveref} %
\crefname{line}{\text{line}}{\text{lines}} %

\newcommand{\defProblem}[3]
{%
\noindent\fcolorbox{black}{blue!15}{
	\begin{minipage}{.95\columnwidth}
		\textbf{#1 problem:}\\
		\textsc{Input}: #2\\
		\textsc{Problem}: #3
	\end{minipage}
}
	
	\smallskip
	
}

\newcommand{\recallResult}[2]
{%
	\smallskip

	\noindent\fcolorbox{black}{green!15}{
		\begin{minipage}{.95\columnwidth}
			\noindent\textbf{\cref{#1} (recalled).}
			{\em{}#2}
		\end{minipage}
	}
	
	\smallskip
}

\usepackage{tikz}
\usetikzlibrary{arrows,automata,calc} %

\tikzstyle{pta}=[auto, ->, >=stealth']
\tikzstyle{every node}=[initial text=]
\tikzstyle{location}=[rectangle, rounded corners, minimum size=12pt, draw=black, fill=blue!10, inner sep=2pt]
\tikzstyle{invariant}=[draw=black, dotted, inner sep=1pt, yshift=-3em] %
\tikzstyle{final}=[double, fill=blue!50]
\tikzstyle{urgent}=[dotted, draw=red, very thick]

\tikzstyle{methodBox} = [line width=1mm, draw=black!15]
\tikzstyle{axis}=[thick,->]
\tikzstyle{signalA}=[-,draw=red, thick]
\tikzstyle{signalB}=[-,draw=blue,densely dotted, thick]

\tikzstyle{discrete_transition}=[draw=coloract, thick, dashed]

\tikzstyle{urgent}=[fill=yellow, thick, dotted] %

\definecolor{coloract}{rgb}{0.50, 0.70, 0.30}
\definecolor{colorclock}{rgb}{0.4, 0.4, 1}
\definecolor{colordisc}{rgb}{1, 0, 1}
\definecolor{colorloc}{rgb}{0.4, 0.4, 0.65}
\definecolor{colorparam}{rgb}{1, 0.6, 0.0}

\definecolor{loccolor1}{rgb}{1, 0.3, 0.3}
\definecolor{loccolor2}{rgb}{0.3, 1, 0.3}
\definecolor{loccolor3}{rgb}{0.3, 0.3, 1}
\definecolor{loccolor4}{rgb}{1, 0.3, 1}
\definecolor{loccolor5}{rgb}{1, 1, 0.3}
\definecolor{loccolor6}{rgb}{0.3, 1, 1}
\definecolor{loccolor7}{rgb}{0.9, 0.6, 0.2}
\definecolor{loccolor8}{rgb}{0.7, 0.4, 1}
\definecolor{loccolor9}{rgb}{0.5, 1, 0.75}
\definecolor{loccolor10}{rgb}{0.8, 0.7, 0.6}
\definecolor{loccolor11}{rgb}{0.6, 0.7, 0.8}
\definecolor{loccolor12}{rgb}{0.2, 0.5, 0.9}
\definecolor{loccolor13}{rgb}{0.5, 0.9, 0.2}
\definecolor{loccolor14}{rgb}{0.9, 0.2, 0.5}
\definecolor{loccolor15}{rgb}{0.7, 0.7, 0.7}
\definecolor{loccolor16}{rgb}{0.8, 0.8, 0.5}

\newcommand{\styleact}[1]{\ensuremath{\textcolor{coloract}{{#1}}}}
\newcommand{\styleclock}[1]{\ensuremath{\textcolor{colorclock}{{#1}}}}
\newcommand{\styledisc}[1]{\ensuremath{\textcolor{colordisc}{{#1}}}}
\newcommand{\stylesignal}{\styledisc}

\newcommand{\styleparam}[1]{\ensuremath{\textcolor{colorparam}{{#1}}}}
\newcommand{\init}{_0}

\newcommand{\A}{\ensuremath{\mathcal{A}}}

\newcommand{\Actions}{\Sigma}
\newcommand{\action}{\ensuremath{a}}
\newcommand{\actioni}[1]{\ensuremath{\styleact{\action_{#1}}}}
\newcommand{\assign}{\leftarrow}

\newcommand{\BTrue}{\text{true}}
\newcommand{\BFalse}{\text{false}}
\newcommand{\Constraint}{\ensuremath{\mathbf{C}}}

\newcommand{\Clocks}{\mathbb{C}} %
\newcommand{\clock}{c} %
\newcommand{\clockx}{\styleclock{\clock}} %
\newcommand{\compOp}{\bowtie}

\newcommand{\concreterun}{\ensuremath{\rho}}
\newcommand{\concreterunprefix}{\concreterun_\mathit{pref}}
\newcommand{\concreterunsuffix}{\concreterun_\mathit{suf}}
\newcommand{\ConcreteRuns}{\ensuremath{\mathcal{R}}}

\newcommand{\CFalse}{\mathbf{false}}
\newcommand{\edge}{e}
\newcommand{\Edges}{E}
\newcommand{\edgeat}{\ensuremath{\mathit{edgeAt}}}
\newcommand{\longuefleche}[1]{\stackrel{#1}{\longrightarrow}}
\newcommand{\longueflecheRel}[1]{\stackrel{#1}{\mapsto}}
\newcommand{\flecheRel}{{\rightarrow}}
\newcommand{\flow}{f}
\newcommand{\guard}{g}

\newcommand{\Intervals}{\mathcal{I}}
\newcommand{\invariant}{I}
\newcommand{\K}{K}

\newcommand{\loc}{\ensuremath{\ell}} %
\newcommand{\locinit}{\loc \init}
\newcommand{\Loc}{L} %
\newcommand{\LocsFinal}{\ensuremath{F}}

\newcommand{\locTarget}{\ensuremath{\loc_{T}}}
\newcommand{\none}{\ensuremath{\mathbf{none}}}
\newcommand{\Param}{\mathbb{P}} %
\newcommand{\param}{p} %
\newcommand{\paramp}{\ensuremath{\styleparam{\param}}}
\newcommand{\ParamCard}{M} %
\newcommand{\plterm}{\mathit{plt}}
\newcommand{\pval}{\lambda} %
\newcommand{\PZG}{\ensuremath{\mathcal{PZG}}} %
\newcommand{\R}{{\mathbb{R}}}
\newcommand{\resets}{R}
\newcommand{\setN}{\ensuremath{\mathbb N}}
\newcommand{\setQ}{\ensuremath{\mathbb Q}}
\newcommand{\setQplus}{\ensuremath{\setQ_{+}}} %
\newcommand{\setR}{\ensuremath{\mathbb R}}
\newcommand{\setRplus}{\ensuremath{\setR_{+}}} %

\newcommand{\Signals}{\ensuremath{\mathbb{S}}} %
\newcommand{\signal}{\ensuremath{s}} %
\newcommand{\signali}[1]{\ensuremath{\stylesignal{\signal_{#1}}}} %
\newcommand{\Sinit}{S\init} %
\newcommand{\concstate}{\ensuremath{s}} %
\newcommand{\stateat}{\ensuremath{\mathit{stateAt}}}
\newcommand{\States}{S} %
\newcommand{\Succ}{\mathsf{Succ}}
\newcommand{\runconcat}{+}
\newcommand{\timelapse}{\ensuremath{\mathit{te}}}
\newcommand{\Variables}{\mathbb{V}} %
\newcommand{\VariablesInit}{V\init} %
\newcommand{\VariablesCard}{H} %
\newcommand{\variable}{v} %
\newcommand{\variablei}[1]{\ensuremath{\styleclock{\variable_{#1}}}} %
\newcommand{\VariablesZero}{\vec{0}}
\newcommand{\vval}{\mu} %

\newcommand{\algoMain}{\ensuremath{\mathit{exemplify}}}
\newcommand{\computeD}{\ensuremath{\mathit{computeDur}}}
\newcommand{\constructNeg}{\ensuremath{\mathit{constructNeg}}}
\newcommand{\exemplifythree}{\ensuremath{\mathit{exemplify3}}}
\newcommand{\exhibitPred}{\ensuremath{\mathit{exhibitPred}}}
\newcommand{\exhibitPredDiscrete}{\ensuremath{\mathit{exhibitPredDisc}}}
\newcommand{\exhibitPredContinuous}{\ensuremath{\mathit{exhibitPredCont}}}
\newcommand{\exhibitPoint}{\ensuremath{\mathit{exhibitPoint}}}
\newcommand{\findPdeadlock}{\ensuremath{\mathit{findPdeadlock}}}
\newcommand{\findXdeadlock}{\ensuremath{\mathit{findVdeadlock}}}
\newcommand{\hasPdeadlock}{\ensuremath{\mathit{hasPdeadlock}}}
\newcommand{\hasXdeadlock}{\ensuremath{\mathit{hasVdeadlock}}}
\newcommand{\PickSymbRun}{\ensuremath{\mathit{PickSymbRun}}}
\newcommand{\reconstructPos}{\ensuremath{\mathit{reconstructPos}}}

\newcommand{\hasminimum}{\ensuremath{\mathit{hasMinimum}}}
\newcommand{\minimum}{\ensuremath{\mathit{minimum}}}
\newcommand{\infimum}{\ensuremath{\mathit{infimum}}}
\newcommand{\supremum}{\ensuremath{\mathit{supremum}}}

\newcommand{\styleSymbStatesSet}[1]{\ensuremath{\mathbf{#1}}}

\newcommand{\symbrun}{\ensuremath{\styleSymbStatesSet{r}}} %
\newcommand{\symbstate}{\ensuremath{\styleSymbStatesSet{s}}} %
\newcommand{\SymbState}{\ensuremath{\styleSymbStatesSet{S}}} %
\newcommand{\symbstateinit}{\symbstate\init} %
\newcommand{\symbtimelapse}[2]{\ensuremath{\styleSymbStatesSet{te}(#1, #2)}} %
\newcommand{\symbtimepast}[2]{\ensuremath{\styleSymbStatesSet{tp}(#1, #2)}} %
\newcommand{\symbtrans}{{\Rightarrow}} %

\newcommand{\project}[2]{\ensuremath{#1{\downarrow_{#2}}}}
\newcommand{\projectP}[1]{\ensuremath{\project{#1}{\Param}}}
\newcommand{\reset}[2]{\ensuremath{[#1]_{#2}}}
\newcommand{\valuate}[2]{\ensuremath{#2(#1)}}
\newcommand{\wv}[2]{#1|#2} %

\newcommand{\checkYes}{\textbf{$\textcolor{green!50!black}{\surd}$}}
\newcommand{\checkNo}{\textbf{$\textcolor{red!50!black}{\times}$}}

\newtheorem{assumption}{Assumption}

\ifdefined\VersionWithComments
	\usepackage[colorinlistoftodos,textsize=footnotesize]{todonotes}
\else
	\usepackage[disable]{todonotes}
\fi

\ifdefined\VersionWithComments
	\newcommand{\gennote}[3]{\mbox{}\\\fcolorbox{#2!50!black}{#2!5}{%
		\begin{minipage}{.96\columnwidth}%
			{\small\color{#2}{\textbf{#3}: #1}\xspace}%
		\end{minipage}%
	}\\}
\else
	\newcommand{\gennote}[3]{}
\fi

\ifdefined \VersionWithComments
	\newcommand{\todoinline}[1]{\mbox{}{\color{red}{\textbf{TODO}\ifx#1\\\else:\ \fi #1}}} %
\else
	\newcommand{\todoinline}[1]{}
\fi

\ifdefined \VersionWithComments
	
\else
	
\fi
\usepackage{verbatim} %

\newcommand{\dimension}{\ensuremath{\mathit{dim}}}
\newcommand{\Dimensions}{\ensuremath{\mathit{Dim}}}
\newcommand{\imitator}{\textsf{IMITATOR}}
\ifdefined \VersionWithComments
 	\definecolor{colorok}{RGB}{80,80,150}
\else
	\definecolor{colorok}{RGB}{0,0,0}
\fi

\newcommand{\eg}{\textcolor{colorok}{e.\,g.,}\xspace}

\newcommand{\ie}{\textcolor{colorok}{i.\,e.,}\xspace}

\newcommand{\wrt}{\textcolor{colorok}{w.r.t.}\xspace}

\title{\ourTitle{}\todo{This is the version with comments. To disable comments, comment out line~3 in the \LaTeX{} source.}\thanks{%
	\LongVersion{%
		This is the author (and extended) version of the manuscript of the same name published in the proceedings of the 14th {NASA} Formal Methods Symposium (\href{https://nfm2022.caltech.edu/}{NFM 2022}).
		The final authenticated version is available at %
			\href{https://www.springer.com}{\nolinkurl{springer.com}}.
	}%
	This work is partially supported by ERATO HASUO Metamathematics for Systems Design Project (No.\ JPMJER1603), JST
		and
	by the ANR-NRF French-Singaporean research program \href{https://www.loria.science/ProMiS/}{ProMiS} (ANR-19-CE25-0015).
}
}
\author{\'Etienne Andr\'e\inst{1}%
	\orcidID{0000-0001-8473-9555}%
${}^{\href{https://www.loria.science/andre/}{\text{\Letter}}}$
\and
Masaki Waga\inst{2}\orcidID{0000-0001-9360-7490}
\and
Natuski Urabe\inst{3}\orcidID{0000-0002-1554-6618}
\and
Ichiro Hasuo\inst{3,4}\orcidID{0000-0002-8300-4650}
}
\institute{Université de Lorraine, CNRS, Inria, LORIA, F-54000 Nancy, France
\and
Kyoto University, Kyoto, Japan
\and
National Institute of Informatics, Tokyo, Japan
\and
The Graduate University for Advanced Studies, SOKENDAI, Tokyo, Japan
}

\begin{document}

\sloppy

\pagestyle{plain}

\maketitle{}
\thispagestyle{plain}

\begin{abstract}
	\ourAbstract{}
	
	\keywords{\ourKeywords{}}
\end{abstract}
\section{Introduction}\label{section:introduction}
Model checking has had a lot of successes in the last decades (see,
\eg{} \cite{Kurshan18}).
Still, its use in the industry can be seen as slightly disappointing, considering its high advantages in providing system designers with formal guarantees in the correctness of their system.
This is especially true for \emph{quantitative} model checking, that considers systems extended with quantities such as probabilities, time, costs…
Among the explanations, one reason is the high expertise required by model checking users to master the model, the specification and their semantics.
Even domain experts may do manual errors, leading to specifications with a completely different behaviors from the expectations.
These issues may then only be solved using a tedious debugging phase.

\begin{figure}[tb]

	\begin{subfigure}[b]{0.59\textwidth}
		\centering
		\footnotesize

		\scalebox{.85}{
		\begin{tikzpicture}[pta, scale=1, xscale=1.2, yscale=1.5]
	
			\node[location, initial] at (.4, 0) (l1) {$\loc_1$};
	
			\node[location] at (2, 0) (l2) {$\loc_2$};
			\node[invariant, below=of l2, yshift=+6em] {$\clockx \leq 15$};
	
			\node[location] at (4, 0) (l3) {$\loc_3$};
			\node[invariant, below=of l3, yshift=+6em] {$\clockx \leq 20$};
	
			\node[location, final] at (6.5, 0) (lT) {$\locTarget$};

			\path (l1) edge node[align=center]{%
				$\signali{1} > 50$
				\\
				$\styleact{\mathit{larger}}$} node[below]{$\clockx \assign 0$} (l2);

			\path (l2) edge node[align=center, yshift=+.0em]{%
				$\clockx \leq 15 \land \signali{1} \geq 3 \times \signali{2}$
				\\
				$\styleact{\mathit{check}}$
			} node[below]{$\clockx \assign 0$} (l3);

			\path (l3) edge node[align=center, xshift=0em, yshift=-.2em]{%
				$\clockx \leq 20 \land \signali{1} = \signali{2}$
			} node[below] {$\styleact{\mathit{satisfied}}$} (lT);

		\end{tikzpicture}
		}
		\caption{Non-parametric specification}
		\label{figure:example-PTAS:motivating}
	\end{subfigure}
	\hfill
	\begin{subfigure}[b]{0.4\textwidth}
		\centering
		\footnotesize

		\scalebox{.85}{
		\begin{tikzpicture}[pta, scale=1, xscale=1.25, yscale=1.5]
	
			\node[location, initial] at (0.6, 0) (l1) {$\loc_1$};
			\node[invariant, yshift=+6em, below=of l1] {$\clockx \leq \paramp$};

			\node[location] at (2, 0) (l2) {$\loc_2$};
			\node[invariant, yshift=+6em, below=of l2] {$\clockx \leq \paramp$};

			\node[location, final] at (4, 0) (lT) {$\locTarget$};
	
			\path (l1) edge node[align=center]{%
				$\clockx \geq 5 \land \signali{1} = \signali{2}$
				\\
				$\styleact{\mathit{sense}}$} node[below]{$\clockx \assign 0$} (l2);

			\path (l2) edge node[align=center]{%
				$\clockx \geq 5 \land \signali{1} < \frac{\signali{2}}{2}$} node [below]{$\styleact{\mathit{sense}}$} (lT);

		\end{tikzpicture}
		}
		\caption{Parametric specification}
		\label{figure:example-PTAS:param}
	\end{subfigure}

	\caption{Examples of PTAS}

\end{figure}
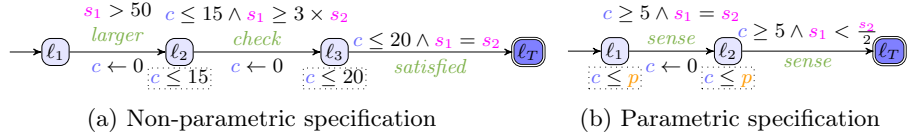

\paragraph{Contribution}
In this work, we propose an approach to exemplify concrete continuous evolutions of signals over time, according to a specification.
We introduce as a specification formalism \emph{parametric timed automata with signals (PTASs)} as an extension of (parametric) timed automata~\cite{AD94,AHV93}: our PTASs use the full power of timed automata, with clocks compared to constants, and add the possibility to specify \emph{signal (linear) constraints}, such as ``$\signali{1} \geq 3 \times \signali{2}$''.
This allows us to easily express specifications of the form ``whenever signal $\signali{1}$ is larger than 50, then within at most 15 time units, it holds that $\signali{1} \geq 3 \times \signali{2}$ and then, within at most 20 more time units, both signals are equal ($\signali{1} = \signali{2}$)''.
\cref{figure:example-PTAS:motivating} depicts the PTAS encoding this specification (where $\clockx$ is a clock, while $\signali{1}$ and~$\signali{2}$ are signals), \ie{} $\locTarget$ is reachable whenever the specification is met for some execution.
\LongVersion{

}%
In addition, we allow for \emph{timing parameters} (unknown constants), thus enabling parametric specifications mixing discrete actions, signal constraints and timing parameters all together, such as ``after a first sensing (action \styleact{\mathit{sense}}) occurring within $[5, \paramp]$, it holds that $\signali{1} = \signali{2}$, and after a second sensing occurring within $[5, \paramp]$, it holds that  $\signali{1} < \frac{\signali{2}}{2}$'', where $\paramp$ is a timing parameter.
The PTAS encoding this specification is given in \cref{figure:example-PTAS:param}.
In this latter case, the exemplification comes in the form of a concrete valuation for~$\paramp$ \emph{and} an evolution of the signals satisfying the specification.

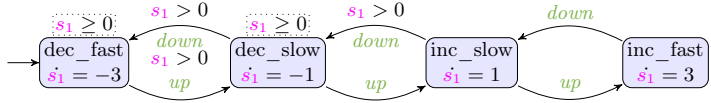
\begin{figure}[tb]
 
	\centering
	 \footnotesize

	 \scalebox{.85}{
	\begin{tikzpicture}[pta, scale=1, xscale=1.2, yscale=1.5, bend angle=22]
 
		\node[location, initial] at (0, 0) (decreasing_fast) [align=center] {dec\_fast\\$\dot{\signali{1}} = -3$};
 
		\node[location] at (2.5, 0) (decreasing_slow) [align=center] {dec\_slow\\$\dot{\signali{1}} = -1$};
 
		\node[location] at (5, 0) (increasing_slow) [align=center] {inc\_slow\\$\dot{\signali{1}} = 1$};
 
		\node[location] at (7.5, 0) (increasing_fast) [align=center] {inc\_fast\\$\dot{\signali{1}} = 3$};
 
		\node[invariant, above=of decreasing_fast, yshift=-0] {$\signali{1} \geq 0$};
		
		\node[invariant, above=of decreasing_slow, yshift=-0] {$\signali{1} \geq 0$};
		
		\path
			(decreasing_fast) edge[bend right] node[above, align=center]{$\signali{1} > 0$ \\ \styleact{\mathit{up}}} (decreasing_slow)

			(decreasing_slow) edge[bend right] node[above]{$\signali{1} > 0$} node[below]{\styleact{\mathit{down}}} (decreasing_fast)
			(decreasing_slow) edge[bend right] node[above]{\styleact{\mathit{up}}} (increasing_slow)
			
			(increasing_slow) edge[bend right] node[above]{$\signali{1} > 0$} node[below]{\styleact{\mathit{down}}} (decreasing_slow)
			(increasing_slow) edge[bend right] node[above]{\styleact{\mathit{up}}} (increasing_fast)
			
			(increasing_fast) edge[bend right] node[above]{\styleact{\mathit{down}}} (increasing_slow)
		;

	\end{tikzpicture}
	}
	\caption{An example of SBA}
	\label{figure:example-SBA-fastslow}

\end{figure}

In order to bound the possible signal behaviors, we introduce as additional input \emph{signal bounding automata} (SBA), \ie{} automata bounding the admissible behaviors of the signals.
These SBAs can be gathered from a (rough) knowledge from the system under consideration;
they can also be used to search among the widely variety signals satisfying the specification\LongVersion{; for example, we may want to search signals for each scenario,} \eg{} driving with/without acceleration/deceleration.
In addition, thanks to the SBAs, we avoid generating irrelevant signals, \eg{} signals with unrealistically large value change even in the negative example generation.
Our SBAs assign signals an arbitrary (but piecewise constant) derivative, according to some guards.
For example, an SBA \LongVersion{could allow a signal~$\signal$ to move from stopped ($\dot{\signal} = 0$) to slowly growing ($\dot{\signal} = 1$) and, provided its value is large enough (\eg{} $\signal > 100$), to growing fast ($\dot{\signal} = 2$).
Or another SBA }could allow signal~$\signal$ to alternate between slowly ($\dot{\signal} = 1$) and rapidly ($\dot{\signal} = 3$) growing---or decreasing; this latter SBA is depicted in \cref{figure:example-SBA-fastslow}.

We generate not only \emph{positive} (``correct'') exemplifications, but also \emph{negative} (``incorrect'', \ie{} that do \emph{not} match the specification).
The crux behind this is that, in order to illustrate a specification, we may need both positive and negative examples that are close to the boundary. 
See \cref{figure:concrete-runs:param} for an example.

\begin{example}\label{example:motivating}
	Let $\A$ be the PTAS in \cref{figure:example-PTAS:motivating};
	let $\A_1$ be the SBA in \cref{figure:example-SBA-fastslow},
	and let $\A_2$ be the SBA in \cref{figure:example-SBA-fastslow} where $\signali{1}$ is replaced with~$\signali{2}$.
	We assume initially $\signali{1}, \signali{2} \in [0,10]$.
	Given the PTAS $\A$ and the 2 SBAs $\A_1$ and~$\A_2$ bounding the behavior of~$\signali{1}$ and~$\signali{2}$,
	our framework automatically generates several signal evolutions satisfying the specification; we give 3 of them in \cref{figure:concrete-runs:running}.
	Observe that they present 3 very different evolutions of the signals, with different initial valuations, evolution rates, and final valuations.
\end{example}
\begin{figure}[tb]
	\begin{subfigure}[b]{0.31\textwidth}

		\scriptsize
		\begin{tikzpicture}[scale=.035, xscale=1.2, yscale=.45]
		\draw[axis] (0,0) -- (80, 0) node[anchor=south] {$t$};
		\draw[axis] (0,0) -- (0, 75) node[anchor=south east] {};

		\foreach \x in {0,10,...,80}
			\draw[draw=black] (\x, 1) -- (\x, -1) node[anchor=north] {$\x$};
		\foreach \y in {0,20,...,70}
			\draw (1, \y) -- (-1, \y) node[anchor=east] {$\y$};

		\draw[signalA] (0, 10) -- (35/2, 55/2) -- (45, 55) -- (50, 60) -- (55, 55) -- (245/4, 195/4) -- (155/2, 65/2) node[above]{$\signali{1}$};

		\draw[signalB] (0, 10) -- (35/2, 55/2) -- (45, 0) -- (50, 5) -- (55, 10) -- (245/4, 65/4) -- (155/2, 65/2) node[below]{$\signali{2}$};
		\end{tikzpicture}
		\caption{Possible run}
		\label{figure:concrete-runs:running:1}
	\end{subfigure}
	\begin{subfigure}[b]{0.38\textwidth}

		\scriptsize
		\begin{tikzpicture}[scale=.035, xscale=1.2, yscale=.45]
		\draw[axis] (0,0) -- (100, 0) node[anchor=south] {$t$};
		\draw[axis] (0,0) -- (0, 75) node[anchor=south east] {};

		\foreach \x in {0,10,...,90}
			\draw[draw=black] (\x, 1) -- (\x, -1) node[anchor=north] {$\x$};
		\foreach \y in {0,20,...,70}
			\draw (1, \y) -- (-1, \y) node[anchor=east] {$\y$};

		\draw[signalA] (0, 0) -- (265/6, 265/6) -- (295/4, 295/4) -- (355/4, 115/4) -- (295/3, 0) node[above, yshift=2em]{$\signali{1}$};

		\draw[signalB] (0, 10) -- (265/6, 325/6) -- (295/4, 295/12) -- (355/4, 115/12) -- (295/3, 0) node[above, xshift=-2em]{$\signali{2}$};
		\end{tikzpicture}
		\caption{Possible run}
		\label{figure:concrete-runs:running:2}
	\end{subfigure}
	\begin{subfigure}[b]{0.25\textwidth}

		\scriptsize
		\begin{tikzpicture}[scale=.035, xscale=1.2, yscale=.45]
		\draw[axis] (0,0) -- (65, 0) node[anchor=south] {$t$};
		\draw[axis] (0,0) -- (0, 75) node[anchor=south east] {};

		\foreach \x in {0,10,...,60}
			\draw[draw=black] (\x, 1) -- (\x, -1) node[anchor=north] {$\x$};
		\foreach \y in {0,20,...,70}
			\draw (1, \y) -- (-1, \y) node[anchor=east] {$\y$};

		\draw[signalA] (0, 10) -- (247/36, 607/36) -- (337/9, 427/9) -- (247/6, 307/6) -- (253/6, 313/6) -- (130/3, 160/3) -- (133/3, 163/3)  -- (190/3, 220/3) node[left, xshift=-.5em]{$\signali{1}$};

		\draw[signalB] (0, 10) -- (247/36, 367/12) -- (337/9, 0) -- (247/6, 67/6) -- (253/6, 85/6) -- (130/3, 46/3) -- (133/3, 49/3)  -- (190/3, 220/3) node[below, yshift=-2em]{$\signali{2}$};
		\end{tikzpicture}
		\caption{Possible run}
		\label{figure:concrete-runs:running:3}
	\end{subfigure}

	\caption{Concrete runs for \cref{figure:example-PTAS:motivating,figure:example-SBA-fastslow}}
	\label{figure:concrete-runs:running}
\end{figure}
\begin{figure}[tb]
	\begin{minipage}[b]{\ifdefined\VersionLong\textwidth\else{}0.6\linewidth\fi}
	\centering
		\begin{subfigure}[b]{0.34\textwidth}

			\scriptsize
			\begin{tikzpicture}[scale=.08, xscale=1.2, yscale=.55]
			\draw[axis] (0,0) -- (22, 0) node[anchor=south] {$t$};
			\draw[axis] (0,0) -- (0, 32) node[anchor=south east] {};

			\foreach \x in {0,5,...,20}
				\draw[draw=black] (\x, 1) -- (\x, -1) node[anchor=north] {$\x$};
			\foreach \y in {0,10,...,30}
				\draw (1, \y) -- (-1, \y) node[anchor=east] {$\y$};

			\draw[discrete_transition] (10, 30) -- (10, 0) node[above left, xshift=.3em] {\tiny$\styleact{\mathit{sense}}$};
			\draw[discrete_transition] (20, 30) -- (20, 0) node[above left, xshift=.3em] {\tiny$\styleact{\mathit{sense}}$};

			\draw[signalA] (0, 10) -- (1/2, 21/2) -- (1, 11) -- (10, 20) -- (20, 10) node[above]{$\signali{1}$};

			\draw[signalB] (0, 11) -- (1/2, 23/2) -- (1, 11) -- (10, 20) -- (20, 30) node[below]{$\signali{2}$};
			\end{tikzpicture}
			\caption{Pos\LongVersion{itive run} for $\paramp = 10$}
			\label{figure:concrete-runs:param:pos}
		\end{subfigure}
		\begin{subfigure}[b]{0.26\textwidth}

			\scriptsize
			\begin{tikzpicture}[scale=.08, xscale=1.2, yscale=.55]
			\draw[axis] (0,0) -- (15, 0) node[anchor=south] {$t$};
			\draw[axis] (0,0) -- (0, 32) node[anchor=south east] {};

			\foreach \x in {0,5,...,10}
				\draw[draw=black] (\x, 1) -- (\x, -1) node[anchor=north] {$\x$};
			\foreach \y in {0,10,...,30}
				\draw (1, \y) -- (-1, \y) node[anchor=east] {$\y$};

			\draw[discrete_transition] (5, 30) -- (5, 0) node[above left, xshift=.3em] {\tiny$\styleact{\mathit{sense}}$};
			\draw[discrete_transition] (6, 30) -- (6, 0) node[above right, xshift=-.2em] {\tiny$\styleact{\mathit{sense}}$};

			\draw[signalA] (0, 18) -- (1, 19) -- (5, 15) -- (6, 14)  node[above right]{$\signali{1}$};

			\draw[signalB] (0, 10) -- (1, 11) -- (5, 15) -- (6, 16) node[below right]{$\signali{2}$};
			\end{tikzpicture}
			\caption{Neg\LongVersion{ run for} $\paramp = 5$}
			\label{figure:concrete-runs:param:neg5}
		\end{subfigure}
		\begin{subfigure}[b]{0.28\textwidth}

			\scriptsize
			\begin{tikzpicture}[scale=.08, xscale=1.2, yscale=.55]
			\draw[axis] (0,0) -- (15, 0) node[anchor=south] {$t$};
			\draw[axis] (0,0) -- (0, 32) node[anchor=south east] {};

			\foreach \x in {0,5,...,10}
				\draw[draw=black] (\x, 1) -- (\x, -1) node[anchor=north] {$\x$};
			\foreach \y in {0,10,...,30}
				\draw (1, \y) -- (-1, \y) node[anchor=east] {$\y$};

			\draw[discrete_transition] (10, 30) -- (10, 0) node[above left, xshift=.3em] {\tiny$\styleact{\mathit{sense}}$};

			\draw[discrete_transition] (40/3, 30) -- (40/3+.2, 0) node[above right, xshift=-.2em, yshift=1em] {\tiny$\styleact{\mathit{sense}}$};

			\draw[signalA] (0, 20) -- (40/3, 100/3) node[left, xshift=-.5em]{$\signali{1}$};

			\draw[signalB] (0, 19) -- (1/2, 39/2) -- (1, 21) -- (10, 30) -- (40/3, 100/3) node[below right]{$\signali{2}$};
			\end{tikzpicture}
			\caption{Neg\LongVersion{ run for} $\paramp = 10$}
			\label{figure:concrete-runs:param:neg10}
		\end{subfigure}
		\caption{Positive and negative runs for \cref{figure:example-PTAS:param}}
		\label{figure:concrete-runs:param}
	\end{minipage}
	\ifdefined\VersionLong
	
	\else{}\hfill{}\fi
	\begin{minipage}[b]{\ifdefined\VersionLong\textwidth\else{}0.38\linewidth\fi}

		\centering
		\scriptsize

		\begin{tikzpicture}[thick, xscale=.8, yscale=.7]
			\draw [fill=yellow!5, fill opacity=0.5] (0.5, 1.35) rectangle (6.3, 3.65);
			\draw [fill=cyan, fill opacity=0.5] (3, 2.5) ellipse (2.2 and 2.2/2);
			\draw [fill=blue, fill opacity=0.5] (3, 2.5) ellipse (1.2 and 1.2/2);
			\draw [fill=orange, fill opacity=0.5] (4.9, 2.5) ellipse (1.2 and 1.2/2);
			
			\node at (1, 3.4) {RHA};
			\node at (3, 2.5) {SBA};
			\node at (3, 3.3) {PLMA};
			\node at (5.65, 2.5) {PTAS};
		\end{tikzpicture}

		\caption{Formalisms\LongVersion{ manipulated in our approach}}
		\label{figure:formalisms}
	\end{minipage}
\end{figure}

\begin{figure}[tb]
 
	\centering
	 \scriptsize

	\begin{tikzpicture}[pta, scale=1, node distance=1em, xscale=2.5, yscale=1.5]

		\node[methodBox] at (0, 0) (PTAS) [align=center] {%
			\textbf{Specification}
			\\
			A PTAS~$\A$ with $n$ signals
			\\
			\scalebox{.3}{
			\begin{tikzpicture}[pta, scale=1, xscale=2.5, yscale=1.5]
		
				\node[location, initial] at (0, 0) (l1) {$\loc_1$};
		
				\node[location] at (1, 0) (l2) {$\loc_2$};
		
				\node[location, final] at (2, 0) (lT) {$\locTarget$};
		
				\node[invariant, above=of l2, yshift=-5] {$\clockx \leq \paramp$}; %
				
				\path (l1) edge node[align=center]{$\signali{1} > \signali{2}$ \\ $\clockx \assign 0$} (l2);

				\path (l2) edge node[align=center]{%
					$\clockx = \paramp$ %
					\\
					$\land \signali{1} < \signali{2}$
				} (lT);

			\end{tikzpicture}
			}
		};
		\node[methodBox, anchor=north, below=of PTAS.south, yshift=+.8em] (SBAs) [align=center] {%
			\textbf{Bounding behavior}
			\\
			$n$ SBAs $\A_i, i \in \{ 1, \dots , n \}$
			\\
			\scalebox{.3}{
			\begin{tikzpicture}[pta, scale=1, xscale=2.5, yscale=1.5]

			\node[location, initial] at (0, 0) (increasing) [align=center] {increasing\\$\dot{\signali{1}} = 1$};
		
				\node[location] at (1, 0) (stable) [align=center] {stable\\$\dot{\signali{1}} = 0$};
		
				\node[location] at (2, 0) (decreasing) [align=center] {decreasing\\$\dot{\signali{1}} = -1$};
		
				\node[invariant, above=of decreasing, yshift=-7] {$\signali{1} \geq 0$};
				
				\path
					(increasing) edge[bend right] (stable)
					(stable) edge[bend right] (increasing)
					(stable) edge[bend right] node[below]{$\signali{1} > 0$} (decreasing)
					(decreasing) edge[bend right] (stable)
				;
			
			\end{tikzpicture}
			}
			\vspace{-.5em}
			\\
			\vspace{-.5em}
			$\vdots$
			\\
			\scalebox{.3}{
			\begin{tikzpicture}[pta, scale=1, xscale=2.5, yscale=1.5]

			\node[location, initial] at (0, 0) (increasing) [align=center] {increasing\\$\dot{\stylesignal{\signal_n}} = 1$};
		
				\node[location] at (1, 0) (stable) [align=center] {stable\\$\dot{\stylesignal{\signal_n}} = 0$};
		
				\node[location] at (2, 0) (decreasing) [align=center] {decreasing\\$\dot{\stylesignal{\signal_n}} = -1$};
		
				\node[invariant, above=of decreasing, yshift=-7] {$\stylesignal{\signal_n} \geq 0$};
				
				\path
					(increasing) edge[bend right] (stable)
					(stable) edge[bend right] (increasing)
					(stable) edge[bend right] node[below]{$\stylesignal{\signal_n} > 0$} (decreasing)
					(decreasing) edge[bend right] (stable)
				;
			
			\end{tikzpicture}
			}
		};
		\node[methodBox, xshift=13em] (PLMA) at ($(PTAS)!0.5!(SBAs)$) [align=center] {%
			A PLMA \\
			$\A \parallel \A_1 \parallel \cdots \parallel \A_n $
		};
		\node[below=of SBAs.south, yshift=+1.2em] (inputs) [align=center] {\textbf{Inputs}};
		\node[methodBox, right=of PLMA.east, xshift=1.5em] (signals) [align=center] {%
			\textbf{Exemplification}
			\\
			Set of concrete runs
			\\
			{\scriptsize
			\begin{tabular}{c c c}
				$\pval_1$
				&
				\checkYes{}
				& 
				\scalebox{.3}{
					\begin{tikzpicture}[scale=1]

					\draw[axis] (0,0) -- (6, 0) node[anchor=north west] {$t$};
					\draw[axis] (0,0) -- (0, 1.5) node[anchor=south east] {};

					\draw[signalA] (0, 0.2) -- (2, 0.3) -- (3, 1) -- (3.5, 0.2) -- (5, 0.2) -- (6, 1.2); 

					\draw[signalB] (0, 0.0) -- (1.5, 0.2) -- (2, 1) -- (3, 1.2) -- (4.5, 1.1) -- (6, 0.5);

					\end{tikzpicture}
				}
				\\
				&
				&
				{$\vdots$}
				\\
				$\pval_n$
				&
				\checkNo{}
				& 
				\scalebox{.3}{
					\begin{tikzpicture}[scale=1]

					\draw[axis] (0,0) -- (6, 0) node[anchor=north west] {$t$};
					\draw[axis] (0,0) -- (0, 1.5) node[anchor=south east] {};

					\draw[signalA] (0, 0.5) -- (3, 0.5) -- (5, 0.2) -- (6, 0.1); 

					\draw[signalB] (0, 1.0) -- (1.5, 0.2) -- (1.5, 1) -- (4, 0) -- (4, 1.1) -- (6, 0.5); 

				\end{tikzpicture}
				}
			\end{tabular}
			}
			
		};
		\node[] (outputs) at (inputs -| signals)  [align=center] {\textbf{Outputs}};

		\draw[] (PTAS.east) -- ++(+0.1, 0) |-  (PLMA.west);
		\draw[] (SBAs.east) -- ++(+0.1, 0) |-  (PLMA.west);	
		\draw[] (PLMA.east) -- (signals);	

	\end{tikzpicture}
	
	\caption{Our general approach}
	\label{figure:framework}

\end{figure}
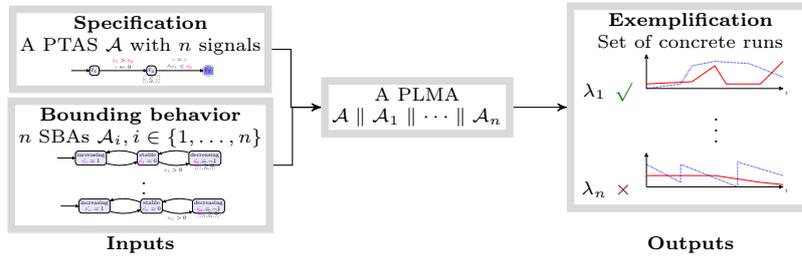

Our approach is \LongVersion{summarized}\ShortVersion{given} in \cref{figure:framework}. %
More specifically, our contributions are\LongVersion{ as follows}:

\begin{enumerate}
	\item We introduce three formalisms, all being subclasses of rectangular hybrid automata~\cite{Henzinger96}, namely \emph{parametric timed automata with signals} (PTASs) to express specifications, \emph{signal bounding automata} (SBAs) to bound the behavior of each signal, and \emph{parametric linear multi-rate automata} (PLMAs) that will be used for the parallel composition of the aforementioned formalisms; the relationship between these classes is given in \cref{figure:formalisms};
	\item We equip PLMAs with both a concrete and a symbolic semantics;
	\item We propose an exemplification algorithm for PLMAs, yielding concrete parameter valuations together with positive and negative runs;
	\item We implement our framework into 
	the \imitator{} model checker~\cite{Andre21};
	\item We show the applicability of our approach on a set of specifications.
\end{enumerate}

\paragraph{Outline}
\cref{section:related} reviews related works.
\cref{section:preliminaries} recalls the necessary preliminaries.
Then, \cref{section:PLMA} introduces the general class of parametric linear multirate automata (PLMAs), as well as two subclasses used in the subsequent approach.
\cref{section:problem} formally defines our specification exemplification problem.
\cref{section:approach} 
is the core of our contribution, proposing to exemplify specifications using %
	techniques
to exhibit parameter valuations and concrete runs for reachability properties in PLMAs.
\cref{section:experiments} exemplifies our approach on a set of specifications.
\cref{section:conclusion} concludes and proposes \LongVersion{possible }future works.

\section{Related works}\label{section:related}
There are several works~\cite{HMF15,RHM17,PLK18,BDGMN21} to visualize counterexamples of a formal specification.
One of the closest works to ours is STLInspector~\cite{RHM17}.
Given a signal temporal logic (STL)~\cite{MN04} formula $\varphi$,
STLInspector generates a signal~$s$ differentiating~$\varphi$ and a mutated formula~$\varphi'$.
Similarly, in~\cite{PLK18}, concrete traces are automatically generated, that satisfy or violate an STL formula.
Such signals are generated by SMT.
A difference between \cite{PLK18} and~\cite{RHM17} is that \cite{PLK18} considers linear (as opposed to rectangular) predicates.
Another related work is ShapEx~\cite{BDGMN21}.
Given a shape expression~\cite{NQFMD19} $\varphi$,
ShapEx generates signals represented by $\varphi$ based on a sampling-based algorithm.
Compared with most of these related works, the main difference with our approach is the use of \emph{signal bounding automata}: 
since most of the existing techniques generate a signal without bounding the admissible behaviors,
an unrealistic signal may be generated.
Another difference, especially from SMT-based approaches, is that it is easy for our automata-based approach to generate various signals by covering various paths of the automaton.
In contrast, for example, \cite{PLK18} requires an additional constraint, called a \emph{blocking constraint}, to generate various signals.
Nevertheless, the use of SMT in the analysis of an automaton (much like nuXmv~\cite{CGMRT19}) is future work.
In addition, most of these works utilize MITL~\cite{MNP06}, STL~\cite{MN04}, or an extension of regular expressions.
Our approach takes as input a more general, automata-based formalism (using notably timing parameters and multi-rate variables), not restricted to a given logic.
We note that one can translate a formula in most of these logics to a timed automaton, which our formalism captures.
See \eg{}~\cite{BGHM17,ACM02} for translation of such logical expressions to timed automata.

In~\cite{PBV18}, a method is proposed for \emph{visualizing} counterexamples for function block diagrams, of properties expressed in~LTL.\@
Both the model and the property can be animated.
In~\cite{DR19}, the focus is \emph{explaining} the violation of a property against a concrete run.
The property is given in the low-level ``control flow temporal logic'' (CFTL).
An originality is the notion of \emph{severity}, explaining by how much a timing constraint is violated (which shares similarities with \emph{robustness}\LongVersion{~\cite{DM10}}).
The approach is implemented into \href{http://cern.ch/vypr}{\textsc{VyPR2}}\LongVersion{~\cite{DRFPG19}}.
A main difference with our approach is that~\cite{DR19} targets the explanation of \emph{one} particular run violation, whereas we seek arbitrary exemplifications of a property (both positive and negative), independently of a run.
Visualization of specifications was also considered, \eg{} for Z specification~\cite{KC99} and for a DSL based on Event-B~\cite{TMB16}.

Another direction to tackle the difficulty of specification writing is translation of a natural language description to a temporal logic formula, \eg{}~\cite{HBNIG21}.

Finally, our new notion of \emph{signal bounding automaton}, used to bound the possible behavior of the signals, can be reminiscent of the recent model-bounded monitoring framework, which we introduced in~\cite{WAH21}.
In that paper, we used a rough over-approximation to bound the possible behaviors while performing monitoring of a black-box system.
Similar idea is also used in~\cite{BBDDKY20} to bound the signal space in the falsification problem by a timed automaton~\cite{AD94}.

The main originality of our work is
\begin{ienumerate}%
	\item the use of quantitative specifications (involving notably continuous time, timing parameters and signals),
	and
	\item the use of signal bounding automata to bound the admissible behaviors.
\end{ienumerate}%
\section{Preliminaries: Constraints and Rect\ShortVersion{.\ }\LongVersion{angular }Hybrid Automata}\label{section:preliminaries}

\LongVersion{%
\subsection{Clocks, parameters and guards}\label{ss:clock_parameters_guards}
}

We assume a set~$\Variables = \{ \variable_1, \dots, \variable_\VariablesCard \} $ of real-valued continuous variables.
Different from timed automata ``clocks''~\cite{AD94}, our variables (closer to hybrid systems' ``continuous variables'') can have different rates, and turn negative.
A variable valuation is\LongVersion{ a function}
$\vval : \Variables \to \setR$.
We write $\VariablesZero$ for the variable valuation assigning $0$ to all variables.
Given $d \in \setR$, %
and a flow (or rate) function $ \flow : \Variables \to \setQ$ assigning each variable with a flow (\ie{} the value of its derivative), we define the time elapsing function $\timelapse$ as follows:
$\timelapse(\vval, \flow, d)$ is the valuation such that
\(\forall \variable \in \Variables : \timelapse(\vval, \flow, d)(\variable) = \vval(\variable) + \flow(\variable) \times d \).
Given $\resets \subseteq \Variables$, we define the \emph{reset} of a valuation~$\vval$, denoted by $\reset{\vval}{\resets}$, as follows: $\reset{\vval}{\resets}(\variable) = 0$ if $\variable \in \resets$, and $\reset{\vval}{\resets}(\variable)=\vval(\variable)$ otherwise.

We assume a set~$\Param = \{ \param_1, \dots, \param_\ParamCard \} $ of \emph{(timing) parameters}\LongVersion{, \ie{} unknown constants}.
A parameter {\em valuation} $\pval$ is\LongVersion{ a function}
$\pval : \Param \to \setQplus$.
We assume ${\compOp} \in \{<, \leq, =, \geq, >\}$.
A \emph{parametric linear term} over $\Variables \cup \Param$ is of the form $\sum_{1 \leq i \leq \VariablesCard} \alpha_i \variable_i + \sum_{1 \leq j \leq \ParamCard} \beta_j \param_j + d$, with
	$\variable_i \in \Variables$,
	$\param_j \in \Param$,
	and
	$\alpha_i, \beta_j, d \in \setQ$.
A \emph{parametric linear inequality} is $\plterm \compOp 0$, where $\plterm$ is a parametric linear term.
A \emph{parametric linear constraint} $\Constraint$ (\ie{} a convex polyhedron) over $\Variables \cup \Param$ is a conjunction of parametric linear inequalities.
Given~$\Constraint$, we write~$\vval \models \pval(\Constraint)$ if %
the expression obtained by replacing each~$\variable$ with~$\vval(\variable)$ and each~$\param$ with~$\pval(\param)$ in~$\Constraint$ evaluates to~true.

\LongVersion{
\subsection{Rectangular hybrid automata}
}

Let $\Intervals(\setR)$ denote the set of all intervals over~$\setR$.
We first recall rectangular hybrid automata (RHAs)\LongVersion{\footnote{%
	We use a slightly different definition of RHAs when compared to, \eg{} \cite{Henzinger96}: in that latter work, RHAs use \emph{bounded} rectangular regions for invariants and flows.
	In addition, the definition of the variable reset is also different in~\cite{Henzinger96} (they use a \emph{rectangular update}).
	These definitions have no impact on the correctness nor applicability of our approach.
}}%
, a subclass of hybrid automata.
Our definition involves (timing) parameters; parameters could be seen as syntactic sugar for a subset of variables (\ie{} variables of arbitrary initial value and of zero rate throughout the automaton), but we still add them explicitly as they will explicitly appear in subsequent subclasses of RHAs.

\begin{definition}[RHA]\label{def:RHA}
	A \emph{rectangular hybrid automaton} (RHA) $\A$ is a tuple \mbox{$\A = (\Actions, \Loc, \locinit, \LocsFinal, \Variables, \VariablesInit, \Param, \invariant, \flow, \Edges)$}, where:
	\begin{oneenumerate}
		\item $\Actions$ is a finite set of actions,
		\item $\Loc$ is a finite set of locations,
		\item $\locinit \in \Loc$ is the initial location,
		\item $\LocsFinal \subseteq \Loc$ is the set of accepting locations,
		\item $\Variables$ is a finite set of variables,
		\item $\VariablesInit : \Variables \to \Intervals(\setR)$ is the initial set of variable valuations,
		\item $\Param$ is a finite set of parameters,
		\item $\invariant$ is the invariant, assigning to every $\loc \in \Loc$ a parametric linear constraint $\invariant(\loc)$ over $\Variables \cup \Param$,
		\item $\flow$ is the flow (or rate), assigning to every $\loc \in \Loc$ and $\variable \in \Variables$ a flow $\flow(\loc, \variable) \in \Intervals(\setR)$,
		\item $\Edges$ is a finite set of edges $\edge = (\loc,\guard,\action,\resets,\loc')$
		where~$\loc,\loc'\in \Loc$ are the source and target locations, $\action \in \Actions$, $\resets \subseteq \Variables$ is a set of variables to be reset, and $\guard$ is a parametric linear constraint over $\Variables \cup \Param$.
	\end{oneenumerate}%
\end{definition}

\paragraph{Parallel composition}
RHAs can be \emph{composed} using synchronized product (see \eg{} \cite[Definition~4]{Raskin05}) in a way similar to finite-state automata.
The synchronized product of $n$ RHAs $\A_i, i \in \{ 1, \dots, n \}$, denoted by $\A_1 \parallel \A_2 \parallel \cdots \parallel \A_n$, is\LongVersion{ known to be} an RHA~\cite{HPR94}.
Of importance is that, in a composed location, the global flow constraint is the \emph{intersection} of the local component flow constraints.

\LongVersion{\paragraph{Concrete semantics}}
We do not give the concrete semantics of this formalism, as we will manipulate a subclass called parametric linear multi-rate automaton (PLMA).

\section{Parametric linear multi-rate automata}\label{section:PLMA}

Timed automata extend finite-state automata with clocks (\ie{} real-valued variables evolving at the same constant rate~1), that can be compared with integer constants along transitions (``guards'') or within locations (``invariants'').
Parametric timed automata (PTAs) extend TAs with parameters within guards and invariants in place of integer constants~\cite{AHV93}, \ie{} allowing inequalities of the form $\variable \compOp \param$ (simple guards) or sometimes $\variable - \variable' \compOp \param$ (diagonal constraints), where $\variable, \variable' \in \Variables$ and $\param \in \Param$.
Here, we extend PTAs notably with:
\begin{ienumerate}%
	\item multi-rate clocks (called \emph{variables}), \ie{} each clock can have an arbitrary (but constant) rational rate in each location; and
	\item linear constraints over variables and parameters, instead of the usual definition $\variable \compOp \param$.
\end{ienumerate}%
We first define parametric linear multi-rate automata (PLMA) with their syntax (\cref{ss:syntax}) and semantics (\cref{ss:semantics});
we then propose two other subformalisms of RHAs (\cref{ss:subclasses}) used subsequently in this paper.

\subsection{Syntax}\label{ss:syntax}

We extend (P)TAs with (constant) \emph{flows}; in the absence of timing parameters, this formalism is usually called \emph{multi-rate timed automata}~\cite{ACHHHNOSY95,DY95}.
Also note that, different from TA clocks, our variables can possibly turn \emph{negative}.
In addition, we extend the usual syntax of clock guards to our aforementioned definition of \emph{parametric linear constraints}.

\begin{definition}[PLMA]\label{def:PLMA}
	An RHA \mbox{$\A = (\Actions, \Loc, \locinit, \LocsFinal, \Variables, \VariablesInit, \Param, \invariant, \flow, \Edges)$} is a \emph{parametric linear multi-rate automaton} (PLMA) if:
$\forall \loc \in \Loc, \forall \variable \in \Variables : \flow(\loc, \variable) \in \setQ$.
\end{definition}

That is, a PLMA is an RHA where %
all flows are constant.
Observe that the flow is taken in~$\setQ$, which includes negative rates and zero-rates (also called \emph{stopwatches}~\cite{CL00}).
\LongVersion{

}%
A PLMA is \emph{strongly deterministic} if
$\forall \loc \in \Loc, \forall \action \in \Actions, |\{ (\loc_1',\guard',\action',\resets',\loc_2') \in \Edges \mid \loc_1' = \loc \land \action' = \action \}| \leq 1$.
\begin{figure*}[tb]
 
	\centering
	 \footnotesize

	 \scalebox{.8}{
	\begin{tikzpicture}[pta, scale=1, xscale=2, yscale=1.5]
 
		\node[] at (-1.5, 0) (preinit) {};
 
		\node[location] at (0, 0) (l1) [align=center] {$\loc_1$
			\\$\dot{\variablei{1}} = 2$
			\\$\dot{\variablei{2}} = 3$
			};
 
		\node[location] at (2, 0) (l2) [align=center] {$\loc_2$
			\\$\dot{\variablei{1}} = 1$
			\\$\dot{\variablei{2}} = 0$
			};
 
		\node[location, accepting] at (4, 0) (l3) [align=center] {$\loc_3$
			\\$\dot{\variablei{1}} = 1$
			\\$\dot{\variablei{2}} = 1$
			};
 
		\node[invariant, above=of l1, yshift=0] {$\variablei{1} \leq 10$};
		
		\node[invariant, above=of l2, yshift=0] {$\variablei{1} \leq 3$};
		
		\path (preinit) edge[] node[above]{$\variablei{1} \assign 0$} node[below]{$\variablei{2} \assign [-2, 2]$} (l1);
		
		\path
			(l1) edge[] node[above,align=center]{$2 \times \variablei{1} > \variablei{2} + 2$\\$\actioni{1}$} node[below]{$\variablei{1} \assign 0$} (l2)
			(l2) edge[] node[above,align=center]{$\variablei{1} = 3 $\\$\land \paramp - 3 \leq \variablei{2} \leq \paramp + 1$} node[below]{$\actioni{2}$} (l3)
		;

	\end{tikzpicture}
	}
	\caption{A PLMA example}
	\label{figure:example-PLMA}

\end{figure*}
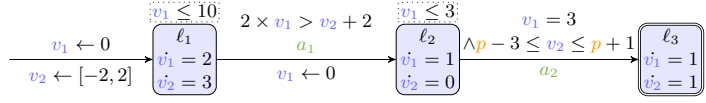
\begin{example}
	Consider the PLMA in \cref{figure:example-PLMA}.
	In the PLMA figures, we use notation $\dot{\variable_i} = d$ in location~$\loc_j$ to denote $\flow(\loc_j, \signal_i) = d$.
	This PLMA contains two variables $\variable_1$ and~$\variable_2$, and one parameter~$\param$.
	$\loc_1$ is the initial location, while~$\loc_3$ is the (only) accepting location.
	We have $\VariablesInit(\variablei{1}) = [0,0]$ and $\VariablesInit(\variablei{2}) = [-2,2]$.
\end{example}

Given\LongVersion{ a parameter valuation}~$\pval$, we denote by $\valuate{\A}{\pval}$ the non-parametric structure where all occurrences of a parameter~$\param_i$ have been replaced by~$\pval(\param_i)$.
We call such a structure a \emph{linear multi-rate automaton} (LMA).
Note that, whenever
	all rates are~1 ($\forall \loc \in \Loc, \forall \variable \in \Variables, \flow(\loc, \variable) = 1$),
	all guards and invariants are of the form $\variable \compOp d$, $d \in \setQplus$,
	and
	all variables are initially~0 (\ie{} $\forall \variable \in \Variables : \VariablesInit(\variable) = \{ 0 \}$),
then
the resulting structure is a \emph{timed automaton}~\cite{AD94}.\LongVersion{\footnote{%
	Strictly speaking, a TA requires $d \in \setN$; however, using an appropriate rescaling of the constants (by multiplying all constants in $\valuate{\A}{\pval}$ by the least common multiple of their denominators), we obtain an equivalent (integer-valued) TA.
}}
\subsection{Semantics}\label{ss:semantics}
\subsubsection{Concrete semantics of LMAs}
The semantics of LMAs is close to that of multi-rate automata, extended with linear constraints over variables.

\begin{definition}[Semantics of an LMA]
	Given a PLMA $\A = (\Actions, \Loc, \locinit, \LocsFinal, \Variables, \VariablesInit, \Param, \invariant, \flow, \Edges)$,
	and a parameter valuation~\(\pval\),
	the semantics of $\valuate{\A}{\pval}$ is given by the timed transition system (TTS) $(\States, \Sinit, \flecheRel)$, with%
	\begin{itemize}
		\item $\States = \{ (\loc, \vval) \in \Loc \times \setR^\VariablesCard \mid \vval \models \valuate{\invariant(\loc)}{\pval} \}$, %

		\item $\Sinit = \{(\locinit, \vval) \mid \vval \models \valuate{\invariant(\locinit)}{\pval} \land \forall \variable : \vval(\variable) \in \VariablesInit(\variable) \} $,
		
		\item  $\flecheRel$ consists of the discrete and (continuous) delay transition relations:
		\begin{enumerate}
			\item discrete transitions: $(\loc,\vval) \longueflecheRel{\edge} (\loc',\vval')$, %
				if $(\loc, \vval) , (\loc',\vval') \in \States$, and there exists $\edge = (\loc,\guard,\action,\resets,\loc') \in \Edges$, such that $\vval'= \reset{\vval}{\resets}$, and $\vval \models \valuate{\guard}{\pval}$.
			\item delay transitions: $(\loc,\vval) \longueflecheRel{d} (\loc, \timelapse(\vval, \flow(\loc), d))$, with $d \in \setRplus$, if $\forall d' \in [0, d], (\loc, \timelapse(\vval, \flow(\loc), d')) \in \States$.
		\end{enumerate}
	\end{itemize}
\end{definition}

Moreover we write $(\loc, \vval)\longuefleche{(d, \edge)} (\loc',\vval')$ for a delay transition followed by a discrete transition if
		$\exists  \vval'' :  (\loc,\vval) \longueflecheRel{d} (\loc,\vval'') \longueflecheRel{\edge} (\loc',\vval')$.

Given an LMA~$\valuate{\A}{\pval}$ with concrete semantics $(\States, \Sinit, \flecheRel)$, we refer to the states of~$\States$ as the \emph{concrete states} of~$\valuate{\A}{\pval}$.
A \emph{concrete run} of~$\valuate{\A}{\pval}$ is an alternating sequence of concrete states of $\valuate{\A}{\pval}$ and pairs of edges and delays starting from an initial state $\concstate_0 \in \Sinit$ of the form
$\concstate_0, (d_0, \edge_0), \concstate_1, \cdots$
with
$i = 0, 1, \dots$, $\edge_i \in \Edges$, $d_i \in \setRplus$ and
	$\concstate_i \longuefleche{(d_i, \edge_i)} \concstate_{i+1}$.
Given\LongVersion{ a state}~$\concstate=(\loc, \vval)$, we say that $\concstate$ is \emph{reachable} in~$\valuate{\A}{\pval}$ if $\concstate$ appears in a run of $\valuate{\A}{\pval}$.
By extension, we say that $\loc$ is reachable.
A run~$\concreterun$ is said to be \emph{accepting} if there exists $\loc \in \LocsFinal$ such that $\loc$ is reachable along~$\concreterun$.

A \emph{negative run} of~$\valuate{\A}{\pval}$ is an alternating sequence of states $(\loc_i, \vval_i)$ and pairs of edges and delays of the form
$(\loc_0, \vval_0), (d_0, \edge_0), (\loc_1, \vval_1), \cdots$
with
$i = 0, 1, \dots$, $\edge_i \in \Edges$ and $d_i \in \setRplus$,
which is not a concrete run of~$\valuate{\A}{\pval}$.
That is, there exists some~$i$ such that $(\loc_i, \vval_i)$ is not a concrete state of~$\valuate{\A}{\pval}$, or $(\loc_i, \vval_i) \longuefleche{(d_i, \edge_i)} (\loc_{i+1}, \vval_{i+1})$ does not belong to the semantics of~$\valuate{\A}{\pval}$.
To distinguish from negative runs, we will sometimes refer to concrete runs as \emph{positive} runs.

\begin{example}\label{example:runs}
	Consider again the PLMA~$\A$ in \cref{figure:example-PLMA}, and let $\pval$ be such that $\pval(\param) = 12$.
	Consider the following run~$\concreterun$ of $\valuate{\A}{\pval}$:
	\((\loc_1, (0, -2)) ,
		(\edge_1, 3.8) ,
	(\loc_2, (0, 9.4)) ,
		(\edge_2, 3) ,
	(\loc_3, (3, 9.4))\), where
		$\edge_1$ is the edge from $\loc_1$ to~$\loc_2$ in \cref{figure:example-PLMA},
		and
		$\edge_2$ is the edge from $\loc_2$ to~$\loc_3$.
	(As an abuse of notation, we write $(\loc_0, (0, -2))$ for $(\loc_0, \vval_0)$ where $\vval_0(\variable_1) = 0$ and $\vval_0(\variable_2) = -2$.)
	Observe that, after $3.8$ time units in~$\loc_1$, we have $\variable_1 = 2 \times 3.8 = 7.6$ (which satisfies invariant~$\variable_1 \leq 10$) while $\variable_2 = 9.4$; therefore, guard $2 \times \variable_1 > \variable_2 + 2$ evaluates to $15.2 > 11.4$, and therefore the transition to~$\loc_2$ can be taken.
	After 3 time units in~$\loc_2$, not modifying the value of~$\variable_2$ as $\flow(\loc_2, \variable_2) = 0$, the guard to~$\loc_3$ is satisfied as $9.4 \in [9, 13]$ (recall that~$\pval(\param) = 12$).
	\LongVersion{%
	
	}%
	$\concreterun$ is accepting as it ends in\LongVersion{ the accepting location}~$\loc_3$.

	Now consider the following alternative sequence~$\concreterun'$:
	\((\loc_1, (0, 0)) ,
		(\edge_1, 5) ,
	(\loc_2, (0, 15)) ,
		(\edge_2, 3) ,
	(\loc_3, (3, 15))\).
	This sequence is a \emph{negative} run of $\valuate{\A}{\pval}$, as the transition via~$\edge_2$ cannot be taken for this valuation ($15 \notin [9, 13]$).
	However, $\concreterun'$ is a positive run of~$\valuate{\A}{\pval'}$, where $\pval'(\param) = 14.5$.
	
\end{example}

A graphical representation of (positive and negative) runs focusing on the \emph{evolution of the variables over time} can be obtained directly from the runs.
This graphical representation is made of $\VariablesCard$ lines (where $\VariablesCard$ denotes the variables cardinality) obtained as follows:
given a (positive or negative) run $(\loc_0, \vval_0), (d_0, \edge_0), (\loc_1, \vval_1), \cdots$,
given a variable~$\variable$,
the initial point is $(0, \vval_0(\variable))$.
\LongVersion{%
	Then, for each $i \geq 0$, we add a point
	$\big(\tau_i + d_i, \vval_i(\variable) + d_i \times \flow(\loc_i)(\variable)\big)$
	and a point 
	$\big(\tau_{i+1}, \vval_{i+1}(\variable)\big)$,
		where $\tau_i$ is the absolute date at which $\loc_i$ is entered, \ie{} $\tau_i = \sum_{j = 0}^{j < i}d_j$.
}%
That is, each variable~$\variable$ defines graphically a non-necessarily differentiable piecewise linear function.

\begin{example}
	Consider the first run~$\concreterun$ from \cref{example:runs}.
	Its associated graphical representation is given in \cref{figure:run-graphical}.
\end{example}
\begin{figure}[tb]
	\begin{minipage}[b]{0.26\linewidth}

		\centering
		\scriptsize

		\scalebox{.8}{
		\begin{tikzpicture}[scale=1, xscale=.33, yscale=.12]

			\draw[axis] (0,0) -- (8, 0) node[anchor=north west] {$t$};
			\draw[axis] (0, -2.5) -- (0, 10) node[anchor=south east] {};
			\foreach \x in {1,2,...,7}
				\draw[draw=black] (\x, 1em) -- (\x, -1em) node[anchor=north] {$\x$};
			\foreach \y in {-2,0,2,...,8}
				\draw (.5em, \y) -- (-.5em, \y) node[anchor=east] {$\y$};

			\draw[signalA] (0, 0) -- (3.8, 7.6) -- (3.8, 0) -- (6.8, 3) node[right]{$\variablei{1}$};

			\draw[signalB] (0, -2) -- (3.8, 9.4) -- (6.8, 9.4) node[right]{$\variablei{2}$};

		\end{tikzpicture}
		}
		\caption{Graphical run}
		\label{figure:run-graphical}

	\end{minipage}
	\begin{minipage}[b]{0.72\linewidth}

		\centering
		\scriptsize

		\scalebox{.9}{
		\begin{tikzpicture}[pta, scale=1, xscale=1.5, yscale=1.5]
	
			\node[location, initial] at (0, 0) (l1) [align=center] {$\loc_1$
				\\$0 \leq \variablei{1} \leq 10 \land $
				\\$-4 \leq 3 \times \variablei{1} - 2 \times \variablei{2} \leq 4$
				\\$\land \paramp \geq 0$
				};
	
			\node[location] at (2, 0) (l2) [align=center] {$\loc_2$
				\\$0 \leq \variablei{1} \leq 3 \land $
				\\$-2 < \variablei{2} \leq 17$
				\\$\land \paramp \geq 0$
				};

	\node[location, accepting] at (4, 0) (l3) [align=center] {$\loc_3$
				\\$\variablei{1} \geq 3$
				\\$\land -5 < \variablei{2} - \variablei{1} \leq 14 \land$
				\\$ \paramp - 6 \leq \variablei{2} - \variablei{1} \leq \paramp - 2$
				\\$\land \param \geq 0$
				};
	
			\path
				(l1) edge[] node[above]{$\edge_1$} (l2)
				(l2) edge[] node[above]{$\edge_2$} (l3)
			;

		\end{tikzpicture}
		}
		\caption{A parametric zone graph}
		\label{figure:example-PLMA:PZG}

	\end{minipage}
\end{figure}
\subsubsection{Symbolic semantics}\label{sss:symbolic}
Let us now define the symbolic semantics of PLMAs, as an extension of the semantics of PTAs (see \eg{} \cite{HRSV02,ACEF09,JLR15}) to multi-rates and linear constraints.

\paragraph{Constraints}
We first need to define operations on constraints.
Given a parameter valuation $\pval$ and a variable valuation $\vval$, we denote by $\wv{\vval}{\pval}$ the valuation over $\Variables \cup \Param$ such that
for all variables $\variable$, $\valuate{\variable}{\wv{\vval}{\pval}}=\valuate{\variable}{\vval}$
and
for all parameters $\param$, $\valuate{\param}{\wv{\vval}{\pval}}=\valuate{\param}{\pval}$.
Given a parametric linear constraint~$\Constraint$,
we use the notation $\wv{\vval}{\pval} \models \Constraint$ to indicate that
	$\vval \models \pval(\Constraint)$.
We say that $\Constraint$ is \emph{satisfiable} if $\exists \vval, \pval \text{ s.t.\ } \wv{\vval}{\pval} \models \Constraint$.
We will often use geometrical concepts for constraints;
in particular, whenever $\wv{\vval}{\pval} \models \Constraint$, then the valuation $\wv{\vval}{\pval}$ can be seen as a \emph{point belonging to the polyhedron}~$\Constraint$.

We define the \emph{time elapsing} of~$\Constraint$ \wrt{} flow~$\flow : \Variables \to \setQ
$, denoted by $\symbtimelapse{\Constraint}{\flow}$, as the constraint over $\Variables$ and $\Param$ obtained from~$\Constraint$ by delaying all variables by an arbitrary amount of time according to~$\flow$.
That is,
\(\wv{\vval'}{\pval} \models \symbtimelapse{\Constraint}{\flow} \text{ iff } \exists \vval : \Variables \to \setR, \exists d \in \setR \text { s.t.\ } \wv{\vval}{\pval} \models \Constraint \land \vval' = \timelapse(\vval, \flow, d) \text{.}\)

Given $\resets \subseteq \Variables$, we define the \emph{reset} of~$\Constraint$, denoted by $\reset{\Constraint}{\resets}$, as the constraint obtained from~$\Constraint$ by resetting to~0 the variables in~$\resets$, and keeping the other variables unchanged.
We denote by $\projectP{\Constraint}$ the projection of~$\Constraint$ onto~$\Param$, \ie{} obtained by eliminating the variables not in~$\Param$ (\eg{} using Fourier-Motzkin\LongVersion{~\cite{Schrijver86})}.
The application of these operation to a linear constraint yields a linear constraint; this can be computed efficiently using operations on polyhedra~\cite{BHZ08}.

\LongVersion{%
\begin{definition}[Symbolic state]
}
	A symbolic state is a pair $(\loc, \Constraint)$ where $\loc \in \Loc$ is a location, and $\Constraint$ is a linear constraint called a \emph{parametric zone}.
\LongVersion{%
\end{definition}
}
\begin{definition}[Symbolic semantics]\label{def:PLMA:symbolic}
	Given a PLMA $\A = (\Actions, \Loc, \locinit, \LocsFinal, \Variables, \VariablesInit, \Param, \invariant, \flow, \Edges)$,
	the symbolic semantics of~$\A$ is the labeled transition system called \emph{parametric zone graph}
	$ \PZG(\A) = ( \Edges, \SymbState, \symbstateinit, \symbtrans )$, with
	\begin{itemize}
		\item $\SymbState = \{ (\loc, \Constraint) \mid \Constraint \subseteq \invariant(\loc) \}$, %
		\LongVersion{\item} $\symbstateinit = \big(\locinit, \symbtimelapse{(\bigwedge_{1 \leq i\leq\VariablesCard} %
			\variable_i \in \VariablesInit(\variable_i)
		)}{\flow(\locinit)} \land \invariant(\locinit) \big)$,
				\LongVersion{and}
		\item $\big((\loc, \Constraint), \edge, (\loc', \Constraint')\big) \in \symbtrans $ if $\edge = (\loc,\guard,\action,\resets,\loc') \in \Edges$ and
			\(\Constraint' = \symbtimelapse{\big(\reset{(\Constraint \land \guard)}{\resets}\land \invariant(\loc')\big )}{\flow(\loc')} \land \invariant(\loc')\)
			with $\Constraint'$ satisfiable.
	\end{itemize}

\end{definition}

That is, in the parametric zone graph, nodes are symbolic states, and arcs are labeled by edges of the original PLMA.
Observe that, as in PTAs, a symbolic state contains all the valuations after time elapsing (instead of just the valuations after a discrete transition).

If $\big((\loc, \Constraint), \edge, (\loc', \Constraint')\big) \in \symbtrans $, we write $\Succ(\symbstate, \edge) = (\loc', \Constraint')$, where $\symbstate = (\loc, \Constraint)$.
By extension, we write $\Succ(\symbstate)$ for $\cup_{\edge \in \Edges} \Succ(\symbstate, \edge)$.

A \emph{symbolic run} $\symbrun$ of~$\A$ is an alternating sequence of symbolic states of $\A$ and edges starting from the initial state $\symbstateinit$ of the form
$\symbstate_0, \edge_0, \symbstate_1, \cdots$
with
$i = 0, 1, \dots$, $\edge_i \in \Edges$, and
	$\Succ(\symbstate_i, \edge_i) = \symbstate_{i+1}$.
(The symbolic runs of~$\A$ are the runs of~$\PZG(\A)$.)
$\edgeat(\symbrun,k)$ denotes %
	$\edge_{k}$,
and
$\stateat(\symbrun, k)$ denotes $\symbstate_{k}$.
When~$\symbrun$ is finite, $|\symbrun|$ denotes its \emph{length}, \ie{} its number of edges (therefore, a finite symbolic run contains $|\symbrun+1|$ symbolic states).
\begin{example}
	Consider again the PLMA~$\A$ in \cref{figure:example-PLMA}.
	Then, $\PZG(\A)$ (limited to its reachable states) is given in \cref{figure:example-PLMA:PZG}.
	The constraints in each location give both the admissible valuations for~$\paramp$ for which this location is reachable, and a condition over the continuous variables $\variablei{1}$ and~$\variablei{2}$ to remain in this location.
	Note that (the reachable part of) this PZG is finite, which is not necessarily the case in general.
\end{example}
\subsection{Two other subclasses of RHAs: PTASs and SBAs}\label{ss:subclasses}

\LongVersion{%
\subsubsection{Parametric timed automata with signals}
}
\begin{definition}\label{definition:PTAS}
	An RHA~$\A = (\Actions, \Loc, \locinit, \LocsFinal, \Variables, \VariablesInit, \Param, \invariant, \flow, \Edges)$ is a \emph{parametric timed automaton with signals} (PTAS) if:
	\begin{enumerate}
		\item the set of variables is partitioned into $\Variables = \Clocks \uplus \Signals$, where $\Clocks$ is a set of standard TA \emph{clocks} (\ie{} variables with rates~1), and $\Signals$ is a set of \emph{signals};

			\item all clock rates are~1, \ie{} $\forall \loc \in \Loc, \forall \clock \in \Clocks, \flow(\loc, \clock) = 1$;

		\item signals satisfy the following constraints:
		\begin{enumerate}
			\item all signal rates are unconstrained, \ie{} $\forall \loc \in \Loc, \forall \signal \in \Signals, \flow(\loc, \signal) = \setR$;
			
			\item a signal cannot be reset, \ie{} $\forall (\loc,\guard, \action, \resets, \loc') \in \Edges, \forall \variable \in \resets : \variable \notin \Signals$; and
			
			\item each parametric linear inequality in guards and invariants cannot involve both a standard clock from~$\Clocks$ and a signal from~$\Signals$ (\ie{} comparisons of the form $\clock \compOp \signal$, with $\clock \in \Clocks$ and $\signal \in \Signals$, are not allowed).
		\end{enumerate}

	\end{enumerate}
\end{definition}

Observe that, since the signal rates are $= \setR$, the formalism of PTAS is not a subclass of PLMAs (see \cref{figure:formalisms}), as this latter formalism imposes $\flow(\loc, \signal) = d$ for some $d \in \setQ$.
However, in practice, a PTAS will always be composed (using synchronized product) with a set of PLMAs (actually SBAs, see below) constraining the rate of signals (see \cref{lemma:product} below). %
\begin{example}
	Consider the PTAS in \cref{figure:example-PTAS:param}.
	Its clock set is $\Clocks = \{ \clock \}$ while its signal set is $\Signals = \{ \signal_1, \signal_2 \}$.
	The set of parameters is $\Param = \{ \param \}$.
	We have $\flow(\loc_1, \clock) = \flow(\loc_2, \clock) = \flow(\locTarget, \clock) = 1$ (not explicitly depicted in \cref{figure:example-PTAS:param}).
\end{example}

\LongVersion{%
\subsubsection{Signal bounding automata}
}
Second, we define a signal bounding automaton as a special LMA used to constrain the admissible behaviors of a signal.
Therefore, it contains a single variable (actually a signal), no parameter, and no reset.

\begin{definition}\label{definition:SBA}
	A PLMA~$\A = (\Actions, \Loc, \locinit, \LocsFinal, \Signals, \Param, \invariant, \flow, \Edges)$ is a \emph{signal bounding automaton} (SBA) if:
	\begin{oneenumerate}%
		\item $\Param = \emptyset$;
		\item $|\Signals| = 1$; and
		\item no resets are allowed, \ie{} $\forall (\loc,\guard, \action, \resets, \loc') \in \Edges, \resets = \emptyset$.
	\end{oneenumerate}
\end{definition}
\begin{example}
	An example of SBA is given in \cref{figure:example-SBA-fastslow}, where $\Signals = \{ \signal_1 \}$.
	In the SBA figures, we use notation $\dot{\signal_1} = d$ in location~$\loc$ to denote $\flow(\loc, \signal_1) = d$.
\end{example}
\newcommand{\lemmaProduct}{%
	Let $\A$ be a PTAS with $n$ signals.
	Let $\A_i, i \in \{ 1, \dots , n \}$ be $n$ SBAs such that $\A_i$ contains a signal variable~$\signal_i$.
	Then $\A \parallel \A_1 \parallel \cdots \parallel \A_n $ is a PLMA.
}
\begin{lemma}\label{lemma:product}
	\lemmaProduct{}
\end{lemma}
\newcommand{\lemmaProductProof}{%
	First note that the only reason why a PTAS is \emph{not} a PLMA is because the flow of signal variables are unconstrained ($\forall \loc \in \Loc, \forall \signal \in \Signals, \flow(\loc, \signal) = \setR$), while the definition of PLMAs (\cref{def:PLMA}) requires all flows to be (arbitrary) \emph{constants}.
	Let $(\loc, \loc_1, \cdots, \loc_n)$ be a location of $\A \parallel \A_1 \parallel \cdots \parallel \A_n $.
	From the parallel composition of RHAs, the flow of this composed location is the intersection of the flows of each of the locations $\loc, \loc_1, \cdots, \loc_n$.
	Given a signal~$\signal_i$, the flow of~$\signal_i$ is unconstrained in~$\A$ (\ie{} $\flow(\loc, \signal) = \setR$ by \cref{definition:PTAS}), and is unconstrained in all SBAs---except in $\A_i$ where it is a constant, since \cref{definition:SBA} defines SBAs as a subclass of PLMAs.
	Therefore, the intersection of all flows for~$\signal_i$ is equal to this constant.
	And therefore, $\A \parallel \A_1 \parallel \cdots \parallel \A_n $ is a PLMA.
}
\LongVersion{%
	\begin{proof}
	\lemmaProduct{}
	\end{proof}
}

In practice, SBAs can also involve \LongVersion{one or more }clocks, \eg{} to mesure time between signal changes.
This is both harmless in theory, and allowed by our implementation.

\section{Problem}\label{section:problem}

\LongVersion{
\subsection{Framework}
}

\subsubsection{Expressing specifications over signals}
In our work, we consider as first input a PTAS featuring a set of~$n$ signals, and acting as a \emph{specification automaton}.
Given a parameter valuation~$\pval$ and a specification expressed as a PTAS~$\A$ with accepting locations~$\LocsFinal$, the specification is satisfied iff $\LocsFinal$ is reachable in $\valuate{\A}{\pval}$.

\LongVersion{%
\begin{example}
	Consider again the PTAS in \cref{figure:example-PTAS:motivating}, featuring one clock~$\clockx$ and two signals~$\signali{1}$ and~$\signali{2}$.
	Location $\locTarget$ is reachable whenever the following property is satisfied: ``whenever signal $\signali{1}$ is larger than 50, then within at most 15 time units, it holds that $\signali{1} \geq 3 \times \signali{2}$ and then, within at most 20 more time units, both signals are equal''.
\end{example}
}

\subsubsection{Bounding signal behaviors}
In order to define the admissible behaviors of the signals, we also consider an SBA for each of the signals used in the PTAS.

\LongVersion{%
\begin{example}
	Consider again the SBA in \cref{figure:example-SBA-fastslow}.
	This SBA constrains the behavior of signal~$\signali{1}$:
		this signal can either increase (with flow~1 or flow~3), or decrease (flow~$-1$ or~$-3$).
		This automaton also constrains $\signali{1}$ to remain non-negative.
\end{example}
}

\LongVersion{
\subsection{Formal problem}
}
Since the specification (given by a PTAS) is parametric, we first aim at deriving concrete parameter valuations for which the specification is valid, \ie{} for which one accepting state is reachable.
Second, for a given concrete valuation, we aim at deriving \LongVersion{a set of }concrete accepting positive runs, as well as negative runs.

\smallskip

\defProblem
	{Specification exemplification}
	{A PTAS~$\A$ featuring $n$ signals, and $n$ SBAs $\A_i, i \in \{ 1, \dots , n \}$}
	{Exhibit a set of parameter valuations~$\pval$ and a set of concrete accepting positive runs and negative runs of $\valuate{(\A \parallel \A_1 \parallel \cdots \parallel \A_n)}{\pval}$}

\LongVersion{
\subsubsection{Assumptions}
}
Recall that our general approach is given in \cref{figure:framework}.
\LongVersion{%
	In \cref{section:approach}, we come to our main approach for exemplifying specifications over signals with a bounded behavior; to this end, we propose a method to derive concrete parameter valuations and concrete runs for a PLMA.

}%
In our approach, we make the following assumption (only required when computing \emph{negative} runs):

\begin{assumption}\label{assumption:strong-determinism}
	The PTAS and SBAs must be strongly deterministic\LongVersion{, and feature no silent transition}.
\end{assumption}

\LongVersion{(Silent actions, also called $\epsilon$-transitions, are unobservable actions---not defined in \cref{def:RHA} anyway.)}

\section{Exemplifying bounded signal specifications}\label{section:approach}
\begin{algorithm}[tb]
	\Input{A PLMA  with symbolic initial state $\symbstateinit$ and accepting locations $\LocsFinal$}
	\Output{A set of negative runs and positive runs}
	
	Explore $\PZG(\A)$ until a state $(\locTarget, \Constraint)$ is found, for some $\locTarget \in \LocsFinal$ and some~$\Constraint$\nllabel{algo:explore:explore}
	
	\tcc{Pick a run $\symbrun$ from $\symbstateinit$ to $(\locTarget, \Constraint)$}
	
	$\symbrun \assign \PickSymbRun(\PZG, \symbstateinit, (\locTarget, \Constraint))$ \nllabel{algo:explore:pick}
	
	\Return{$\exemplifythree(\A, \symbrun)$}

	\caption{Main algorithm $\algoMain(\A)$}
	\label{algo:explore}
\end{algorithm}

We propose in this section a heuristics-based method to exemplify runs for an arbitrary PLMA.
\LongVersion{

}%
The entry point is $\algoMain$ in \cref{algo:explore}.
We first explore the PZG until a target state is found (\cref{algo:explore:explore}).
Then, we exhibit a symbolic run from the initial state $\symbstateinit$ to the target state (\cref{algo:explore:pick}).
Finally, \cref{algo:explore} calls $\exemplifythree$, given in \cref{algo:exhibit}, that returns (up to) 3~concrete runs: one positive run together with a concrete parameter valuation, one negative run for a different parameter valuation, and one negative run for the same parameter valuation.
\LongVersion{

}%
Let us explain these steps\LongVersion{ in more details in the following}.

\subsection{Exploration and symbolic run exhibition}\label{ss:PZG}

The construction of the PZG is made on-the-fly, using \cref{def:PLMA:symbolic}.
In our implementation, this is done using a breadth-first search (BFS) manner\LongVersion{, but any other exploration order can be used}.

Then, the function $\PickSymbRun$ takes as argument the PZG~$\PZG$, the initial state~$\symbstateinit$, and the target state (here $(\locTarget, \Constraint)$), and returns a symbolic run from~$\symbstateinit$ to~$(\locTarget, \Constraint)$ in~$\PZG$.
The actual function (not given in this paper) is implemented in a straightforward manner in our toolkit using a backward analysis in~$\PZG$ from~$(\locTarget, \Constraint)$ to~$\symbstateinit$.
The exhibited symbolic run is not necessarily unique and, as heuristics, we use a shortest run (again, not necessarily unique), with ``shortest'' to be understood as the number of discrete steps.
Alternative definitions could be used (\eg{} minimal-time run~\cite{ABPP19}%
)\LongVersion{---but are not implemented in our toolkit}.

\LongVersion{
\subsection{Deriving a final concrete valuation}\label{ss:valuation}
}

\ShortVersion{\medskip}

After exhibiting a symbolic run, our next step is to derive \emph{concrete} runs from that symbolic run.
This is the purpose of $\exemplifythree(\A, \symbrun)$, given in \cref{algo:exhibit}.

\begin{algorithm}[tb]
	\Input{A PLMA~$\A$, a symbolic run~$\symbrun$ from $\symbstateinit$ to $(\locTarget, \Constraint)$}
	\Output{A set $\ConcreteRuns$ of concrete negative runs and positive runs}
	
	\LongVersion{\BlankLine}
	
	$\ConcreteRuns \assign \emptyset$
	
	\LongVersion{\BlankLine}
	
	\tcc{Part 1: positive run}
	
	$\wv{\vval}{\pval} \assign \exhibitPoint(\Constraint)$ \nllabel{algo:exhibit:pick}
	
	$ \concreterun \assign \reconstructPos(\A, \symbrun, |\symbrun|, (\locTarget, \wv{\vval}{\pval}))$\nllabel{algo:exhibit:reconstructPos}
	\ShortVersion{\ \ ; \ \ }\LongVersion{
	
	}
	$\ConcreteRuns \assign \ConcreteRuns \cup \{ \concreterun \}$

	\LongVersion{\BlankLine}

	\tcc{Part 2a: negative run (different parameter valuation)}
	
	\If{$\hasPdeadlock(\symbrun)$\nllabel{algo:exhibit:ifPD}}{
		$(\pval_i, (\loc_i, \Constraint_i)) \assign \findPdeadlock(\symbrun)$
		\ShortVersion{\ \ ; \ \ }\LongVersion{
	
		}
		$\vval_i \assign \exhibitPoint(\valuate{\Constraint_i}{\pval_i})$
		
		$ \concreterunprefix \assign \reconstructPos(\A, \symbrun, i, (\loc_i, \wv{\vval_i}{\pval_i}))$

		$ \concreterunsuffix \assign \constructNeg(\A, \symbrun, i, |\symbrun|, \wv{\vval_i}{\pval_i})$\nllabel{algo:exhibit:P:constructNeg}
		
		$\concreterun \assign \concreterunprefix \runconcat \concreterunsuffix$
		\ShortVersion{\ \ ; \ \ }\LongVersion{
	
		}
		$\ConcreteRuns \assign \ConcreteRuns \cup \{ \concreterun \}$
		\label{algo:exhibit:ifPDend}
	}

	\LongVersion{\BlankLine}

	\tcc{Part 2b: negative run (same parameter valuation)}

	\If{$\hasXdeadlock(\symbrun)$\nllabel{algo:exhibit:ifXD}}{
		$(\vval_i, (\loc_i, \Constraint_i)) \assign \findXdeadlock(\symbrun, \pval)$
		
		$ \concreterunprefix \assign \reconstructPos(\A, \symbrun, i, (\loc_i, \wv{\vval_i}{\pval}))$

		$ \concreterunsuffix \assign \constructNeg(\A, \symbrun, i, |\symbrun|, \wv{\vval_i}{\pval})$\nllabel{algo:exhibit:X:constructNeg}
		
		$\concreterun \assign \concreterunprefix \runconcat \concreterunsuffix$
		\ShortVersion{\ \ ; \ \ }\LongVersion{
	
		}
		$\ConcreteRuns \assign \ConcreteRuns \cup \{ \concreterun \}$
		\label{algo:exhibit:ifXDend}
	}
	
	\LongVersion{\BlankLine}
	
	\Return{$\ConcreteRuns$}

	\caption{$\exemplifythree(\A, \symbrun)$: Exemplifying 3 concrete runs}
	\label{algo:exhibit}
\end{algorithm}

We first explain \cref{algo:exhibit} as a whole, and then proceed to subfunctions in the following.
The first step in $\exemplifythree$ is to exhibit a ``point'', \ie{} a concrete variable and parameter valuation in the target state constraint~$\Constraint$ (\cref{algo:exhibit:pick}).
Since~$\Constraint$ is a polyhedron, we use a dedicated function $\exhibitPoint(\Constraint)$.
There is no theoretical difficulty in exhibiting a concrete point in a polyhedron; however, our dedicated function must both be efficient and yield a valuation which is as ``human-friendly'' as possible, \ie{} avoiding random rational numbers and avoiding as much as possible to select ``0'' if another suitable valuation exists.
The body of our function $\exhibitPoint$ is given in \LongVersion{\cref{appendix:exhibitPoint}}\ShortVersion{\cite{AWUH22report}}.

\subsection{Exhibiting concrete example runs}

We then reconstruct a concrete positive run (\cref{algo:exhibit:reconstructPos} in \cref{algo:exhibit}) from the point $\wv{\vval}{\pval}$ that was just exhibited in the final constraint.
This function $\reconstructPos$ poses no specific theoretical difficulty, but yields some practical subtleties, discussed in \LongVersion{\cref{ss:algo:constructPos}}\ShortVersion{\cite{AWUH22report}}.
Note that it is always possible to reconstruct a concrete run from a symbolic run.

The second part of \cref{algo:exhibit} (\cref{algo:exhibit:ifPD}--\cref{algo:exhibit:ifPDend})
consists in exhibiting a negative run (based on~\symbrun) for a different parameter valuation than the one ($\pval$) exhibited in the first part of the algorithm.
The heuristics we use is to (try to) exhibit a parameter valuation that \emph{cannot} take one of the transitions of the symbolic run~$\symbrun$: this is a \emph{parametric deadlock}.
If such a valuation exists, then the projection onto the parameters of some constraints along the run~$\symbrun$ is \emph{shrinked}, \ie{} this run is possible for some parameter valuations up to some state, and then possible for less parameter valuations.

\LongVersion{%
\subsubsection{Parametric deadlocks}\label{sss:pdeadlocks}
}
Parametric deadlock checking was studied in, \eg{} \cite{Andre16,ALime17}, and $\findPdeadlock$ is basically based on these former works, except that we used the symbolic semantics of PLMAs instead of PTAs.
$\findPdeadlock$ attempts at exhibiting a parameter valuation that cannot pass one of the edges of a symbolic run~$\symbrun$.
\LongVersion{%
	In other words, it tries to exhibit a parameter valuation that is a member of a polyhedron at state~$i$, but not anymore at~$i+1$; therefore, there exists a concrete run in $\valuate{\A}{\pval}$ equivalent to~$\symbrun$ up to position~$i$, but this does not hold for~$i+1$.
}%
$\findPdeadlock$ is given in \LongVersion{\cref{appendix:algo:findPdeadlock}}\ShortVersion{\cite{AWUH22report}}.

The third part of \cref{algo:exhibit} (\cref{algo:exhibit:ifXD}--\cref{algo:exhibit:ifXDend}) consists in exhibiting a negative run for the same parameter valuation as the one ($\pval$) exhibited in the first part of the algorithm.
Our heuristics is as follows: we try to find a transition within~$\symbrun$ for which some variable valuation (for the parameter valuation~$\pval$) cannot take this transition.
This can come from an unsatisfied guard or invariant: this is a \emph{non-parametric deadlock}.

\LongVersion{%
In the following, we explain the subfunctions used in the above description of \cref{algo:exhibit}.
}

\LongVersion{%
\subsubsection{Non-parametric deadlocks}\label{sss:xdeadlocks}
}
$\findXdeadlock$ attempts to exhibit a variable valuation~$\vval$ and a symbolic state~$\symbstate$ of a symbolic run~$\symbrun$ such that there exists a deadlock after~$\symbstate$ for~$\vval$, \ie{} $\vval$ cannot take the edge following $\symbstate$ along~$\symbrun$, even after elapsing some time.
\LongVersion{%
This is typically the case of the following situations:
\begin{ienumerate}%
	\item A valuation ``missing'' the guard: \eg{} a symbolic state over $\variable$ (with $\flow(\variable) = 1$) constraining $\variable \geq 0$, with an outgoing guard $\variable = 2$; any value $\variable > 2$ is deadlocked, as there is no way to take the guard;
	\item A guard restraining the relationship between variables: \eg{} a symbolic state over $\variablei{1}$ and~$\variablei{2}$ (with $\flow(\variablei{1}) = 1$ and $\flow(\variablei{2}) = 2$) constraining $\variablei{1} \in [0, 2] \land \variablei{2} \in[0,2]$ with an outgoing guard $\variablei{1} = \variablei{2} = 2$: for example $\variablei{1} = \variablei{2} = 0$ cannot pass this guard, even after elapsing some time, due to the flow differences of $\variablei{1}$ and~$\variablei{2}$.
\end{ienumerate}%
}%
$\findXdeadlock$ is given in \LongVersion{\cref{appendix:algo:findXdeadlock}}\ShortVersion{\cite{AWUH22report}}.

\subsection{Exhibiting negative concrete example runs}\label{ss:negative}
The reconstruction of a negative run fragment is given in \cref{algo:constructNeg}.
It takes as arguments the start ($i$) and end ($j$) positions of the symbolic run~\symbrun{}, as well as the concrete valuation $\wv{\vval_i}{\pval_i}$ to start from at position~$i$.
\cref{algo:constructNeg} simply starts from the valuation~$\wv{\vval_i}{\pval_i}$, and takes the same discrete actions as in the symbolic run, but with an (arbitrary) duration~1:
that is, for each $k$ from~$i$ to~$j$, we add a transition $(\edgeat(\symbrun, k) , 1)$ (where $1$ denotes the duration), and we add the updated valuation $(\wv{\vval_i}{\pval_i} + (k-i))$, which is equal to $(\wv{\vval_i}{\pval_i}$ incremented by the number of transitions computed so far ($k-i)$).
Note that it would be possible to take any other duration than~1, and apply the resets as in the symbolic run.
The fact that this concrete run is an invalid run comes from the fact that the valuation $\wv{\vval_i}{\pval_i}$ is known to be unable to take the immediately following transition, as it is called at \cref{algo:exhibit:P:constructNeg,algo:exhibit:X:constructNeg} of \cref{algo:exhibit} where a parametric (resp.\ non-parametric) deadlock was exhibited.
\begin{algorithm}[tb]
	\Input{A PLMA~$\A$
		; A symbolic run~$\symbrun$ from $\symbstateinit$ to $(\locTarget, \Constraint)$
		; Start position~$i$ and end position~$j$
		; Starting valuation $\wv{\vval_i}{\pval_i}$}
	\Output{A concrete negative run fragment}
	
	$\concreterun \assign \wv{\vval_i}{\pval_i}$
	
	\lFor{$k = i$ \kwTo{} $j$}{
		$\concreterun \assign \concreterun , \big(\edgeat(\symbrun, k) , 1 \big) , \big(\wv{\vval_i}{\pval_i} + (k-i) \big)$
	}
	
	\Return{$\concreterun$}

	\caption{$\constructNeg(\A, \symbrun, i, j, \wv{\vval_i}{\pval_i})$: Reconstruct a negative run from a symbolic run fragment}
	\label{algo:constructNeg}
\end{algorithm}
\subsection{Formal result}\label{ss:formal}

Exemplifying runs for parametric timed formalisms is a very hard problem, as the mere existence of a parameter valuation for which a location is reachable in a PTA is undecidable~\cite{AHV93}.
While our method is mostly heuristics-based, we prove that, \emph{provided at least one parameter valuation allows to reach an accepting location\LongVersion{~$\locTarget \in \LocsFinal$}}, then our method is able to infer at least one (positive) concrete run.

\newcommand{\propositionTermination}{%
	Let $\A$ be a PLMA with accepting locations~$\LocsFinal$.
	Assume $\exists \pval : \valuate{\A}{\pval}$ reaches some~$\locTarget \in \LocsFinal$.
	Then, assuming a BFS computation of $\PZG(\A)$, $\algoMain(\A)$ terminates, and outputs at least one positive run.
}
\begin{proposition}\label{proposition:termination}
	\propositionTermination{}
\end{proposition}

Our algorithm has no guarantee to exhibit negative runs for several reasons:
notably,
	we use only heuristics, here based on deadlocks: there could be \emph{other} negative runs than those exhibited based on a (parametric\LongVersion{ or non-parametric}) deadlocks. %
Still, one can guarantee the following:

\newcommand{\propMasaki}{%
	Let $\A$ be a PLMA and $\symbrun$ be a symbolic run of $\valuate{\A}{\pval}$ with a parameter valuation $\pval$.
	Assume there is a concrete negative run due to parametric (resp.\ non-parametric) deadlock with the same discrete actions as $\symbrun$.
	Then, assuming a BFS computation of $\PZG(\A)$, $\exemplifythree(\A, \symbrun)$ outputs a concrete negative run due to parametric (resp.\ non-parametric) deadlock.
}
\begin{proposition}\label{proposition:Masaki}
	\propMasaki{}
\end{proposition}
\section{Proof of concept}\label{section:experiments}

We implemented our exemplification algorithm in \imitator{}~\cite{Andre21} (\href{https://github.com/imitator-model-checker/imitator/releases/tag/v3.3.0-alpha}{\nolinkurl{v.3.3-alpha}} ``Cheese Caramel au beurre salé'').\footnote{%
	Source code, models and results are available at \href{https://www.doi.org/10.5281/zenodo.6382893}{\nolinkurl{10.5281/zenodo.6382893}}.
}

All polyhedral operations are implemented using PPL~\cite{BHZ08}.
The approach takes as input a network of PLMAs, and attempts to output a set of runs and parameter valuations.
As a heuristics, we try to call up to 6~times \cref{algo:explore}, \ie{} we try to exhibit up to 6 symbolic runs, and then for each of them, following \cref{algo:exhibit}, we derive one parameter valuation and a concrete run, followed by a negative run for a different parameter valuation (if any) and a negative run for the same \LongVersion{parameter }valuation (if any).
All analyses terminate within a few seconds, including graphics generation.

All outputs are textual (in a JSON-like format); however, \imitator{} also automatically outputs basic graphics.
\ShortVersion{%
	The figures in this paper were however (manually) redrawn using \LaTeX{}.
}%
\LongVersion{%
	While graphics such as in \cref{figure:concrete-runs:running} were (manually) redrawn using \LaTeX{}, those in \cref{fig:ex2signals2:result} (\cref{appendix:ex2signals2:results}) are the exact output by \imitator{}.
}

\subsubsection{Extensions}
Thanks to the expressive power of \imitator{}, we can go beyond the formalism presented here.
Notably, arbitrary updates (not necessarily to~0, but to parameters, or other variables) are allowed;
also, Boolean variables can encode predicates, which can be seen as a simpler setting than signals (see below).

\paragraph{A non-parametric specification over Booleans}
Assume the following specification:
``whenever action $\actioni{1}$ occurs, then following a non-0 time, predicate~$P_1$ must hold;
then, strictly less than 3 time units later, $\actioni{2}$ occurs and predicate~$P_2$ must not hold''.
The PTAS encoding this specification is given in \cref{figure:bool:PTAS}; the SBA in \cref{figure:bool:SBA} simply allows both predicates to switch anytime between \BTrue{} and \BFalse{}.

We give two positive runs in \cref{figure:concrete-runs:bool:positive,figure:concrete-runs:bool:positive:4}
	and one negative run in \cref{figure:concrete-runs:bool:negative:5}.
Observe that the run in \cref{figure:concrete-runs:bool:negative:5} violates the specification because action $\actioni{2}$ occurs exactly in 3 time units (instead of $<3$ time units) after~$\styleact{\mathit{check}}$.

\LongVersion{%
	The full set of runs output by our toolkit is given in \cref{appendix:exBool:results}.
}

\begin{figure}[tb]
 
	\centering
	 \scriptsize

	\begin{subfigure}[b]{0.7\textwidth}
		\begin{tikzpicture}[pta, scale=1, xscale=1.25, yscale=1.5]
	
			\node[location, initial] at (0.6, 0) (l1) {$\loc_1$};
	
			\node[location] at (2, 0) (l2) {$\loc_2$};
	
			\node[location] at (4, 0) (l3) {$\loc_3$};
			\node[invariant, above=of l3] {$\clockx \leq 3$}; 
	
			\node[location, final] at (6, 0) (l4) {$\loc_4$};
				
			\path (l1) edge node[align=center]{%
				$\actioni{1}$} node[below]{$\clockx \assign 0$} (l2);

			\path (l2) edge node[align=center]{%
				$\styledisc{P_1} \land \clockx > 0$
				\\
				$\styleact{\mathit{check}}$} node[below]{$\clockx \assign 0$} (l3);

			\path (l3) edge node[align=center]{%
				$\neg \styledisc{P_2} \land \clockx < 3$ %
				\\
				$\actioni{2}$} (l4);

		\end{tikzpicture}
		\caption{PTAS}
		\label{figure:bool:PTAS}
	\end{subfigure}
	\begin{subfigure}[b]{0.28\textwidth}
		\scalebox{.7}{
		\begin{tikzpicture}[pta, scale=1]
	
			\node[location, initial] at (0, 0) (l1) {$\loc_1$};
	
			\path (l1) edge[loop above] node[align=center]{%
				$\styledisc{P_1}$
				\\
				$\styledisc{P_1} \assign \BFalse$
				} (l1);

			\path (l1) edge[loop below] node[align=center]{%
				$\neg \styledisc{P_1}$
				\\
				$\styledisc{P_1} \assign \BTrue$
				} (l1);

			\path (l1) edge[loop right] node[align=center]{%
				$\styledisc{P_2}$
				\\
				$\styledisc{P_2} \assign \BFalse$
				} (l1);

			\path (l1) edge[loop left] node[align=center]{%
				$\neg \styledisc{P_2}$
				\\
				$\styledisc{P_2} \assign \BTrue$
				} (l1);

		\end{tikzpicture}
		}
		\caption{SBA}
		\label{figure:bool:SBA}
	\end{subfigure}

	\caption{A non-parametric specification over Boolean predicates}
	\label{figure:bool}

\end{figure}
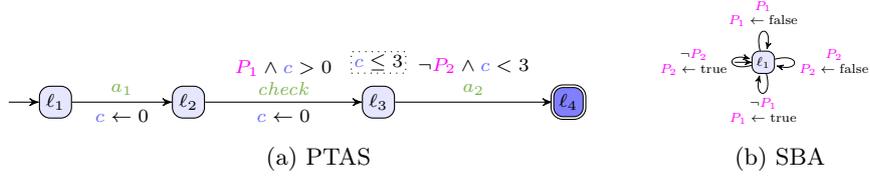
\begin{figure}[tb]
	\begin{subfigure}[b]{0.32\textwidth}
	{\centering
		\begin{tikzpicture}[scale=.8, xscale=.8]

		\draw[axis] (0,0) -- (2.5, 0) node[anchor=west] {$t$};
		\draw[axis] (0,0) -- (0, 1.25) node[anchor=south east] {};

		\foreach \x in {1,2}
			\draw[draw=black] (\x, .4em) -- (\x, -.4em) node[anchor=north] {$\x$};
		\draw (.3em, 0) -- (-.3em, 0) node[anchor=east] {$\BFalse$};
		\draw (.3em, 1) -- (-.3em, 1) node[anchor=east] {$\BTrue$};

		\draw[discrete_transition] (0, 1.1) -- (0, -1.2em) node[anchor=north] {$\actioni{1}$};
		\draw[discrete_transition] (1, 1.1) -- (1, -1.2em) node[anchor=north] {$\styleact{\mathit{check}}$};
		\draw[discrete_transition] (2, 1.1) -- (2, -1.2em) node[anchor=north] {$\actioni{2}$};

		\draw[signalA] (0, 1) -- (0, 0) -- (0.5, 0) -- (0.5, 1) -- (2, 1) node[below left]{$\styledisc{P_1}$};

		\draw[signalB] (0, 1) -- (0.25, 1) -- (0.25, 0) -- (2, 0) node[above left]{$\styledisc{P_2}$}; 

		\end{tikzpicture}
		\caption{Positive run 1}
		\label{figure:concrete-runs:bool:positive}
	
	}

	\end{subfigure}
	\begin{subfigure}[b]{0.32\textwidth}
	{\centering
		\begin{tikzpicture}[scale=.8, xscale=.8]

		\draw[axis] (0,0) -- (2.5, 0) node[anchor=west] {$t$};
		\draw[axis] (0,0) -- (0, 1.25) node[anchor=south east] {};

		\foreach \x in {1,2}
			\draw[draw=black] (\x, .4em) -- (\x, -.4em) node[anchor=north] {$\x$};
		\draw (.3em, 0) -- (-.3em, 0) node[anchor=east] {$\BFalse$};
		\draw (.3em, 1) -- (-.3em, 1) node[anchor=east] {$\BTrue$};

		\draw[discrete_transition] (0, 1.1) -- (0, -1.2em) node[anchor=north] {$\actioni{1}$};
		\draw[discrete_transition] (1, 1.1) -- (1, -1.2em) node[anchor=north] {$\styleact{\mathit{check}}$};
		\draw[discrete_transition] (5/4, -1.2em) -- (5/4, 1.1) node[anchor=south west] {$\actioni{2}$};

		\draw[signalA] (0, 1) -- (0, 0) -- (0.5, 0) -- (0.5, 1) -- (1.5, 1) -- (1.5, 0)-- (2, 0)  node[above]{$\styledisc{P_1}$};

		\draw[signalB] (0, 0) -- (0, 1) -- (0.25, 1) -- (0.25, 0) -- (2, 0) node[below left]{$\styledisc{P_2}$}; 

		\end{tikzpicture}
		\caption{Positive run 2}
		\label{figure:concrete-runs:bool:positive:4}
	
	}
		
	\end{subfigure}
	\begin{subfigure}[b]{0.32\textwidth}
	{\centering
		\begin{tikzpicture}[scale=.8, xscale=.8]

		\draw[axis] (0,0) -- (4.5, 0) node[anchor=west] {$t$};
		\draw[axis] (0,0) -- (0, 1.25) node[anchor=south east] {};

		\foreach \x in {1,2,...,4}
			\draw[draw=black] (\x, .4em) -- (\x, -.4em) node[anchor=north] {$\x$};
		\draw (.3em, 0) -- (-.3em, 0) node[anchor=east] {$\BFalse$};
		\draw (.3em, 1) -- (-.3em, 1) node[anchor=east] {$\BTrue$};

		\draw[discrete_transition] (0, 1.1) -- (0, -1.2em) node[anchor=north] {$\actioni{1}$};
		\draw[discrete_transition] (1, 1.1) -- (1, -1.2em) node[anchor=north] {$\styleact{\mathit{check}}$};
		\draw[discrete_transition] (4, -1.2em) -- (4, 1.1) node[anchor=north east] {$\actioni{2}$};

		\draw[signalA] (0, 1) -- (0, 0) -- (0.5, 0) -- (0.5, 1) -- (4, 1)  node[above]{$\styledisc{P_1}$};

		\draw[signalB] (0, 0) -- (0, 1) -- (0.25, 1) -- (0.25, 0) -- (4, 0) node[below left]{$\styledisc{P_2}$}; 

		\end{tikzpicture}
		\caption{Negative run}
		\label{figure:concrete-runs:bool:negative:5}
	
	}
		
	\end{subfigure}

	\caption{Positive and negative runs for \cref{figure:bool}}
	\label{figure:concrete-runs:bool}
\end{figure}
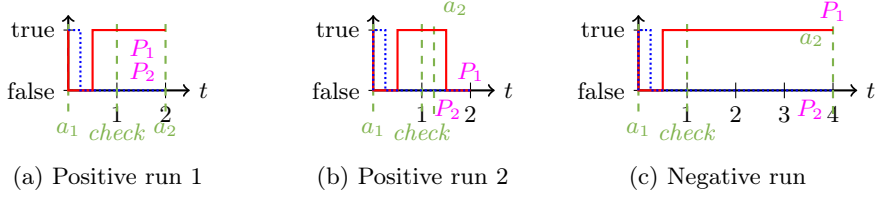

\paragraph{A non-parametric specification over signals}
Recall the motivating specification from \cref{example:motivating} with the PTAS from \cref{figure:example-PTAS:motivating} and the SBA in \cref{figure:example-SBA-fastslow}.
We assume that initially $\signali{1},\signali{2} \in [0,10]$ (such non-deterministic assignment is allowed by our framework, from $\VariablesInit$ in \cref{def:PLMA}).
Three concrete runs are given in \cref{figure:concrete-runs:running},
while all six outputs by \imitator{} are given in \ShortVersion{\cite{AWUH22report}}\LongVersion{\cref{fig:ex2signals2:result} (\cref{appendix:ex2signals2:results})}.

\paragraph{A parametric specification over signals}
Now recall the parametric specification from \cref{figure:example-PTAS:param}.
Our approach derives a parameter valuation $\paramp = 10$, for which this specification can be satisfied, as well as the concrete run in \cref{figure:concrete-runs:param:pos}.
Then, our approach derives a parameter valuation $\paramp = 5$ for which the specification may be violated, with a negative run in \cref{figure:concrete-runs:param:neg5}: this run is not valid because the two \styleact{\mathit{sense}} actions are separated by $< 5$ time units.
Finally, our approach derives a second negative run, this time for $\paramp = 10$, given in \cref{figure:concrete-runs:param:neg10}: again, this run is not valid because two \styleact{\mathit{sense}} actions occur in a time $\frac{10}{3} < 5$.
\LongVersion{%
	The whole set of positive runs is given in \cref{fig:exparam:result}, with the two aforementioned negative runs in \cref{fig:exparam:result:neg} (\cref{appendix:exparam:results}).
}
\section{Conclusion}\label{section:conclusion}

We presented a first approach to exemplify specifications over signals (as real-valued continuous variables with a piecewise-constant rate), also using regular TA clocks and timing parameters.
Our approach's originality is twofold: expressive quantitative specifications (involving notably continuous time, timing parameters and signals),
	and the use of newly introduced \emph{signal bounding automata} to limit the admissible continuous behavior.
Our implementation in \imitator{} makes the process fully automated.

While we do not expect our exemplifying approach to allow for users completely unfamiliar with model checking and timed formalisms to suddenly become experts in these methods, we believe our approach is a first step towards helping users with a low expertise to increase the confidence they have in their specifications.

\paragraph{Future work}
A first future work is to study the theoretical background of our specification formalism, and notably its expressiveness.
Also, we so far considered only reachability properties, and our framework shall be extended to liveness/fairness, \eg{} using the recent liveness synthesis algorithms for PTAs~\cite{NPP18,AAPP21}.
The strong determinism assumption (\cref{assumption:strong-determinism}) is required by our algorithms, but shall eventually be lifted.

Another direction is to allow more flexible formalisms (\eg{} rectangular hybrid automata) to bound the signals.

One of the future directions is to extend our framework to exemplify a more widely used formalism, \eg{} LTL, MITL~\cite{MNP06}, or STL~\cite{MN04}. %
At least theoretically, this would be straightforward thanks to the high expressiveness of PTASs.
In this latter case, we can also benefit from the \emph{positive} run exemplification to exhibit \emph{negative} runs, by taking as input the PTAS corresponding to the \emph{negation} of the original formula.

Further, providing some ``coverage'' guarantees, with a sufficient number of positive and negative runs, is on our agenda.

One longer-term future work is to use and evaluate our framework to teach students or engineers who are not familiar with formal specifications.

\ifdefined\VersionLong
	\newcommand{\CCIS}{Communications in Computer and Information Science}
	\newcommand{\ENTCS}{Electronic Notes in Theoretical Computer Science}
	\newcommand{\FAC}{Formal Aspects of Computing}
	\newcommand{\FI}{Fundamenta Informaticae}
	\newcommand{\FMSD}{Formal Methods in System Design}
	\newcommand{\IJFCS}{International Journal of Foundations of Computer Science}
	\newcommand{\IJSSE}{International Journal of Secure Software Engineering}
	\newcommand{\IPL}{Information Processing Letters}
	\newcommand{\JAIR}{Journal of Artificial Intelligence Research}
	\newcommand{\JLAP}{Journal of Logic and Algebraic Programming}
	\newcommand{\JLAMP}{Journal of Logical and Algebraic Methods in Programming} %
	\newcommand{\JLC}{Journal of Logic and Computation}
	\newcommand{\LMCS}{Logical Methods in Computer Science}
	\newcommand{\LNCS}{Lecture Notes in Computer Science}
	\newcommand{\RESS}{Reliability Engineering \& System Safety}
	\newcommand{\STTT}{International Journal on Software Tools for Technology Transfer}
	\newcommand{\TCS}{Theoretical Computer Science}
	\newcommand{\ToPNoC}{Transactions on Petri Nets and Other Models of Concurrency}
	\newcommand{\TSE}{{IEEE} Transactions on Software Engineering}
\else
	\newcommand{\CCIS}{CCIS}
	\newcommand{\ENTCS}{ENTCS}
	\newcommand{\FAC}{FAC}
	\newcommand{\FI}{FI}
	\newcommand{\FMSD}{FMSD}
	\newcommand{\IJFCS}{IJFCS}
	\newcommand{\IJSSE}{IJSSE}
	\newcommand{\IPL}{IPL}
	\newcommand{\JAIR}{JAIR}
	\newcommand{\JLAP}{JLAP}
	\newcommand{\JLAMP}{JLAMP}
	\newcommand{\JLC}{JLC}
	\newcommand{\LMCS}{LMCS}
	\newcommand{\LNCS}{LNCS}
	\newcommand{\RESS}{RESS}
	\newcommand{\STTT}{STTT}
	\newcommand{\TCS}{TCS}
	\newcommand{\ToPNoC}{ToPNoC}
	\newcommand{\TSE}{TSE}
\fi

\ifdefined\VersionLong
	\renewcommand*{\bibfont}{\small}
	\printbibliography[title={References}]
\else
	\bibliographystyle{splncs04} %
	\bibliography{biblio}
\fi

\ifdefined\VersionLong

\newpage
\appendix

\begin{center}
	\huge{}\bfseries{}Appendix
\end{center}

\section{Proof of \cref{lemma:product}}\label{appendix:lemma:product}

\recallResult{lemma:product}{\lemmaProduct{}}

\lemmaProductProof{}

\section{Exhibiting a concrete valuation in a polyhedron}\label{appendix:exhibitPoint}

We give our algorithm $\exhibitPoint(\Constraint)$ in \cref{algo:exhibitPoint}.

\begin{algorithm}[!htb]
	\Input{A polyhedron~$\Constraint$}
	\Output{A point $\pval \in \Constraint$}
	
	Let $\pval$

	\ForEach{$\dimension \in \Dimensions(\Constraint)$}{
		\eIf{$\dimension$ is not constrained in $\Constraint$}{
			$\pval(\dimension) \assign 1$ \tcc{arbitrary}
		}{
			
			\uIf{$\hasminimum(\Constraint, \dimension)$}{
				
				\tcc{Case 1: constant minimum}
				
				\eIf{$\minimum(\Constraint, \dimension) = 0 \land \supremum(\Constraint, \dimension) > 0$}{
				
					\tcc{Case 1a: supremum $>1$: return 1 to avoid 0}
					\eIf{$\supremum(\Constraint, \dimension) > 1$}{
						$\pval(\dimension) \assign 1$ \tcc{arbitrary to avoid 0}
					}{
					\tcc{Case 1b: supremum in $(0,1]$: return half of it}
						$\pval(\dimension) \assign \frac{\supremum(\Constraint, \dimension)}{2}$
					}
				}{
				\tcc{Case 1c: return the minimum}
					$\pval(\dimension) \assign \minimum(\Constraint, \dimension)$
				
				}
			}

			\uElseIf{$\infimum(\Constraint, \dimension) > -\infty \land \supremum(\Constraint, \dimension) = \infty$}{
				\tcc{Case 2: some finite infimum, but infinite supremum: return infimum + 1}
				
				$\pval(\dimension) \assign \infimum(\Constraint, \dimension) + 1$
			}
			
			\uElseIf{$\infimum(\Constraint, \dimension) = -\infty \land \supremum(\Constraint, \dimension) < \infty$}{
				\tcc{Case 3: no finite infimum but some supremum: return 1 if possible, otherwise supremum - 1}
			
				$\pval(\dimension) \assign \min(
					1
					,
					\supremum(\Constraint, \dimension) - 1
				)$
			}
			
			\Else{
				\tcc{Case 4: finite infimum and supremum: return middle}
			
				$\pval(\dimension) \assign \frac{\infimum(\Constraint, \dimension) , \supremum(\Constraint, \dimension)}{2}$
			}

		}
		
		\tcc{Valuate $\Constraint$ with the new valuation for $\dimension$}

		$\Constraint \assign \Constraint \land \dimension = \pval(\dimension)$
	} %

	\caption{$\exhibitPoint(\Constraint)$: exhibiting a point in a polyhedron}
	\label{algo:exhibitPoint}
\end{algorithm}

Since a polyhedron is represented by a constraint (set of linear inequalities),
as an abuse of notations, we use both constraint-based notations, and polyhedra-based notations.

Let $\Dimensions(\Constraint)$ denote the set of 
\emph{dimensions} (\ie{} of variables) of a polyhedron~$\Constraint$.

We use the following notations:
\begin{itemize}
	\item $\hasminimum(\Constraint, \dimension)$ is a predicate checking whether variable of dimension~$\dimension$ has a concrete minimum valuation in the polyhedron~$\Constraint$.
	That is, $\project{\Constraint}{\{ \variable_\dimension \}}$ (where $\variable_\dimension$ denotes the variable of dimension~$\dimension$) yields an interval $[a, b]$ or $[a, b)$.
	In this case, $a$ is the minimum for dimension~$\dimension$, denoted by $\minimum(\Constraint, \dimension) = a$.
	
	\item As a slight abuse of notation, we denote by $\infimum(\Constraint, \dimension)$ the minimum or infimum value for dimension~$\dimension$, \ie{} considering $\project{\Constraint}{\{ \variable_\dimension \}}$ yields an interval $(a, b]$ or $(a, b)$ or $[a, b]$ or $[a, b)$, then we define $\infimum(\Constraint, \dimension) = a$.
	Dually, we denote by $\infimum(\Constraint, \dimension) = b$ the maximum or supremum value for dimension~$\dimension$.
\end{itemize}

We say that a dimension~$\dimension$ is \emph{constrained} in a polyhedron~$\Constraint$ if~$\dimension$ appears in an inequality in the underlying constraint.
That is, $\dimension$ is constrained if $\project{\Constraint}{\{ \variable_\dimension \}} \neq (-\infty, \infty)$.

We use \emph{infimum} (resp.\ supremum) to denote the greatest (resp.\ smallest) possible value which is less (resp.\ greater) than or equal to all valuations for a dimension in a constraint.
For example, given $\dimension \in (-2, 3]$ then $-2$ is an infimum but not a minimum (there is no minimum in this constraint), while $3$ is here the supremum and the maximum.
We assume the supremum can be~$\infty$ and the infimum can be~$-\infty$.
\section{Computing a concrete (positive) run}\label{appendix:positive}
\subsection{Positive run reconstruction}\label{ss:algo:constructPos}
The reconstruction of a run (fragment) from a symbolic run is given in \cref{algo:constructPos}.
The reconstruction is performed backwards.
First, we cancel time elapsing (\cref{algo:constructPos:cancel}), \ie{} we compute a final valuation from which we can fire (backwards) the last transition.
We then initialize the run to be built from its last (concrete) state (\cref{algo:constructPos:init}).

Given $\edge = (\loc,\guard,\action,\resets,\loc')$,
	we write $\edge.\guard$ and $\edge.\resets$ to denote $\guard$ and $\resets$ respectively.

\begin{algorithm}[!htb]
	\Input{A PLMA~$\A$}
	\Input{A symbolic run~$\symbrun$ of~$\A$} %
	\Input{Last position~$i$} %
	\Input{Last concrete state $(\loc_i, \wv{\vval_i}{\pval})$}
	\Output{A concrete positive run prefix of~$\valuate{\A}{\pval}$ up to~$i$}

	\smallskip

	$(\loc_{i-1}, \Constraint_{i-1}) \assign \stateat(\symbrun, i-1)$

	\tcc{Cancel time elapsing of the last valuation}
	$\vval_i' \assign \exhibitPredContinuous(\Constraint_{i-1}, \edge_{i-1}, \loc_i, \flow, \invariant, \vval_i)$ %
		\nllabel{algo:constructPos:cancel}

	\smallskip
	
	$\concreterun \assign (\loc_i, \vval_i')$ \tcc{initialization}
		\nllabel{algo:constructPos:init}
	
	$\vval_{k+1} \assign \vval_i'$
	
	\For{$k = i-1$ \kwDownTo{} $0$\nllabel{algo:constructPos:for}}{
		$(\loc_k, \Constraint_k) \assign \stateat(\symbrun, k)$
		
		$\edge_k \assign \edgeat(\symbrun, k)$
		
		$(\loc_{k+1}, \Constraint_{k+1}) \assign \stateat(\symbrun, k+1)$

		\tcc{Exhibit a discrete predecessor valuation}
		$\vval_k \assign \exhibitPred(\Constraint_k, \flow(\loc_k), \edge_k.\guard, \edge_k.\resets, \invariant(\loc_{k+1}), \flow(\loc_{k+1}), \Constraint_{k+1}, \vval_{k+1}, \pval)$
		\nllabel{algo:constructPos:discretepred}
		
		\tcc{Compute duration between $\vval_k$ and its successor}
		
		$d \assign \computeD(\vval_k, \vval_{k+1})$
		\nllabel{algo:constructPos:duration}

		\tcc{Update the concrete run}
		$\concreterun \assign (\loc_k, \vval_k), (\edgeat(\symbrun, k), d), \concreterun$\nllabel{algo:constructPos:add} %

		\tcc{Backup for next iteration}
		$\vval_{k+1} \assign \vval_k$
		\nllabel{algo:constructPos:endfor}
	}
	
	\Return{$\concreterun$}

	\caption{$\reconstructPos(\A, \symbrun, i, (\loc_i, \wv{\vval_i}{\pval}))$: Reconstruct a positive run from a symbolic run}
	\label{algo:constructPos}
\end{algorithm}

The main loop (\cref{algo:constructPos:for}--\cref{algo:constructPos:endfor}) computes the run backwards:
first, we exhibit a discrete predecessor valuation (\cref{algo:constructPos:discretepred}); this is achieved thanks to the dedicated function $\exhibitPred$---described in the subsequent \cref{algo:exhibitPred}.
Then, the duration between the two concrete points is computed (\cref{algo:constructPos:duration})---this function (not given) is discussed below.
Finally, the concrete run is updated (\cref{algo:constructPos:add}).
Here, consistently with our definition in \cref{ss:semantics}, we use ``$,$'' for concatenating states or transitions to a run; that is $(\loc_k, \vval_k), (\edgeat(\symbrun, k), d), \concreterun$ (\cref{algo:constructPos:add}) denotes the addition of state $(\loc_k, \vval_k)$ followed by discrete transition $(\edgeat(\symbrun, k), d)$ in front of~$\concreterun$.

\paragraph{Computing durations between points}
We assume that function $\computeD(\vval_k, \vval_{k+1})$ computes the duration between two consecutive variable valuations.
Due to resets (and possible stopwatches), the only way to achieve this computation is to track a \emph{global time clock}, \ie{} a variable initially~0, of rate~1 and never reset.
In our implementation, we directly add such a clock from the beginning of the analysis (construction of the parametric zone graph); however, it is inefficient because
\begin{ienumerate}%
	\item it adds one more variables everywhere (hence one more dimensions in all polyhedra), and
	\item it may diverge, \ie{} its value will never be reset and therefore matches a previously met valuation.
\end{ienumerate}%
Still, to avoid the second drawback, when testing polyhedra for equality (or inclusion), we first remove (by variable elimination through existential quantification) the global time clock in both polyhedra we test.
This comes at an additional cost, as this elimination is usually not a cheap operation.
A more efficient way of handling these issues could be to only add the clock when reconstructing the symbolic run.
This is future work.

\subsection{Reconstructing predecessors}\label{ss:algo:exhibitPred}

\cref{algo:exhibitPred} computes a predecessor, and simply computes a \emph{discrete} predecessor:
	by first firing backwards the transition (\cref{algo:exhibitPred:discrete}) using the dedicated function $\exhibitPredDiscrete$,
	and second by canceling time elapsing, \ie{} computing a valuation from which another transition can be fired backwards (\cref{algo:exhibitPred:continuous}) using the dedicated function $\exhibitPredContinuous$.
Both functions are described in the following.
Note that the correctness of this procedure comes from the fact that the variable valuation passed as an argument is (iteratively) always a valuation from which a discrete predecessor can be computed, without canceling time elapsing (because this was iteratively performed at the previous call).

\begin{algorithm}[!htb]
	\Input{Constraint $\Constraint_{n-1}$}
	\Input{Edge $\edge_{n-1}$}
	\Input{Location $\loc_n$}
	\Input{Constraint $\Constraint_{n}$}
	\Input{Edge $\edge_{n}$}
	\Input{Flow function $\flow$}
	\Input{Invariant function $\invariant$}
	\Input{Valuation $\vval_{n+1}$}
	\Output{A point $\vval_{n}$ before time elapsing}
	
	\tcc{Compute a discrete predecessor, \ie{} fire backwards the transition $n+1$ to $n$}
	$\vval_{n}' \assign \exhibitPredDiscrete(\Constraint_n, \edge_n, \vval_{n+1})$
		\nllabel{algo:exhibitPred:discrete}

	\tcc{Cancel time elapsing in state $n$}
	$\vval_n \assign \exhibitPredContinuous(\Constraint_{n-1}, \edge_{n-1}, \loc_n, \flow, \invariant, \vval_n')$
		\nllabel{algo:exhibitPred:continuous}
	
	\Return{$\vval_n$}

	\caption{$\exhibitPred(\Constraint_{n-1}, \edge_{n-1}, \loc_n, \Constraint_{n}, \edge_{n}, \flow, \invariant, \vval_{n+1})$: Exhibit a predecessor}
	\label{algo:exhibitPred}
\end{algorithm}
\subsubsection{Continuous predecessor}
The function $\exhibitPredContinuous$ canceling time elapsing is given in \cref{algo:continuous-predecessor}.
We give an algorithm almost at the implementation level (our implementation is extremely close to this algorithm).

Dually to the time elapsing,
we define the \emph{time past} of~$\Constraint$ \wrt{} flow~$\flow : \Variables \to \setQ$, denoted by $\symbtimepast{\Constraint}{\flow}$, as the constraint over $\Variables$ and $\Param$ obtained from~$\Constraint$ by ``delaying backwards'' all variables by an arbitrary amount of time according to~$\flow$.
That is,
\(\wv{\vval'}{\pval} \models \symbtimepast{\Constraint}{\flow} \text{ iff } \exists \vval : \Variables \to \setR, \exists d \in \setR \text { s.t.\ } \wv{\vval}{\pval} \models \Constraint \land \vval' = \timelapse(\vval, \flow, -d) \text{.}\)
\begin{algorithm}[!htb]
	\Input{Constraint $\Constraint_{n-1}$ from the state preceding the current one~$\Constraint_n$}
	\Input{$\edge_{n-1}$ incoming edge (between $\Constraint_{n-1}$ and~$\Constraint_{n}$)}
	\Input{Current location $\loc_n$}
	\Input{Flow function $\flow$}
	\Input{Invariant function $\invariant$}
	\Input{A point $\vval_n'$ of some $\Constraint_n$ after time elapsing}
	\Output{A point $\vval_n$ before time elapsing}

	\tcc{apply time past}
	$\Constraint_n' \assign \symbtimepast{\vval_n'}{(\flow(\loc_n))}$

	\tcc{intersect with current invariant}
	$\Constraint_n' \assign \Constraint_n' \land \invariant(\loc_n)$

	\tcc{find the ``initial'' valuations (symbolic set)}
	\eIf{$(\loc_n, \vval_n')$ is the initial state}{
		$\mathit{initvals} \assign \Constraint_n' \land $ initial valuations of clocks and signals %
	}{

		\tcc{intersect with incoming state (incl.\ guard and invariant), to which updates were applied}
		
		$\mathit{initvals} \assign \Constraint_n' \cap (\reset{(\Constraint_{n-1} \cap \edge_{n-1}.\guard)}{\edge_{n-1}.\resets})$; %

	}
	
	\tcc{Pick a point $\vval_n$ in $\mathit{initvals}$}
	
	$\vval_n \assign \exhibitPoint(\mathit{initvals})$
	
	\Return{$\vval_n$}

	\caption{$\exhibitPredContinuous(\Constraint_{n-1}, \edge_{n-1}, \loc_n, \flow, \invariant, \vval_n')$: Exhibiting a point before time elapsing in a parametric zone}
	\label{algo:continuous-predecessor}
\end{algorithm}
\subsubsection{Discrete predecessor}
The function $\exhibitPredDiscrete$ canceling time elapsing is given in \cref{algo:discretes-predecessor}.
Again, we give an algorithm almost at the implementation level.

\begin{algorithm}[!htb]
	\Input{Preceding constraint $\Constraint_n$}
	\Input{An edge $\edge_n$ between some $\loc_n$ and some $\loc_{n+1}$}
	\Input{A point $\vval_{n+1}$ of some~$\Constraint_{n+1}$ ``before time elapsing''}
	\Output{A point $\vval_n$ of $(\loc_n, \Constraint_n)$ that can reach $\vval_{n+1}$ in 0-time via $\edge_n$}

	\tcc{Cancel resets from $\loc_n$ to $\loc_{n+1}$}
	$\Constraint_n' \assign \project{ \{ \vval_{n+1} \} }{\Variables \setminus \edge_n.\resets} $ \nllabel{algo:discretes-predecessor:antiresets}
	
	\tcc{intersect with $\Constraint_n$ (mainly for the invariant $\invariant(\loc_n)$)}
	$\Constraint_n' \assign \Constraint_n' \cap \Constraint_n$
	
	\tcc{intersect with guard between $\loc_n$ to $\loc_{n+1}$}
	$\Constraint_n' \assign \Constraint_n' \cap \edge_n.\guard$
	
	\tcc{Pick a point $\vval$ in $\Constraint_n'$}
	
	$\vval_n \assign \exhibitPoint(\Constraint_n')$
	
	\Return{$\vval_n$}

	\caption{$\exhibitPredDiscrete(\Constraint_n, \edge_n, \vval_{n+1})$: Exhibiting a discrete predecessor}
	\label{algo:discretes-predecessor}
\end{algorithm}

\cref{algo:discretes-predecessor} takes as inputs
\begin{ienumerate}
	\item an edge $\edge_n$ between some $\loc_n$ and some $\loc_{n+1}$,
	\item a symbolic constraint~$\Constraint_n$,
	\item a point $\vval_{n+1}$ of some~$\Constraint_{n+1}$ ``before time elapsing'', \ie{} from which one can take in 0-time the transition $\edge_n$ backwards to~$(\loc_n, \Constraint_n)$ (this condition is ensured inductively),
\end{ienumerate}%
and returns a point $\vval_n$ of $(\loc_n, \Constraint_n)$ that can reach $\vval_{n+1}$ in 0-time via $\edge_n$.
At \cref{algo:discretes-predecessor:antiresets} in \cref{algo:discretes-predecessor}, ``$\{ \vval_{n+1} \}$'' denotes the constraint made of the single valuation $\vval_{n+1}$.

\section{Exhibiting deadlocks}
\subsection{Exhibiting parametric deadlocks}\label{appendix:algo:findPdeadlock}

$\findPdeadlock$ is given in \cref{algo:findPdeadlock}.

\begin{algorithm}[!htb]
	\Input{A symbolic run~$\symbrun$}
	\Output{A pair made of a parameter valuation~$\pval$, and a symbolic state~$\symbstate$ such that $\pval$ cannot take the edge following $\symbstate$ along~$\symbrun$; or $\none$ if no such valuation exists}
	
	\For{$i = 0$ \kwTo{} $|\symbrun| - 1$}{
		$(\loc_i, \Constraint_i) \assign \stateat(\symbrun, i)$

		$(\loc_{i+1}, \Constraint_{i+1}) \assign \stateat(\symbrun, i+1)$

		\tcc{If some valuations cannot pass the edge}
		\If{$\projectP{\Constraint_{i+1}} \subsetneq \projectP{\Constraint_i}$\nllabel{algo:findPdeadlock:test}}{
			
			\tcc{Pick such a valuation}
			$\pval \assign \exhibitPoint(\projectP{\Constraint_i} \setminus \projectP{\Constraint_{i+1}})$\nllabel{algo:findPdeadlock:pick}

			\Return{$(\pval, (\loc_i, \Constraint_i))$}
		}

		\Return{$\none$}
	}

	\caption{$\findPdeadlock(\symbrun)$: Exhibition of a parametric deadlock within a symbolic run}
	\label{algo:findPdeadlock}
\end{algorithm}
$\findPdeadlock$ attempts at exhibiting a parameter valuation that cannot pass one of the edges of a symbolic run~$\symbrun$.
In other words, it tries to exhibit a parameter valuation that is member of a polyhedron at state~$i$, but not anymore at~$i+1$; therefore, there exists a concrete run in $\valuate{\A}{\pval}$ equivalent to~$\symbrun$ up to position~$i$, but this does not hold for~$i+1$.

Precisely, we test whether there is a restriction in the parameter valuations between~$i$ and~$i+1$ (\cref{algo:findPdeadlock:test} in \cref{algo:findPdeadlock}).
If so, we just pick one valuation in the difference (\cref{algo:findPdeadlock:pick}).

\subsection{Exhibiting non-parametric deadlocks}\label{appendix:algo:findXdeadlock}

$\findXdeadlock$ is given in \cref{algo:findXdeadlock}.

\begin{algorithm}[!htb]
	\Input{A symbolic run~$\symbrun$}
	\Output{A pair made of a variable valuation~$\vval$, and a symbolic state~$\symbstate$ such that $\vval$ cannot take the edge following $\symbstate$ along~$\symbrun$; or $\none$ if no such valuation exists}

	\For{$i = 0$ \kwTo{} $|\symbrun| - 1$}{
		$(\loc_i, \Constraint_i) \assign \stateat(\symbrun, i)$

		$(\loc_{i+1}, \Constraint_{i+1}) \assign \stateat(\symbrun, i+1)$
		
		\tcc{Get the transition guard}
		$\guard \assign \edgeat(\symbrun, i).\guard$

		\tcc{If some variable valuations cannot pass the edge}
		\If{$\Constraint_i \setminus (\symbtimepast{\guard}{\flow(\loc_i)} \land \invariant(\loc_i) ) \neq \CFalse$}{

			\tcc{Pick such a valuation}
			$\vval \assign \exhibitPoint(\Constraint_i \setminus (\symbtimepast{\guard}{\flow(\loc_i)} \land \invariant(\loc_i) ))$\nllabel{algo:findXdeadlock:pick}
		
			\Return{$(\vval, (\loc_i, \Constraint_i))$}
		}

		\Return{$\none$}
	}
	\caption{$\findXdeadlock(\symbrun)$: Exhibition of a variable deadlock within a symbolic run}
	\label{algo:findXdeadlock}
\end{algorithm}

The crux of \cref{algo:findXdeadlock} is at \cref{algo:findXdeadlock:pick}, where we select a point in the polyhedron which cannot intersect the guard, even after time elapsing or, put it differently, we compute the difference between the original polyhedron~$\Constraint_i$ and the guard~$\guard$ to which we applied time past, then again intersected with the location invariant.
\section{Proof of \cref{proposition:termination}}\label{appendix:proposition:termination}

\recallResult{proposition:termination}{\propositionTermination{}}

\begin{proof}
	First, let us prove that some state $(\locTarget, \Constraint)$ will indeed be found in $\PZG(\A)$, for some $\locTarget \in \LocsFinal$.
	Assume there exists a concrete run of length~$n$ in $\valuate{\A}{\pval}$ reaching~$\locTarget$.
	Then, since the parametric zone graph of~$\A$ is discrete, sharing similar edges (up to valuation~$\pval$)%
	, then there exists an equivalent symbolic run of~$\A$ of same length~$n$ reaching $(\locTarget, \Constraint)$, for some~$\Constraint$.
	Assuming a breadth-first search exploration of $\PZG(\A)$, this state will eventually be computed.
	(This may not be the case assuming a \emph{depth}-first search exploration, as the exploration could be stuck in a symbolic path of infinite length, therefore never meeting $(\locTarget, \Constraint)$.)
	
	Second, let us prove that $\exemplifythree{}$ terminates---which is essentially easy.
	Part~1 is straightforward: exhibiting a point (\cref{algo:exhibit:pick} in \cref{algo:exhibit}) does not pose any termination problem, and the reconstruction of the positive run (\cref{algo:exhibit:reconstructPos}) is done backwards, with a guarantee of termination.
	So $\exemplifythree{}$ is guaranteed to exhibit a (positive) concrete run.
	
	Concerning parts 2a and~2b of \cref{algo:exhibit}, there is no guarantee that a parametric deadlock (\cref{algo:exhibit:ifPD}), nor a non-parametric deadlock (\cref{algo:exhibit:ifXD}) can be found---this explains the lack of theoretical guarantee for a negative run exhibition.
	However, the search for such a deadlock is a simple analysis (\cref{algo:findPdeadlock,algo:findXdeadlock}) of the symbolic run, of finite length, and therefore terminates.
	In addition, the negative run reconstruction (\cref{algo:constructNeg}) is also guaranteed to terminate, as this is a simple finite iteration over the symbolic run length.
	This guarantees termination of $\exemplifythree{}$.
\end{proof}
\section{Proof sketch of \cref{proposition:Masaki}}

\recallResult{proposition:Masaki}{\propMasaki{}}

\begin{proof}
	From the fact that our method is \emph{symbolically complete}:
		since our PZG contains all continuous (parameter and variables) valuations, if such a concrete negative run due to parametric (resp.\ non-parametric) deadlock with the same discrete actions as $\symbrun$ exists, then it will be found during a backward analysis from the target location.
		Note that the result only holds because of the BFS computation assumption.
		A DFS computation might never meet the target state, and therefore not be able to subsequently find the negative concrete run.
\end{proof}

\section{Detailed raw results}
\subsection{Runs for \cref{figure:bool}}
\label{appendix:exBool:results}

The positive runs are given graphically in \cref{fig:exBool:result:1,fig:exBool:result:2,fig:exBool:result:3,fig:exBool:result:4}.
The negative runs are given graphically in \cref{fig:exBool:result:5,fig:exBool:result:6,fig:exBool:result:7,fig:exBool:result:8}.

In all subsequent graphics, not only the signals, but also the \emph{clocks} are given.
Notably, in \cref{fig:exBool:result:1,fig:exBool:result:2,fig:exBool:result:3,fig:exBool:result:4} the third signal is irrelevant (this corresponds to the global time, which obviously increases linearly at rate~1 without reset), and the last one is the (unique) variable~$\variable$.

\begin{figure}[h]
	\centering
	\begin{subfigure}[b]{0.24\textwidth}
		\includegraphics[width=\textwidth]{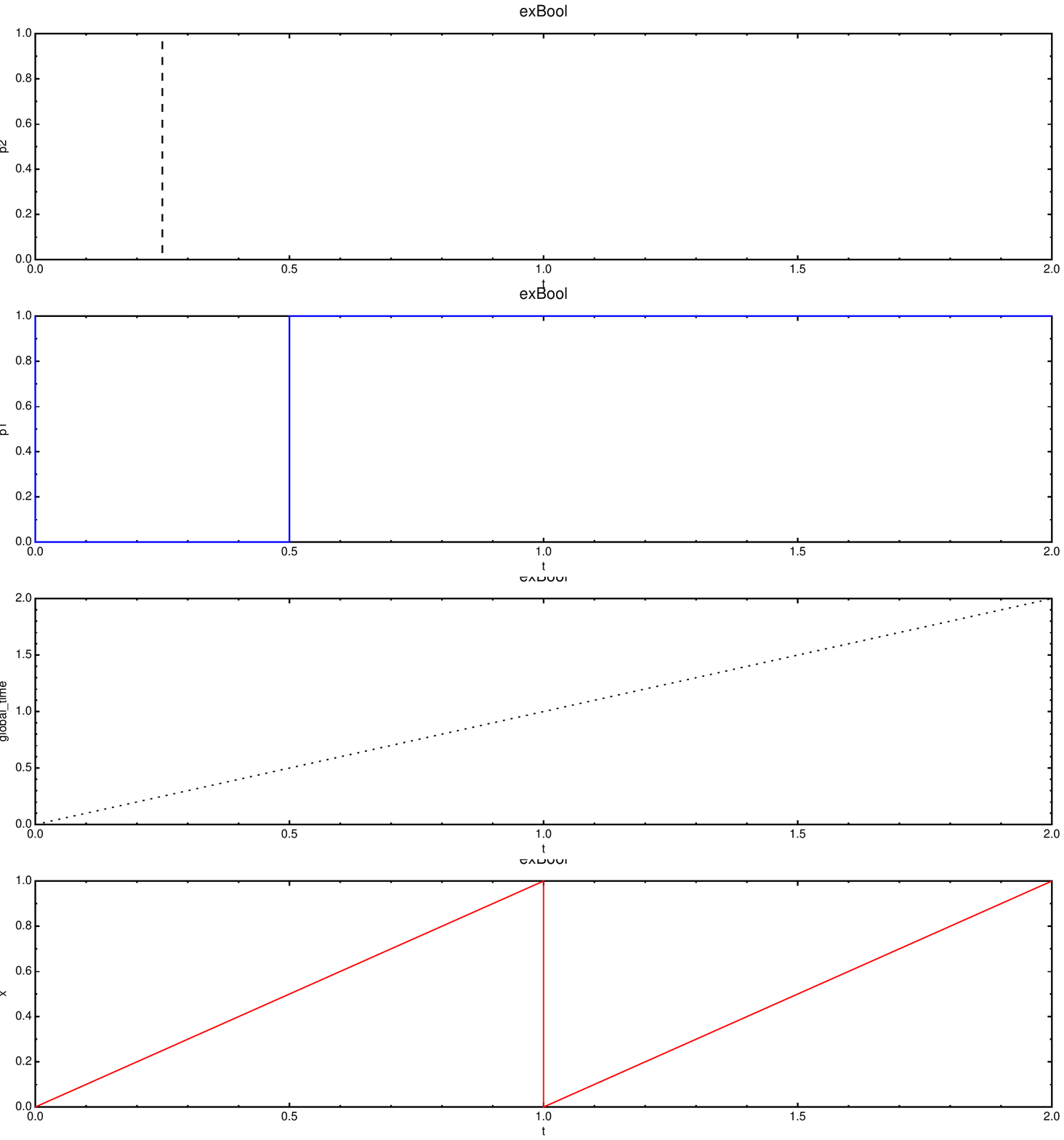}
		\caption{Positive ex 1}
		\label{fig:exBool:result:1}
	\end{subfigure}
	\begin{subfigure}[b]{0.24\textwidth}
		\includegraphics[width=\textwidth]{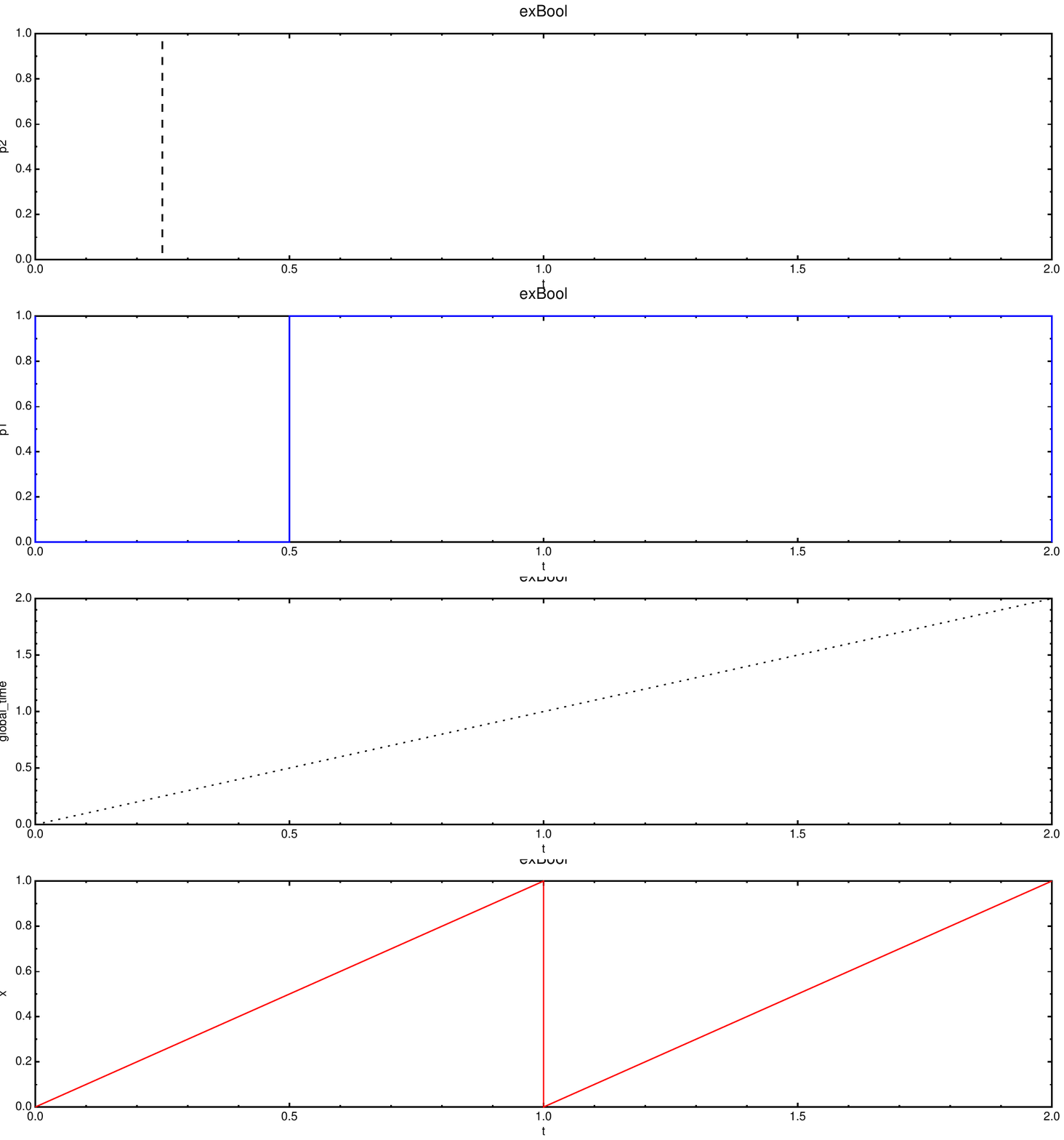}
		\caption{Positive ex 2}
		\label{fig:exBool:result:2}
	\end{subfigure}
	\begin{subfigure}[b]{0.24\textwidth}
		\includegraphics[width=\textwidth]{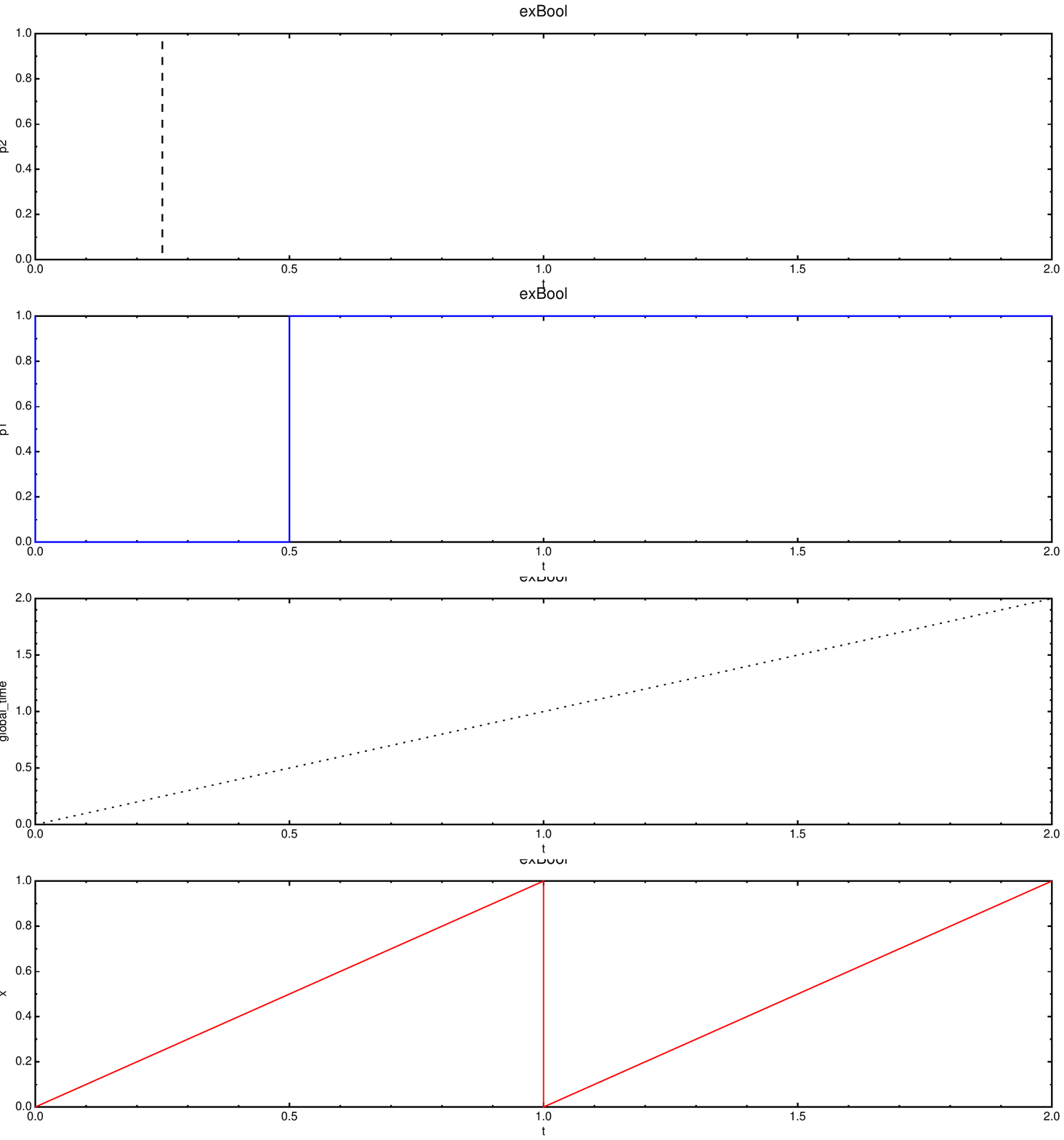}
		\caption{Positive ex 3}
		\label{fig:exBool:result:3}
	\end{subfigure}
	\begin{subfigure}[b]{0.24\textwidth}
		\includegraphics[width=\textwidth]{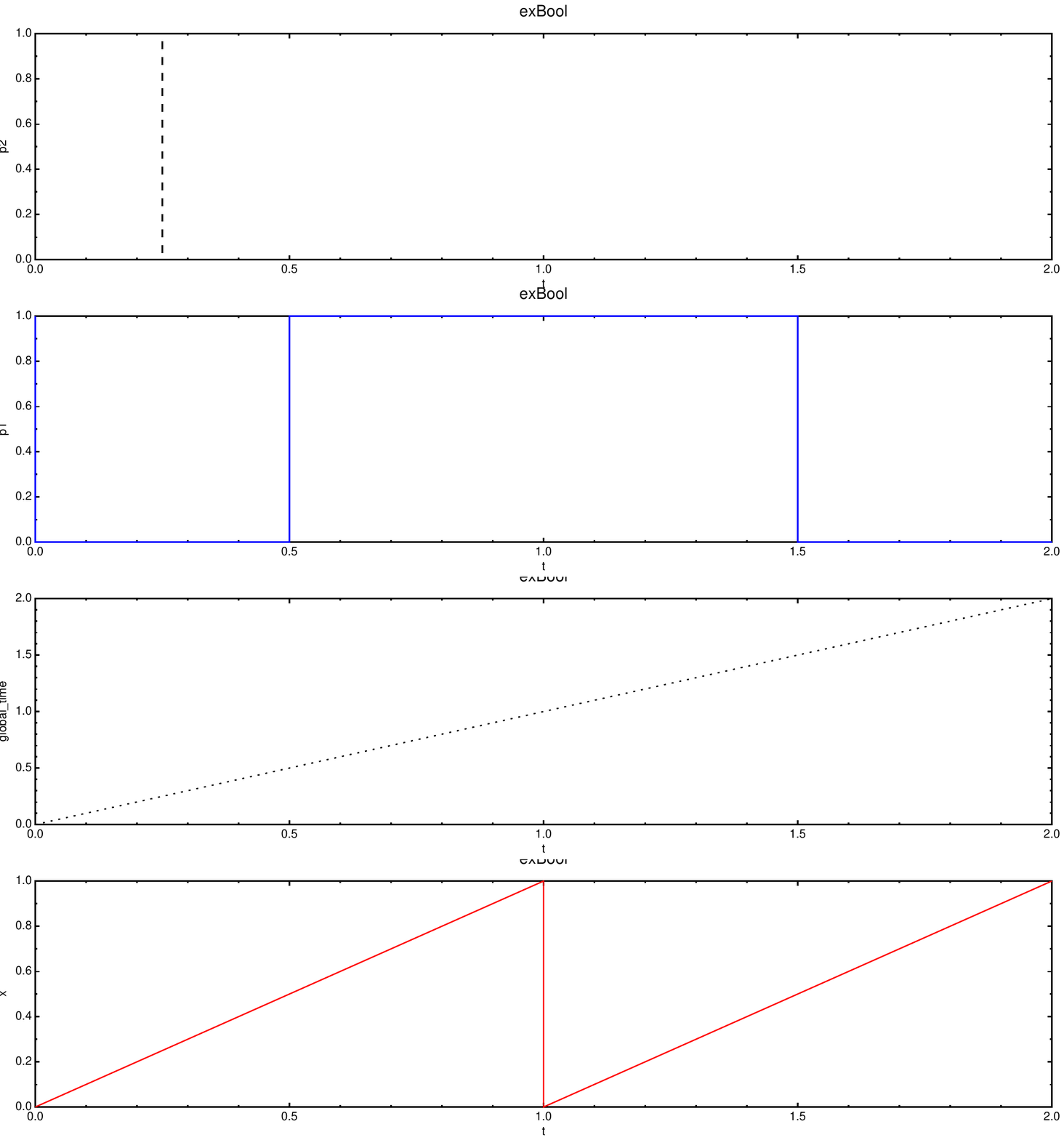}
		\caption{Positive ex 4}
		\label{fig:exBool:result:4}
	\end{subfigure}

	\begin{subfigure}[b]{0.24\textwidth}
		\includegraphics[width=\textwidth]{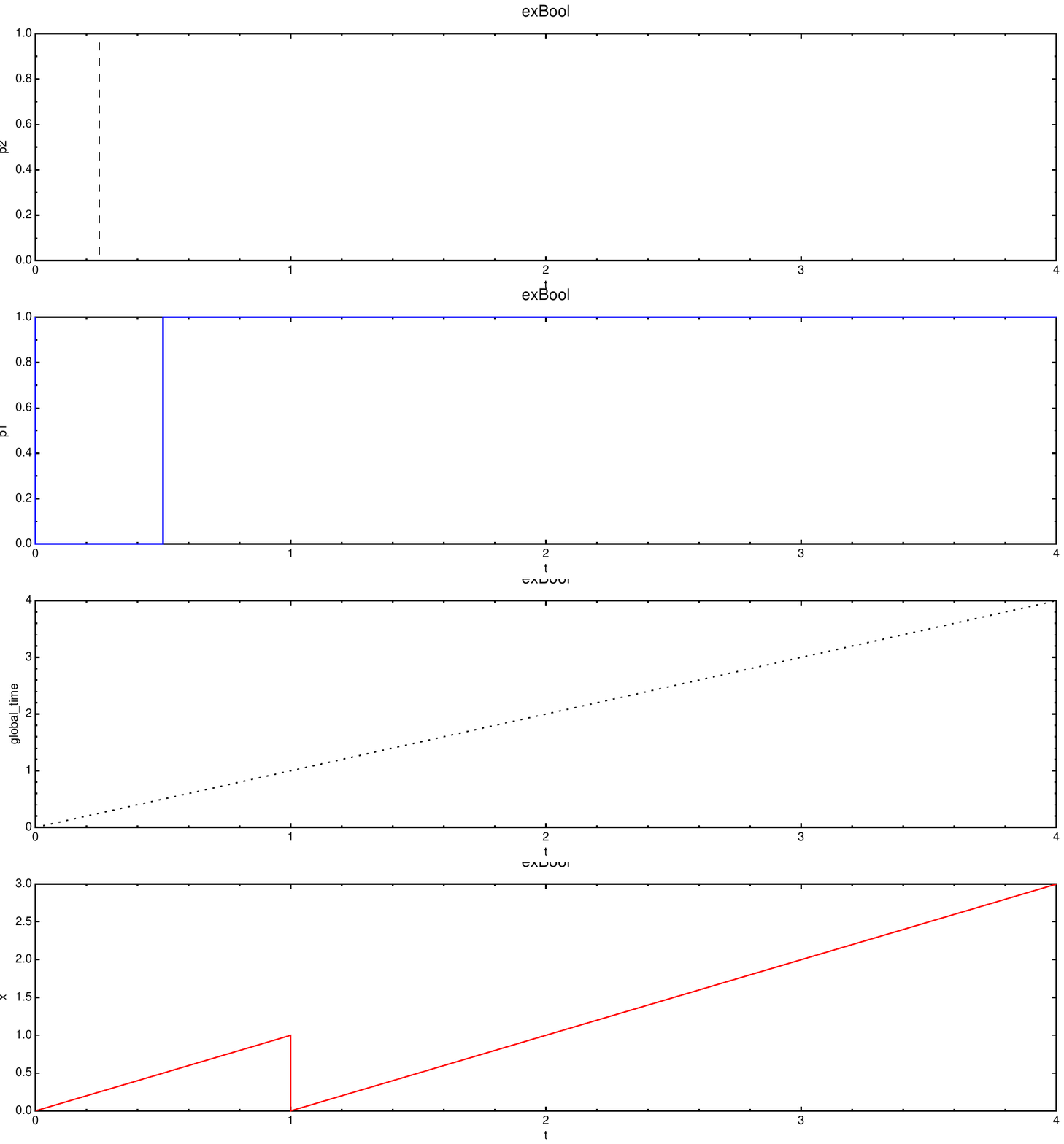}
		\caption{Negative ex 1}
		\label{fig:exBool:result:5}
	\end{subfigure}
	\begin{subfigure}[b]{0.24\textwidth}
		\includegraphics[width=\textwidth]{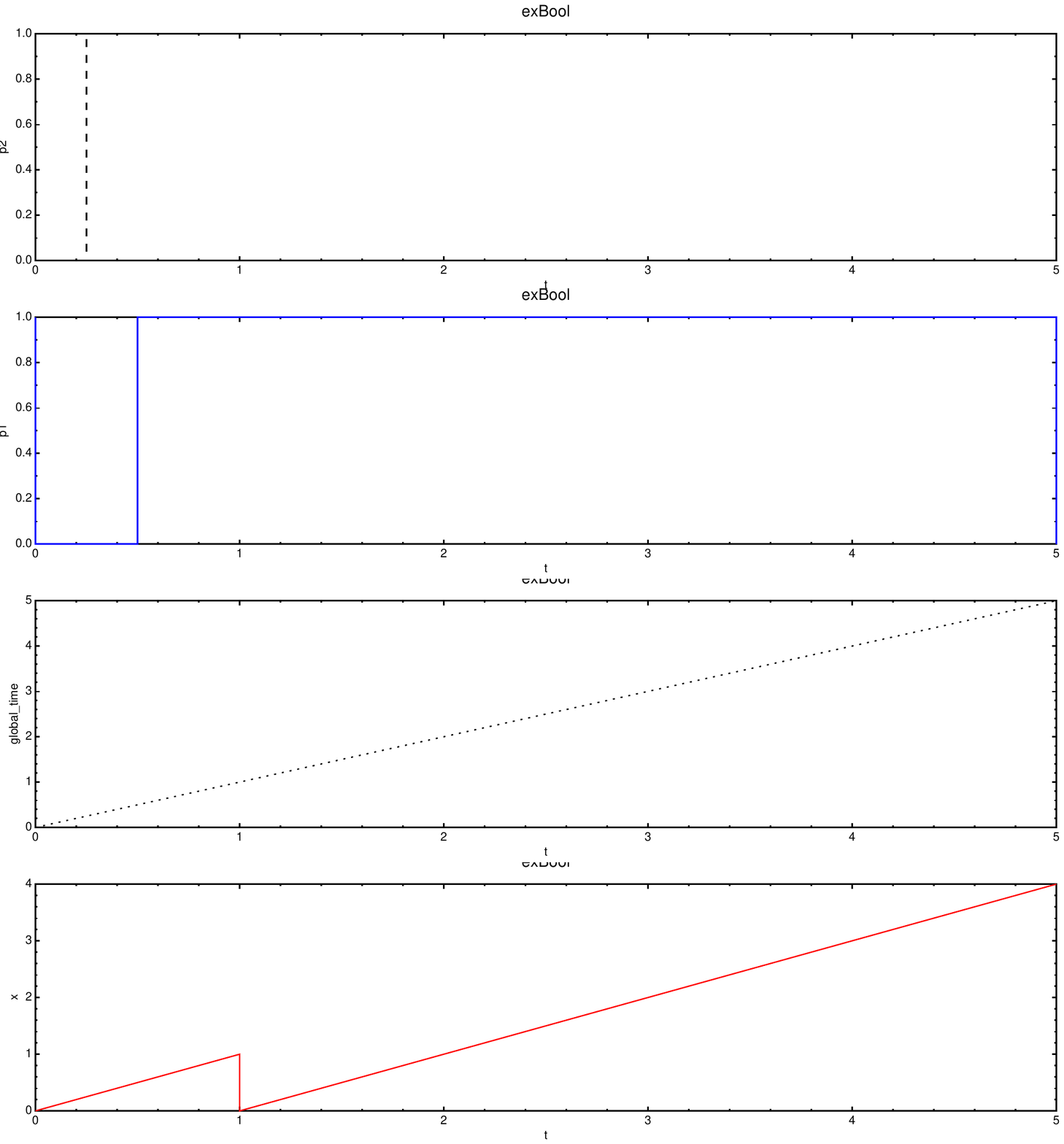}
		\caption{Negative ex 2}
		\label{fig:exBool:result:6}
	\end{subfigure}
	\begin{subfigure}[b]{0.24\textwidth}
		\includegraphics[width=\textwidth]{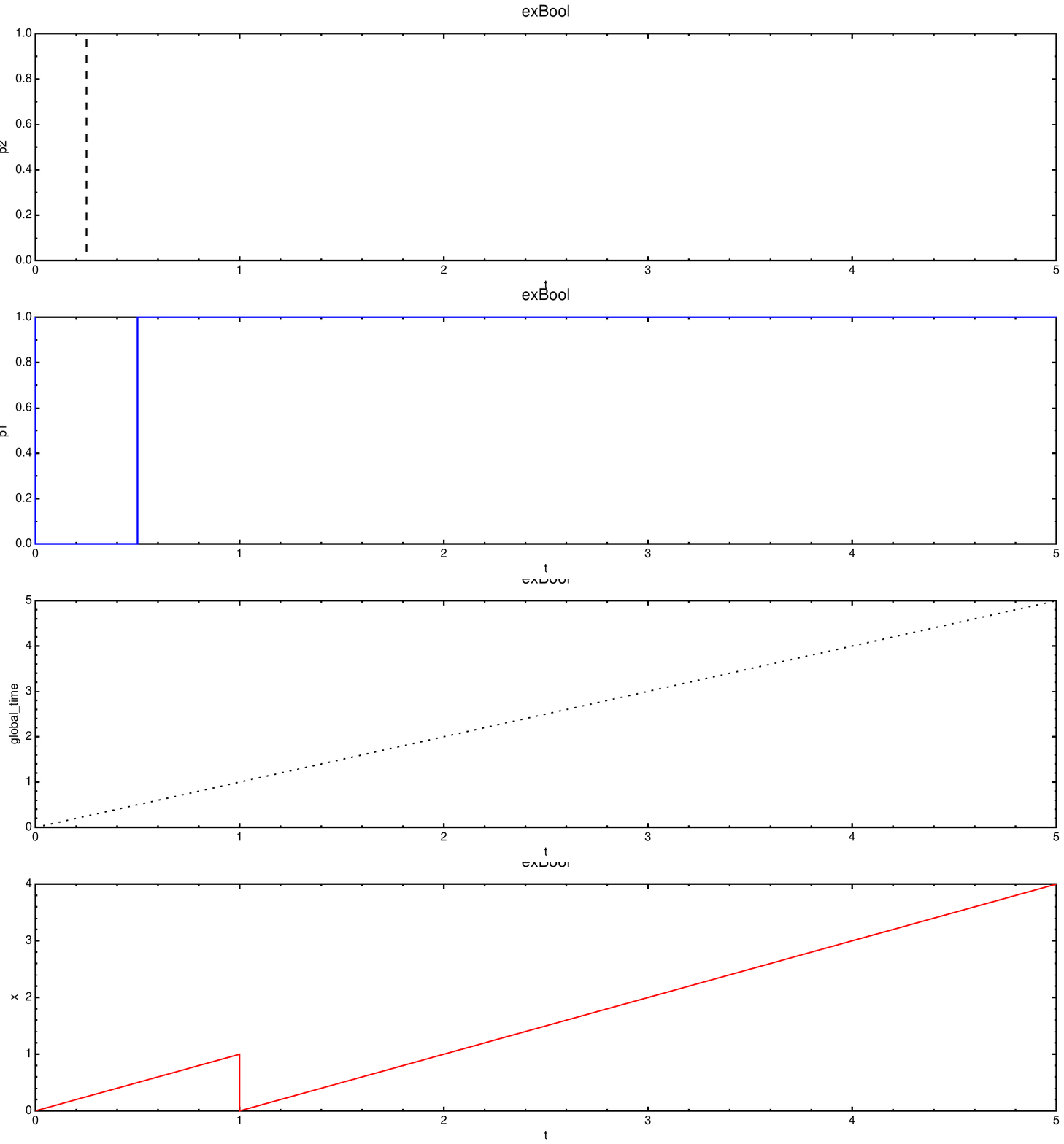}
		\caption{Negative ex 3}
		\label{fig:exBool:result:7}
	\end{subfigure}
	\begin{subfigure}[b]{0.24\textwidth}
		\includegraphics[width=\textwidth]{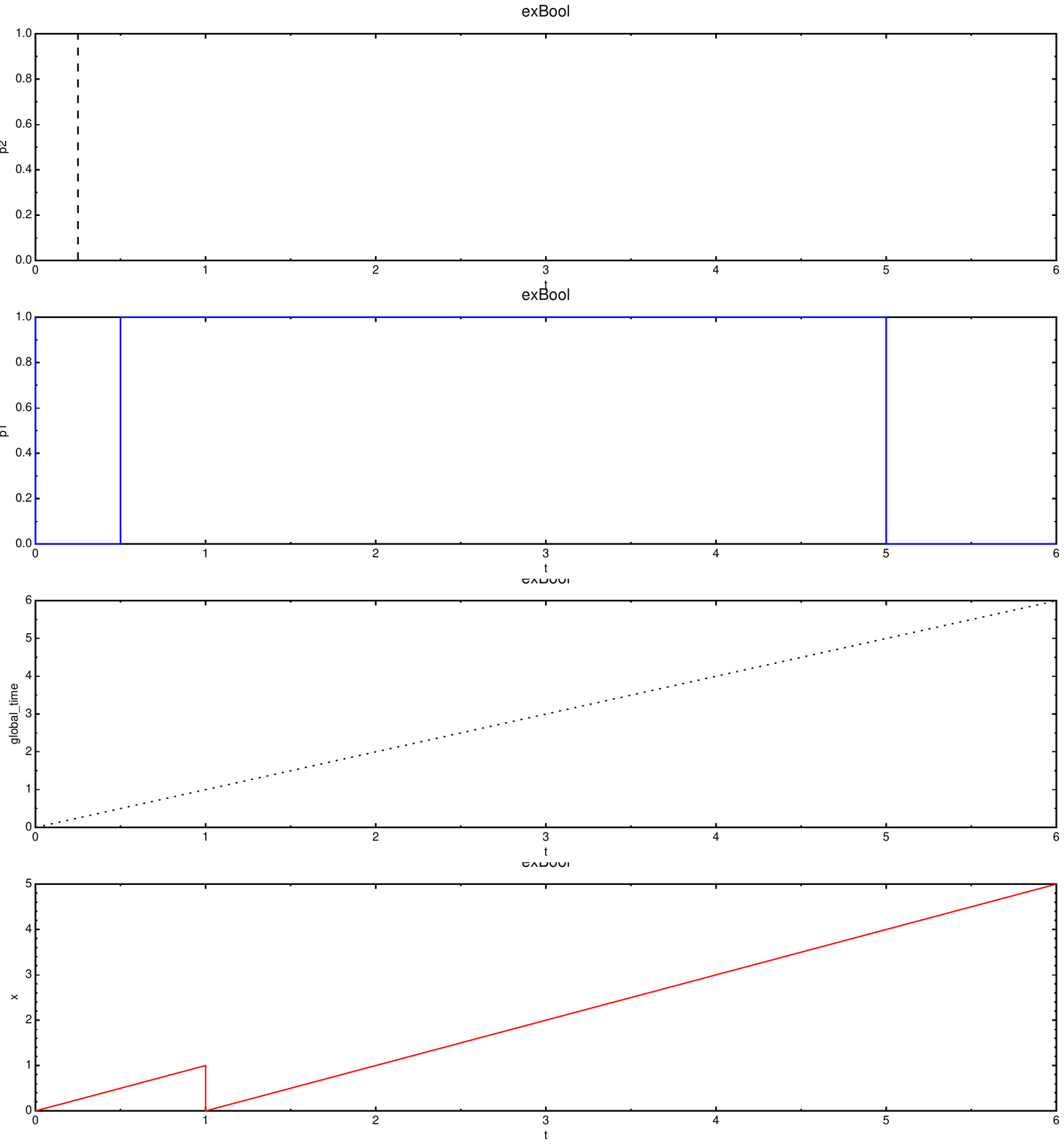}
		\caption{Negative ex 4}
		\label{fig:exBool:result:8}
	\end{subfigure}
	\caption{A non-parametric specification over Booleans (\cref{figure:bool}): examples}
	\label{fig:exBool:result}
\end{figure}
\subsection{Runs for \cref{figure:example-PTAS:motivating,figure:example-SBA-fastslow}}
\label{appendix:ex2signals2:results}

\begin{figure}[h]
	\centering
	\begin{subfigure}[b]{0.31\textwidth}
		\includegraphics[width=\textwidth]{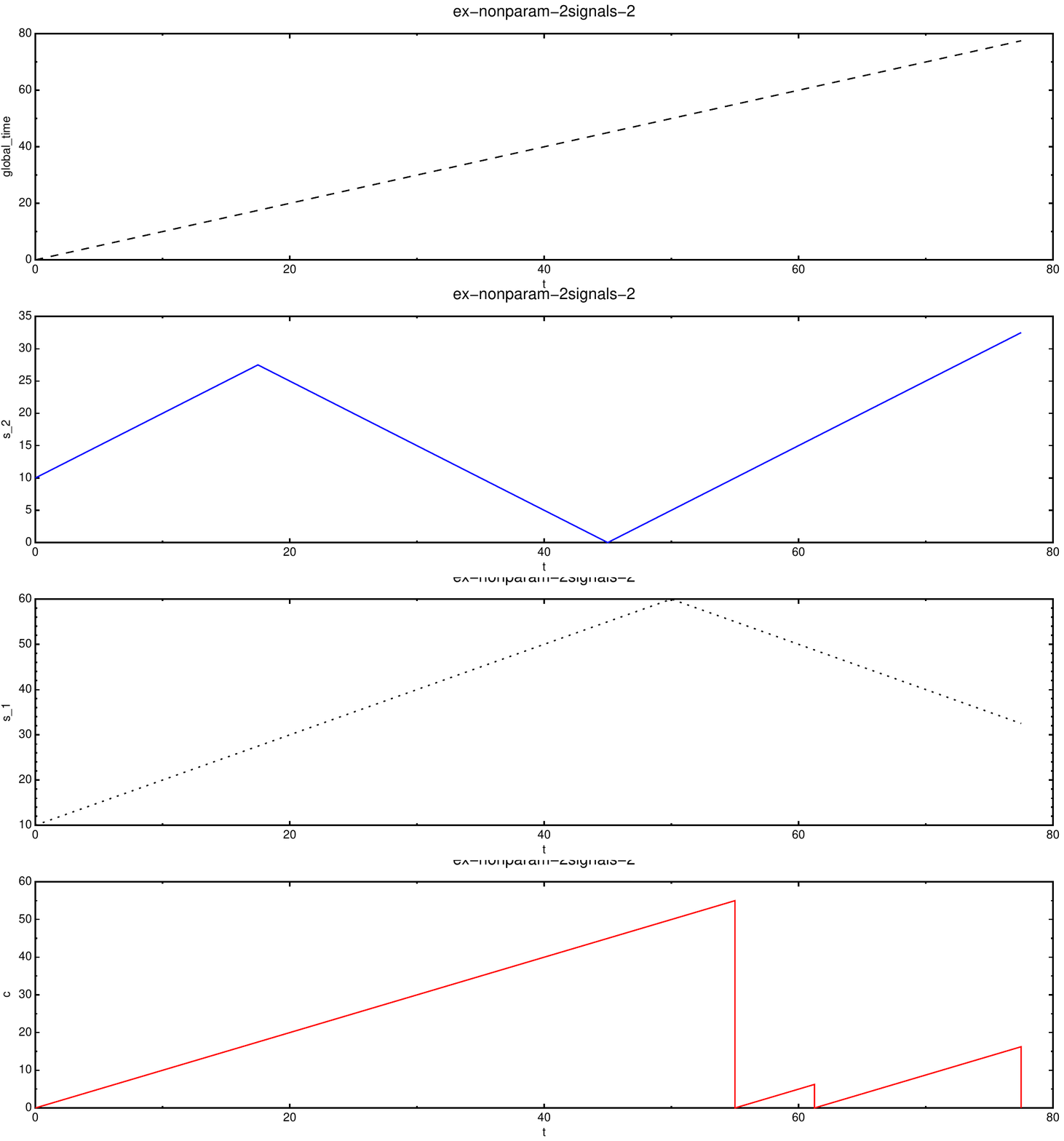}
		\caption{Example 1}
		\label{figex2signals2:1}
	\end{subfigure}
	\begin{subfigure}[b]{0.31\textwidth}
		\includegraphics[width=\textwidth]{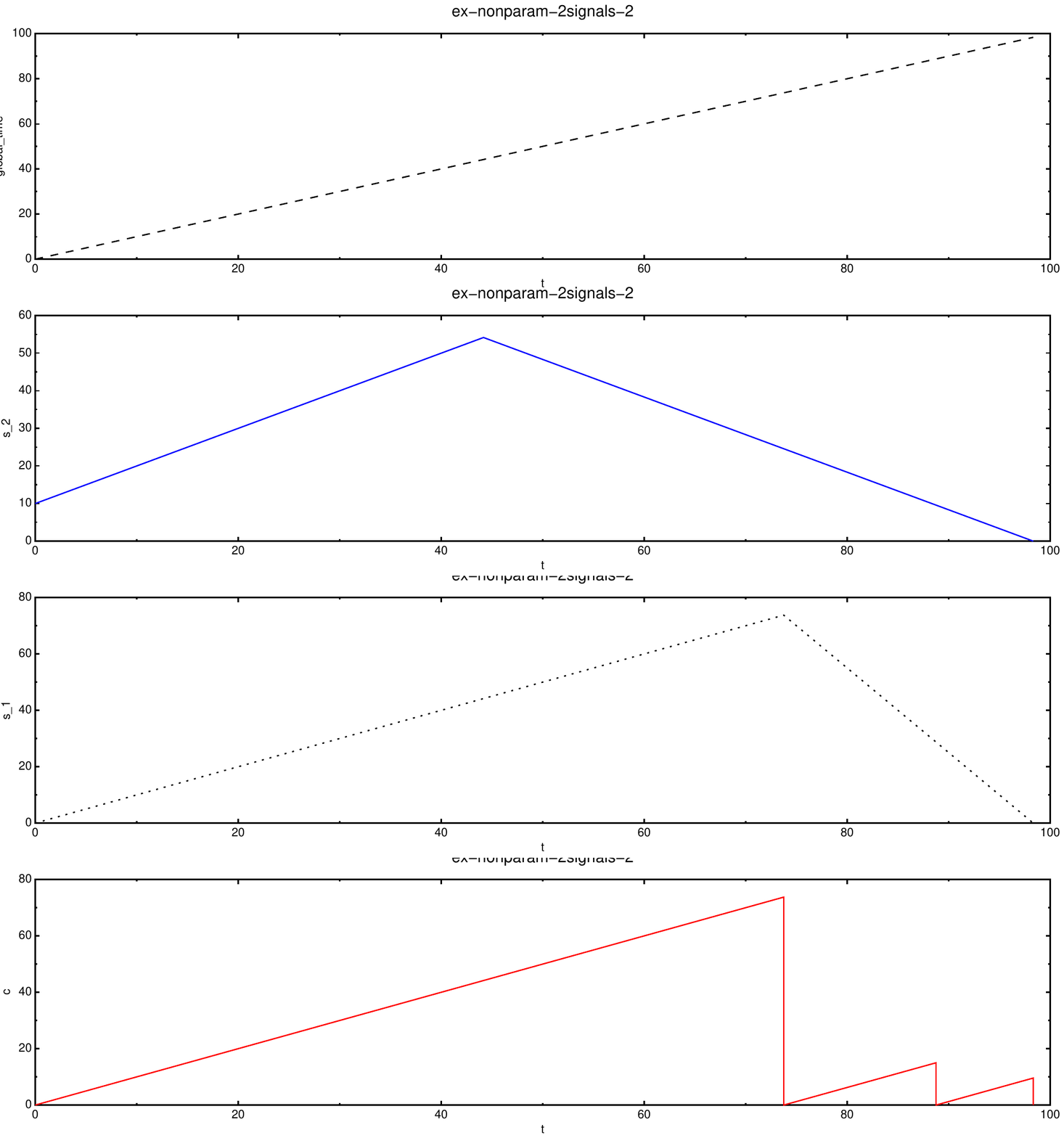}
		\caption{Example 2}
		\label{figex2signals2:2}
	\end{subfigure}
	\begin{subfigure}[b]{0.31\textwidth}
		\includegraphics[width=\textwidth]{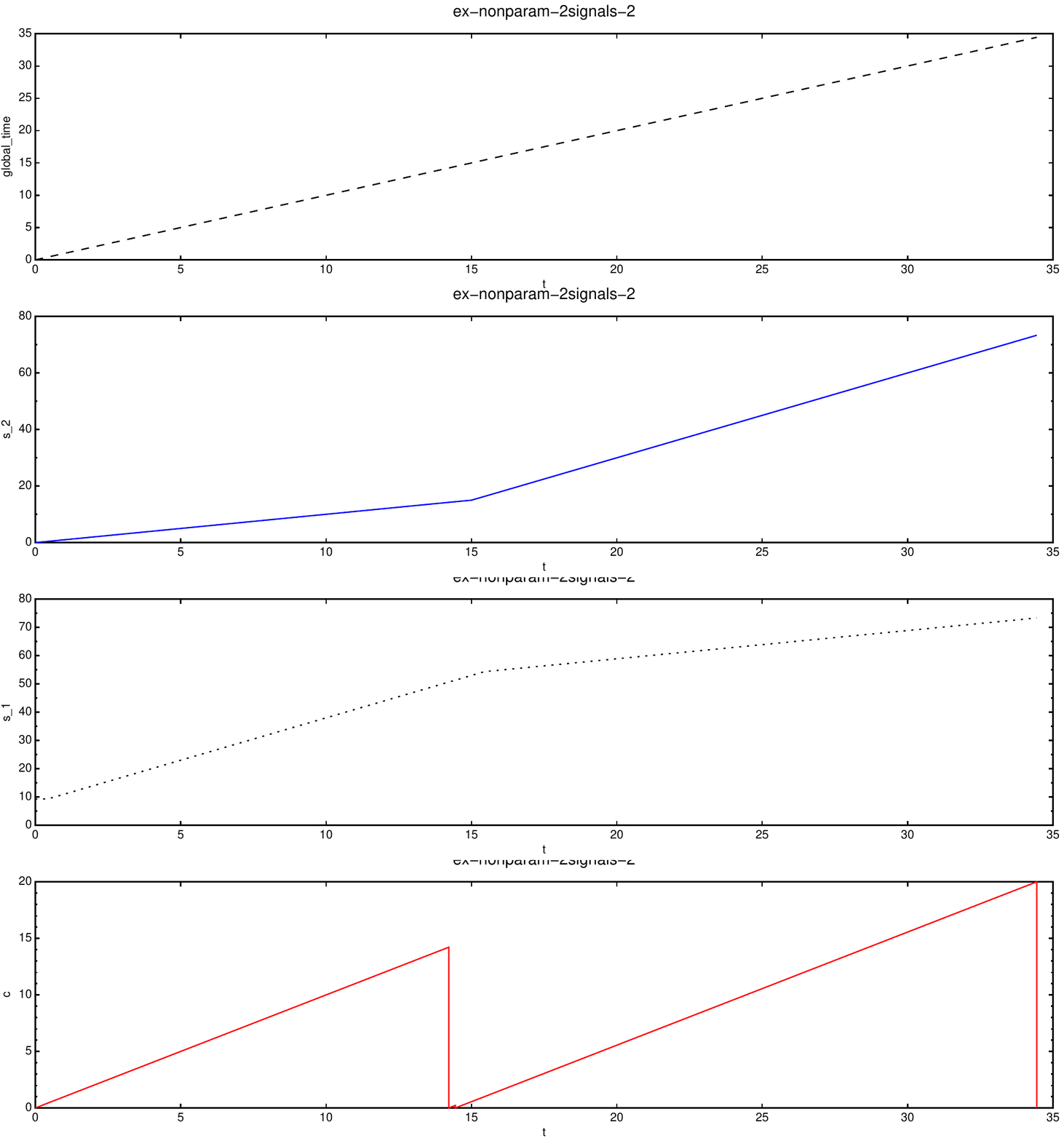}
		\caption{Example 3}
		\label{figex2signals2:3}
	\end{subfigure}
	
	\begin{subfigure}[b]{0.31\textwidth}
		\includegraphics[width=\textwidth]{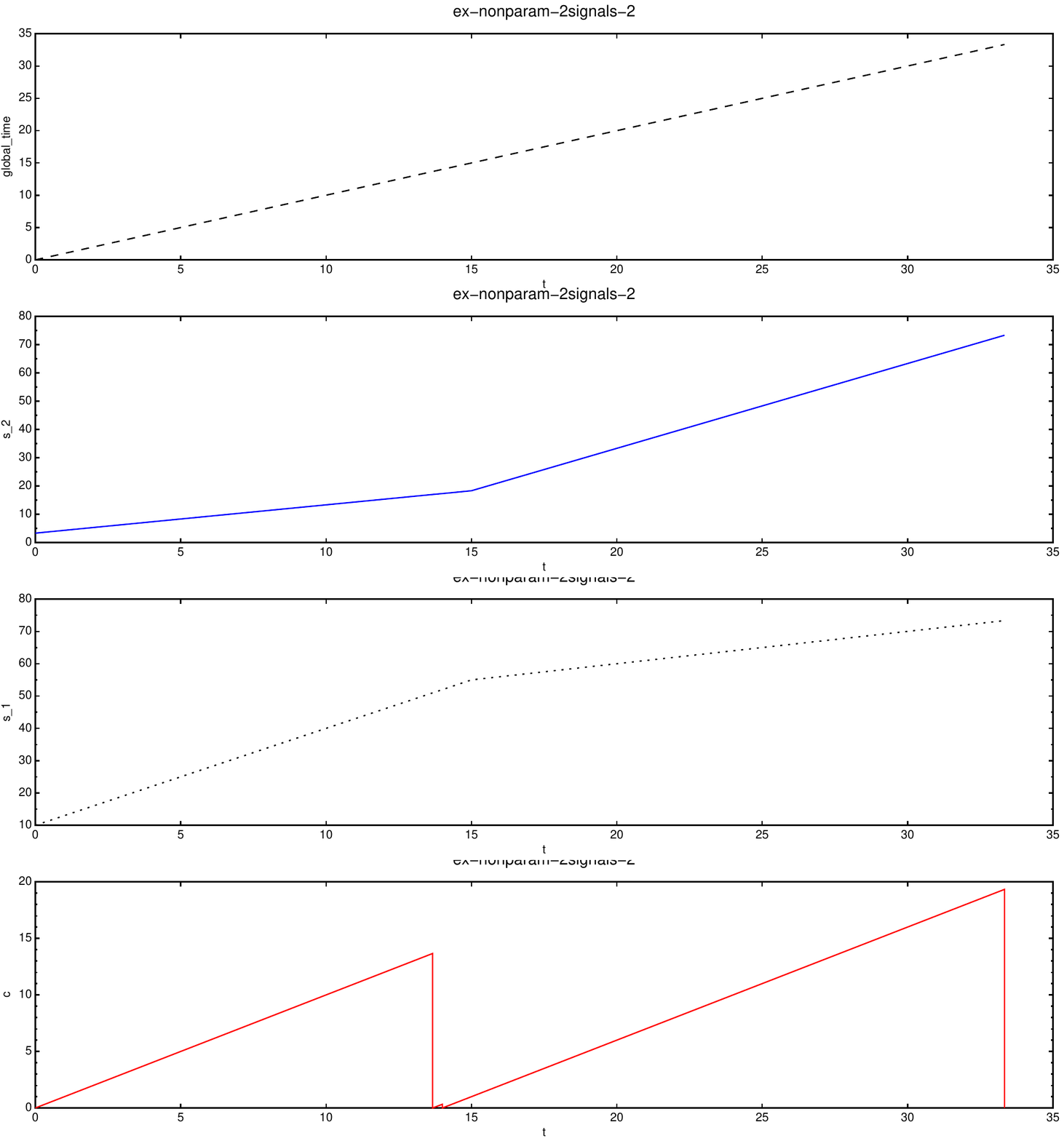}
		\caption{Example 4}
		\label{figex2signals2:4}
	\end{subfigure}
	\begin{subfigure}[b]{0.31\textwidth}
		\includegraphics[width=\textwidth]{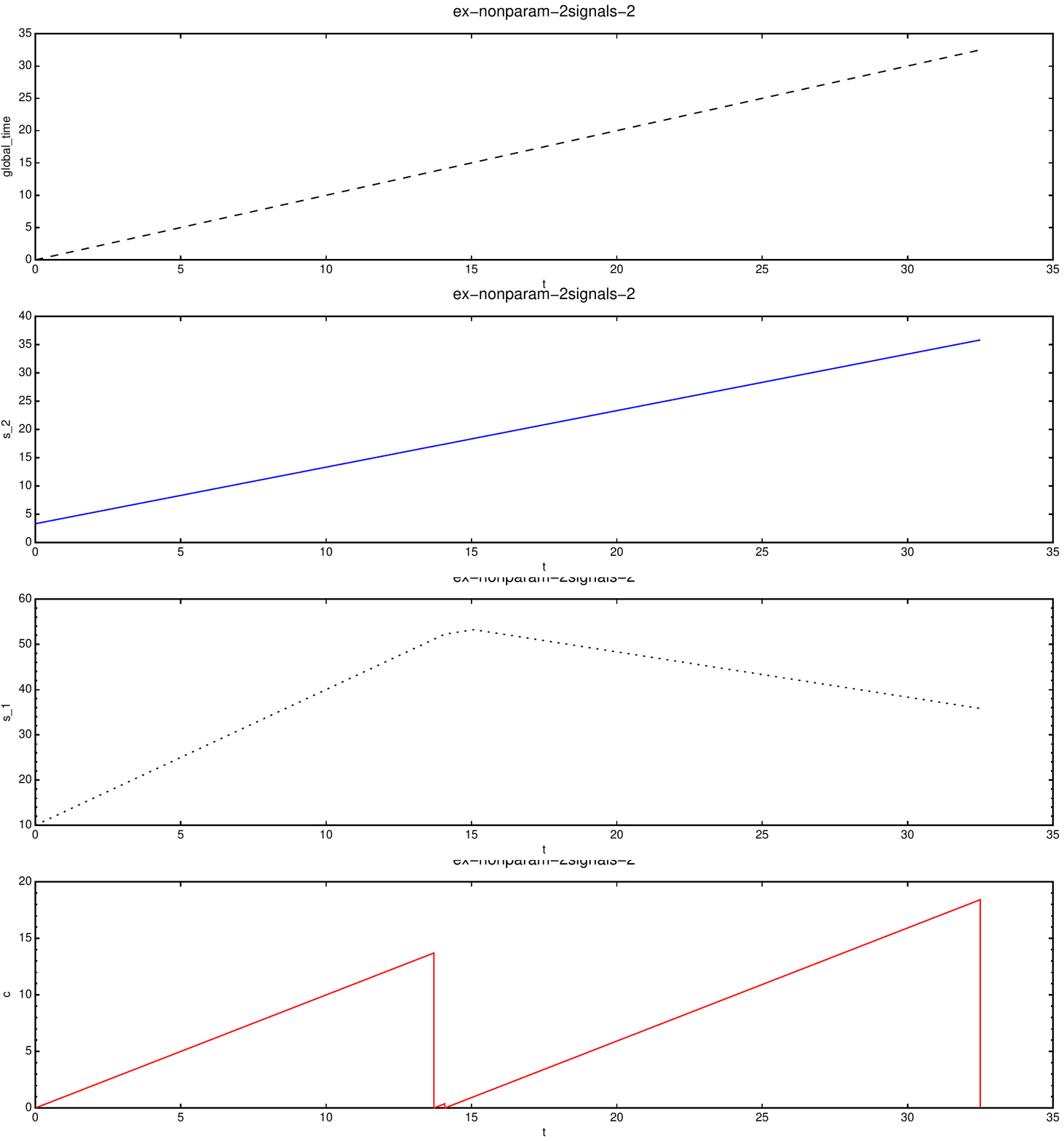}
		\caption{Example 5}
		\label{figex2signals2:5}
	\end{subfigure}
	\begin{subfigure}[b]{0.31\textwidth}
		\includegraphics[width=\textwidth]{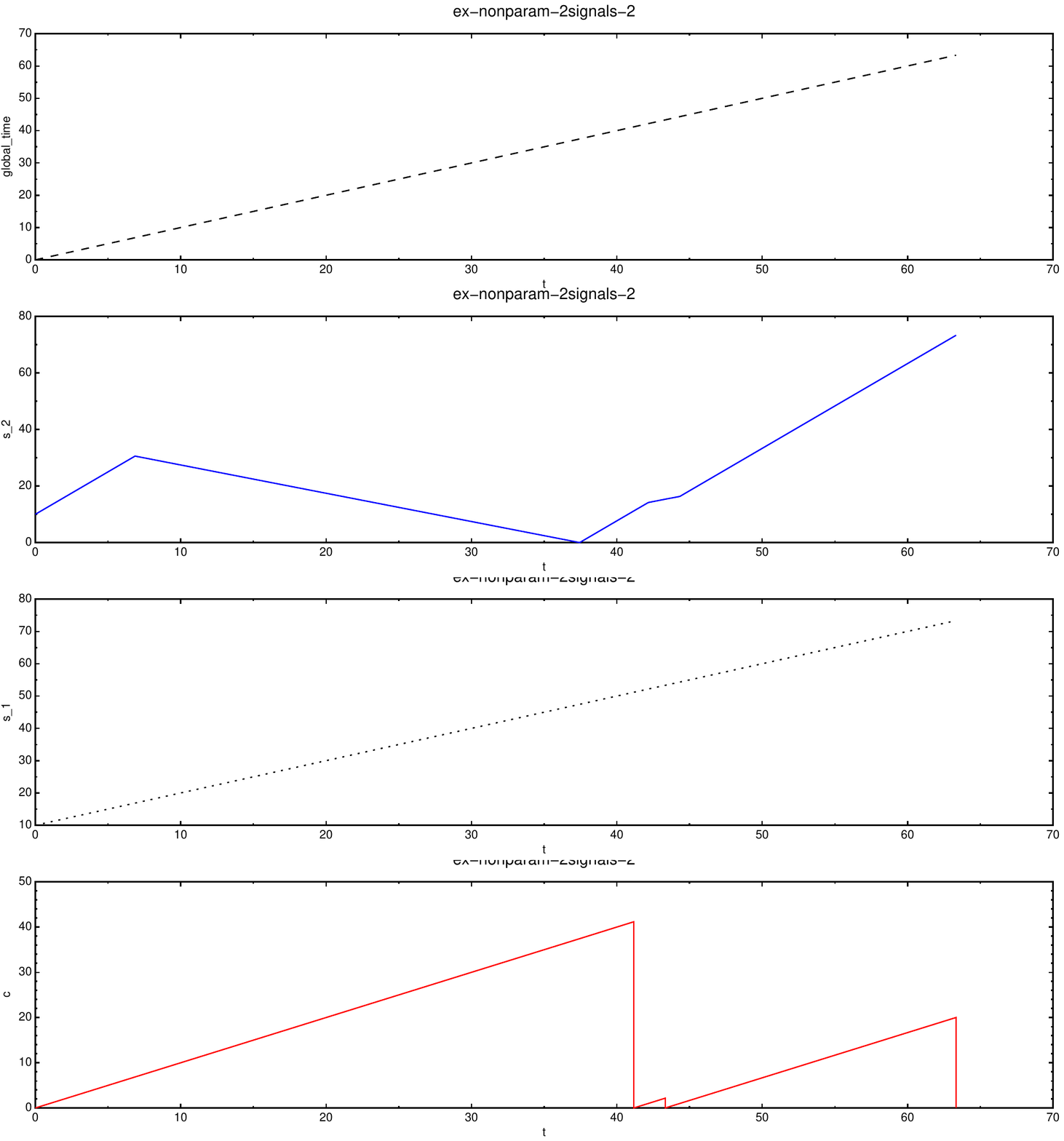}
		\caption{Example 6}
		\label{figex2signals2:6}
	\end{subfigure}
	\caption{Positive runs for \cref{figure:example-PTAS:motivating,figure:example-SBA-fastslow}}
	\label{fig:ex2signals2:result}
\end{figure}

Note that \cref{figure:concrete-runs:running:1,figure:concrete-runs:running:2,figure:concrete-runs:running:3} correspond to a (manual) representation in \LaTeX{} of \cref{figex2signals2:1,figex2signals2:2,figex2signals2:6} respectively.

\subsection{Runs for \cref{figure:example-PTAS:param,figure:example-SBA-fastslow}}
\label{appendix:exparam:results}

A graphical representation of all positive runs is given in \cref{fig:exparam:result}, while two negative runs are given in \cref{fig:exparam:result:neg}.

\begin{figure}[h]
	\centering
	\begin{subfigure}[b]{0.31\textwidth}
		\includegraphics[width=\textwidth]{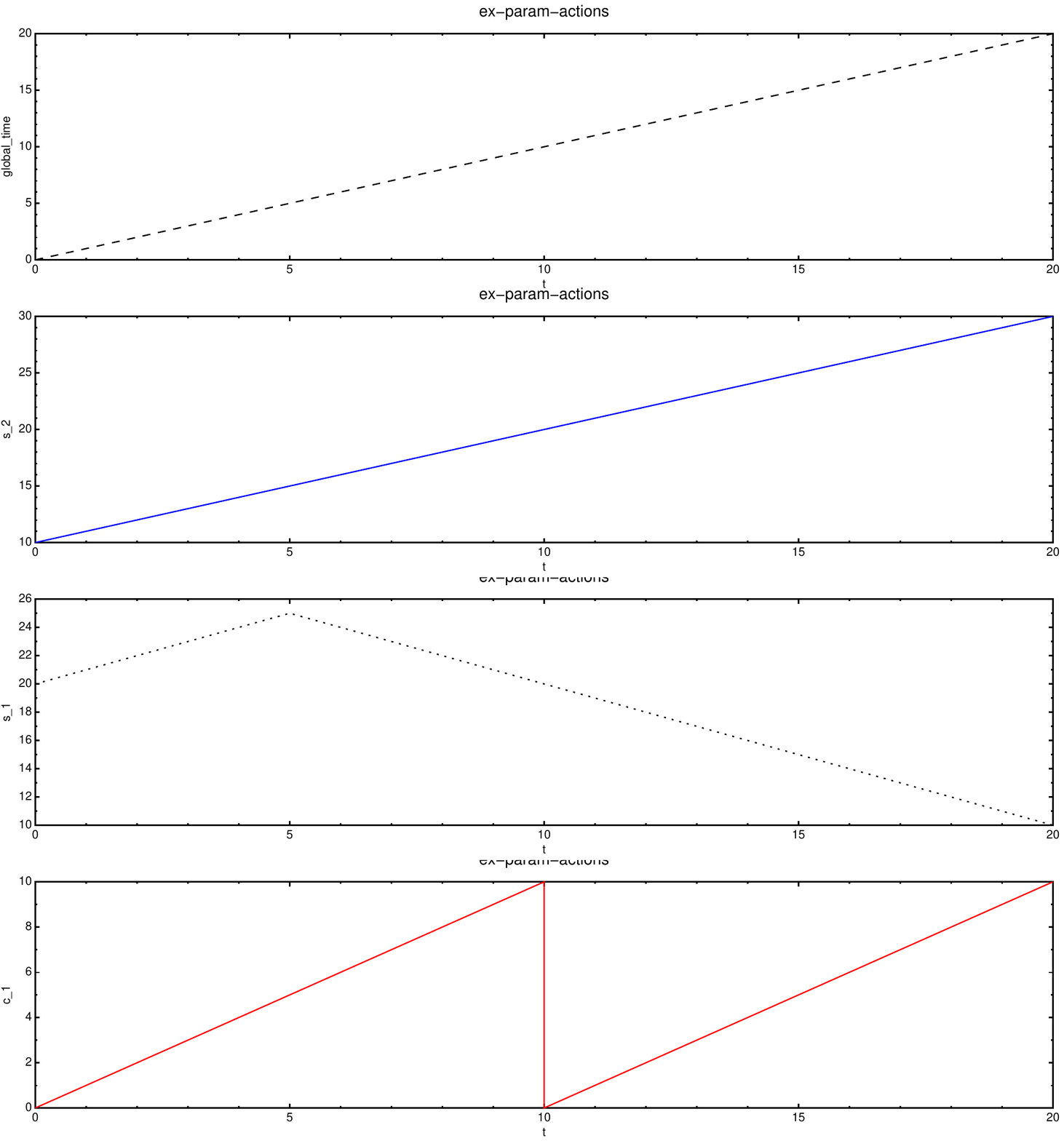}
		\caption{Example 1}
		\label{fig:exparam:1}
	\end{subfigure}
	\begin{subfigure}[b]{0.31\textwidth}
		\includegraphics[width=\textwidth]{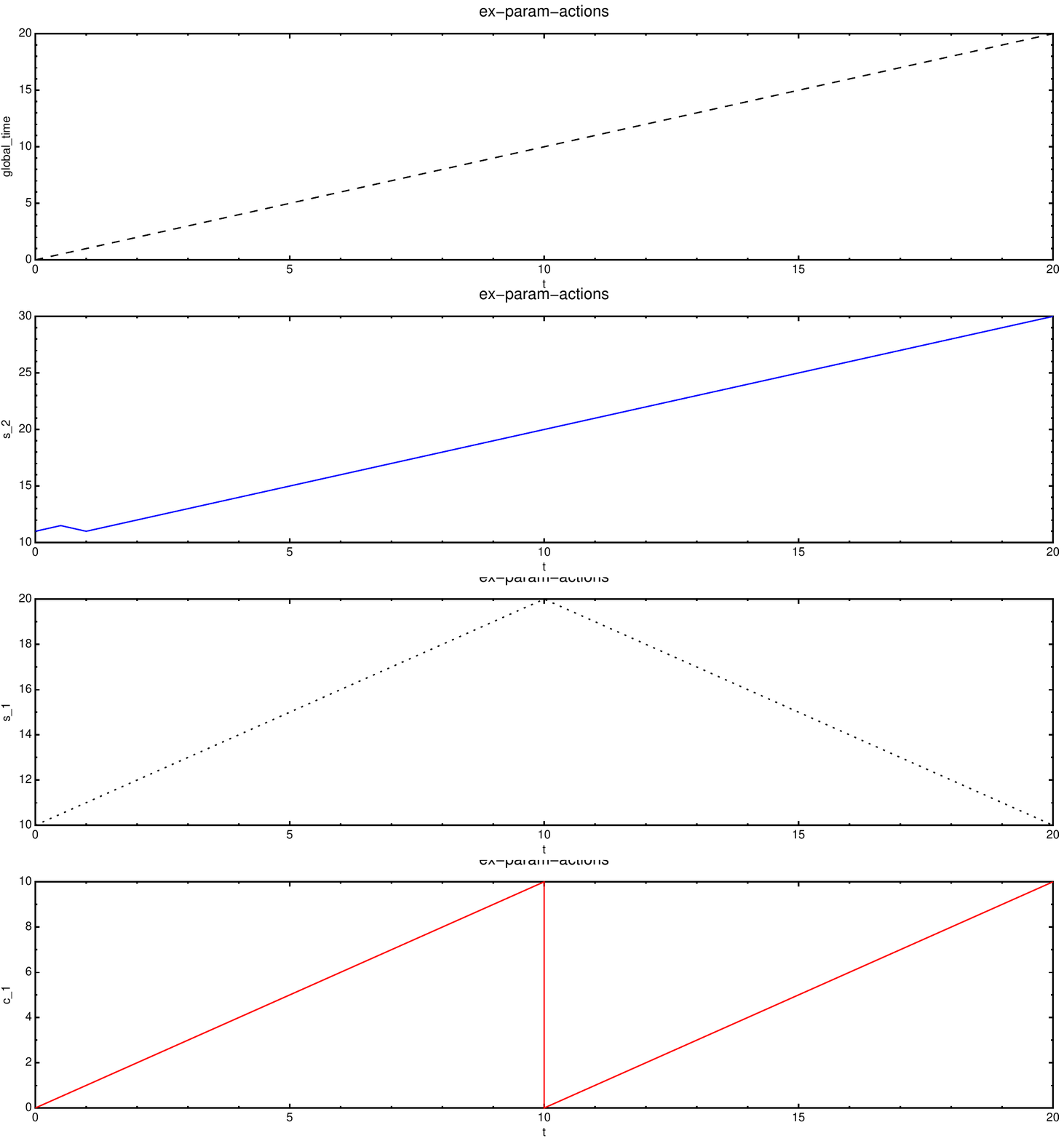}
		\caption{Example 2}
		\label{fig:exparam:2}
	\end{subfigure}
	\begin{subfigure}[b]{0.31\textwidth}
		\includegraphics[width=\textwidth]{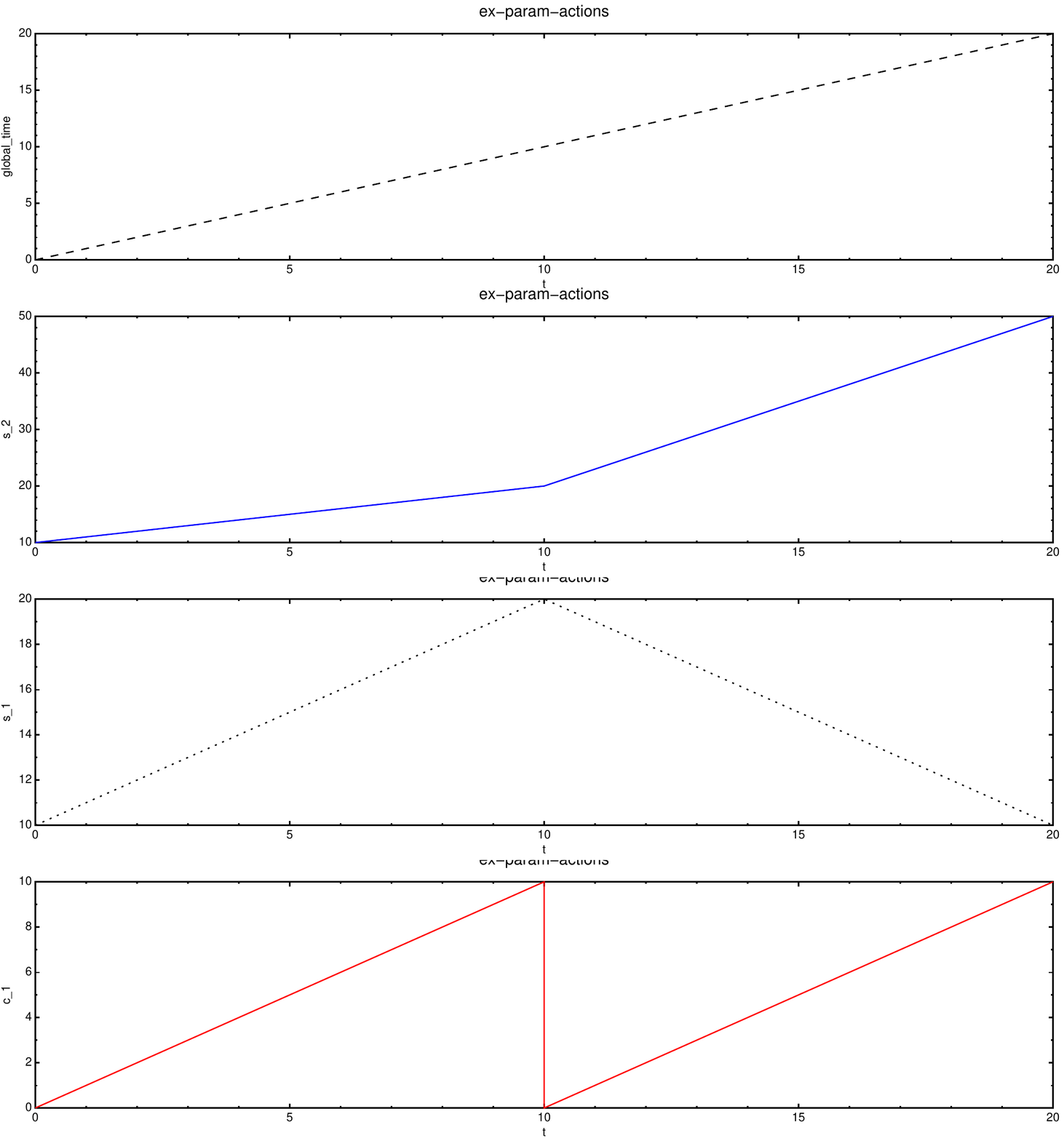}
		\caption{Example 3}
		\label{fig:exparam:3}
	\end{subfigure}
	
	\begin{subfigure}[b]{0.31\textwidth}
		\includegraphics[width=\textwidth]{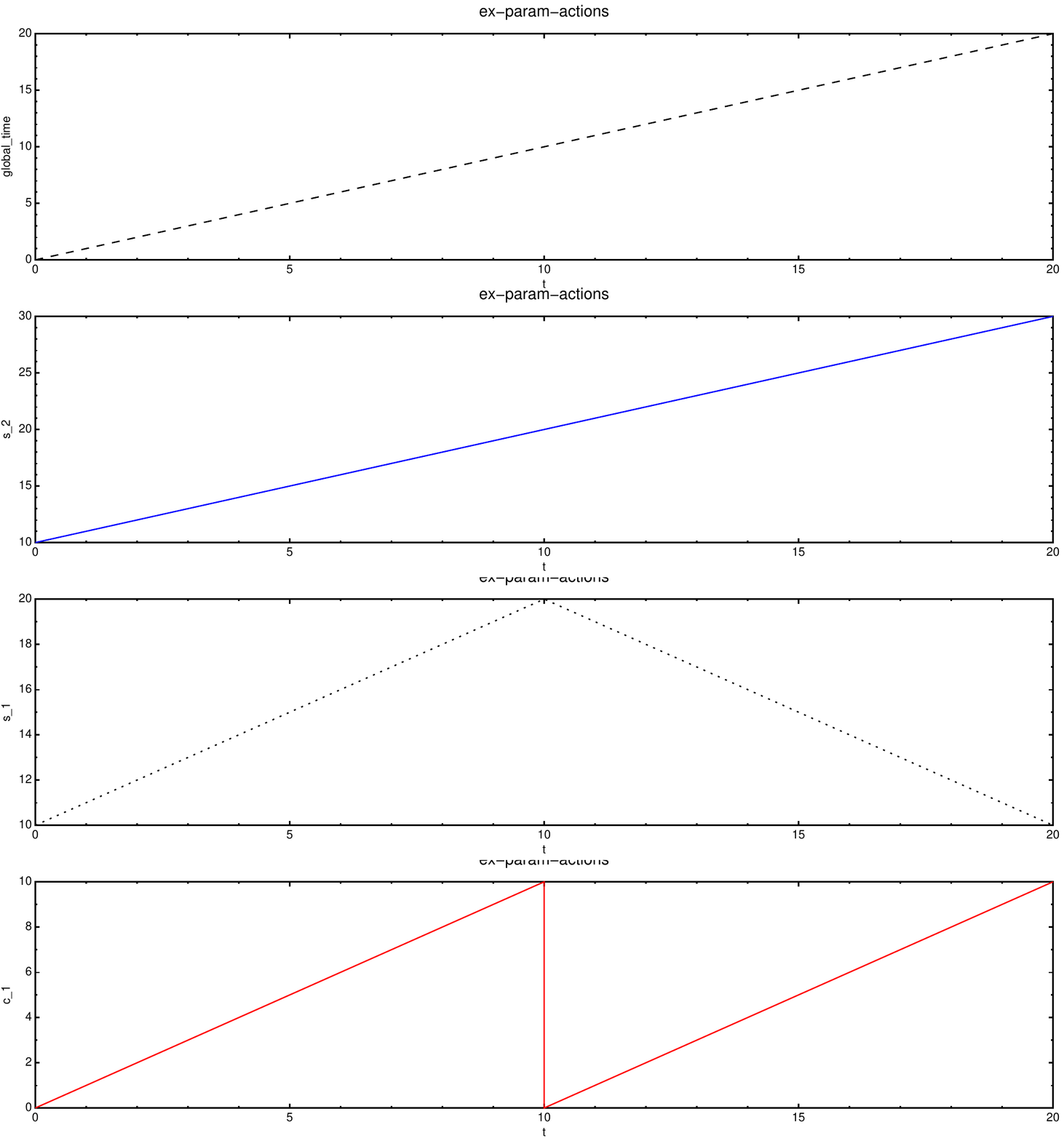}
		\caption{Example 4}
		\label{fig:exparam:4}
	\end{subfigure}
	\begin{subfigure}[b]{0.31\textwidth}
		\includegraphics[width=\textwidth]{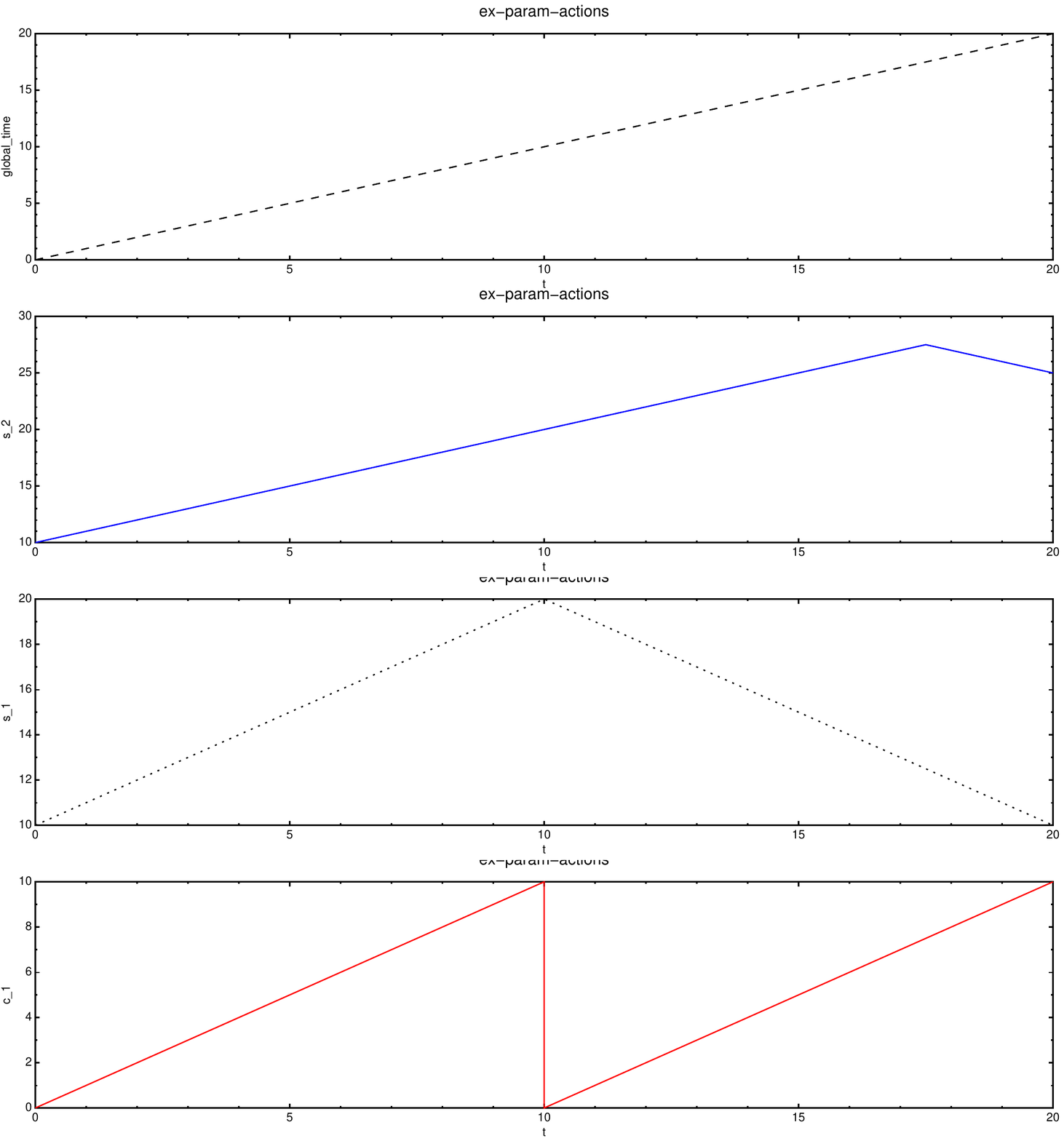}
		\caption{Example 5}
		\label{fig:exparam:5}
	\end{subfigure}
	\begin{subfigure}[b]{0.31\textwidth}
		\includegraphics[width=\textwidth]{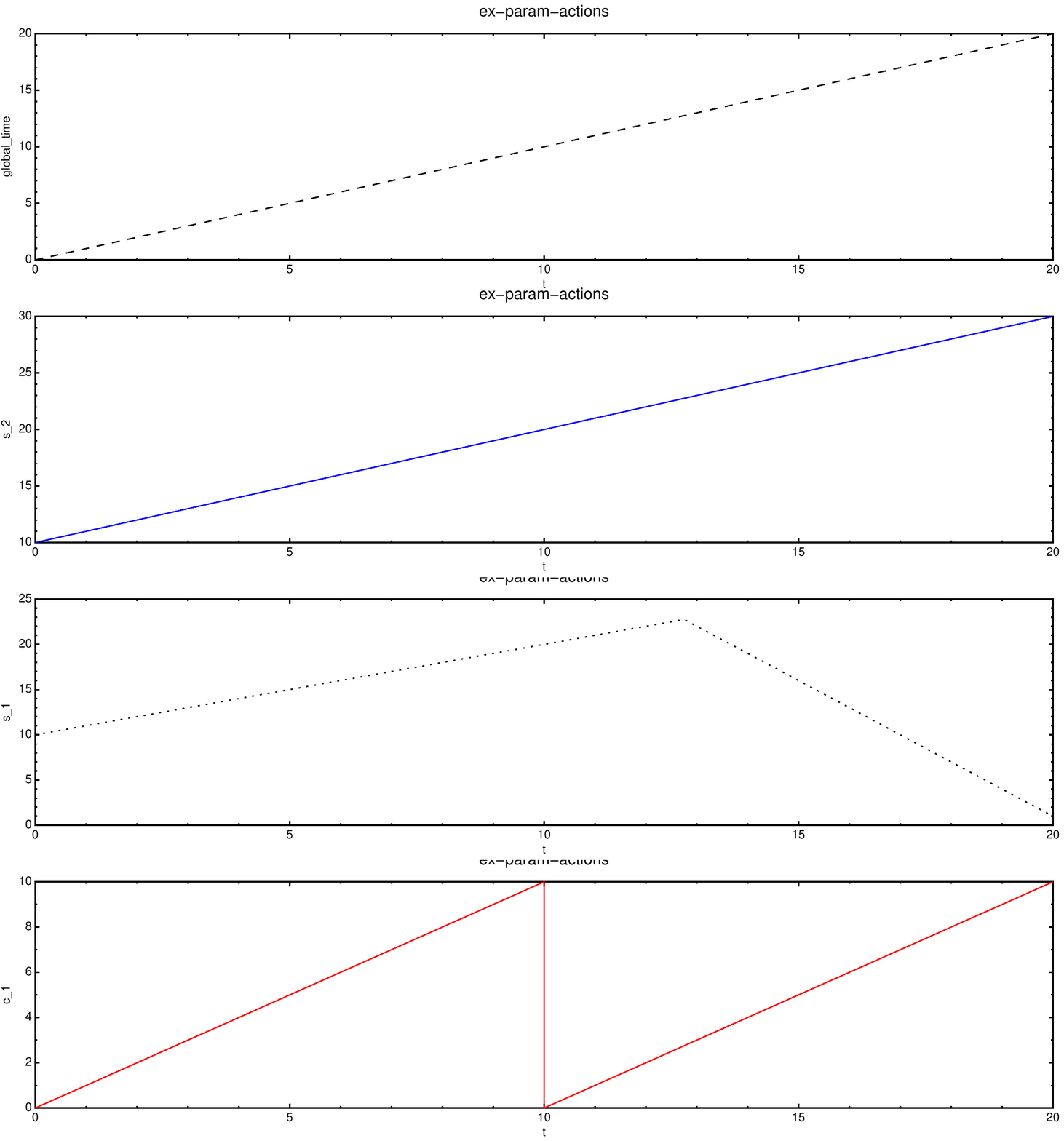}
		\caption{Example 6}
		\label{fig:exparam:6}
	\end{subfigure}
	\caption{Positive runs for \cref{figure:example-PTAS:param,figure:example-SBA-fastslow}}
	\label{fig:exparam:result}
\end{figure}
\begin{figure}[h]
	\centering
	
	\begin{subfigure}[b]{0.31\textwidth}
		\includegraphics[width=\textwidth]{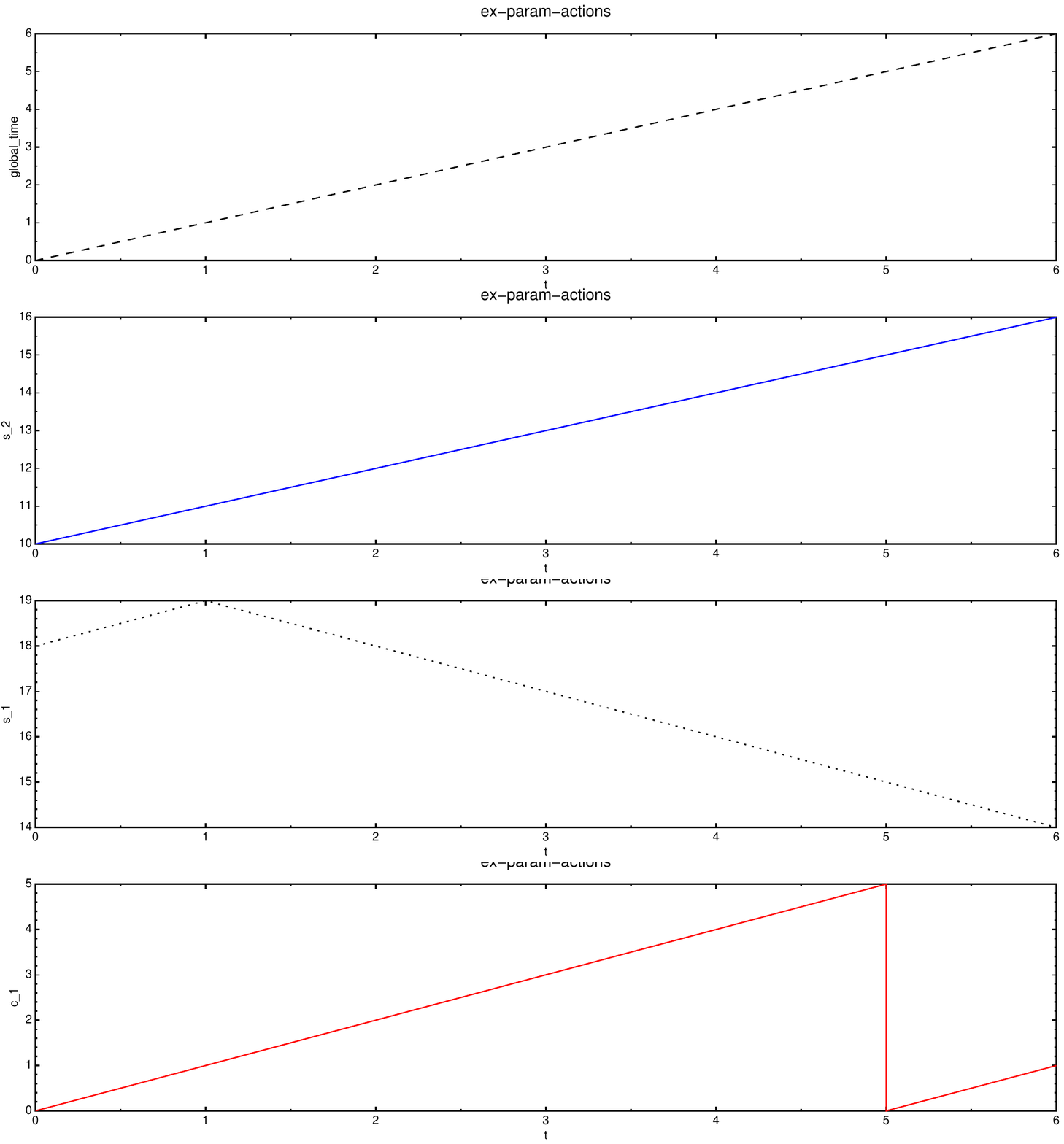}
		\caption{Example 1}
		\label{figex2signals2:neg1}
	\end{subfigure}
	\begin{subfigure}[b]{0.31\textwidth}
		\includegraphics[width=\textwidth]{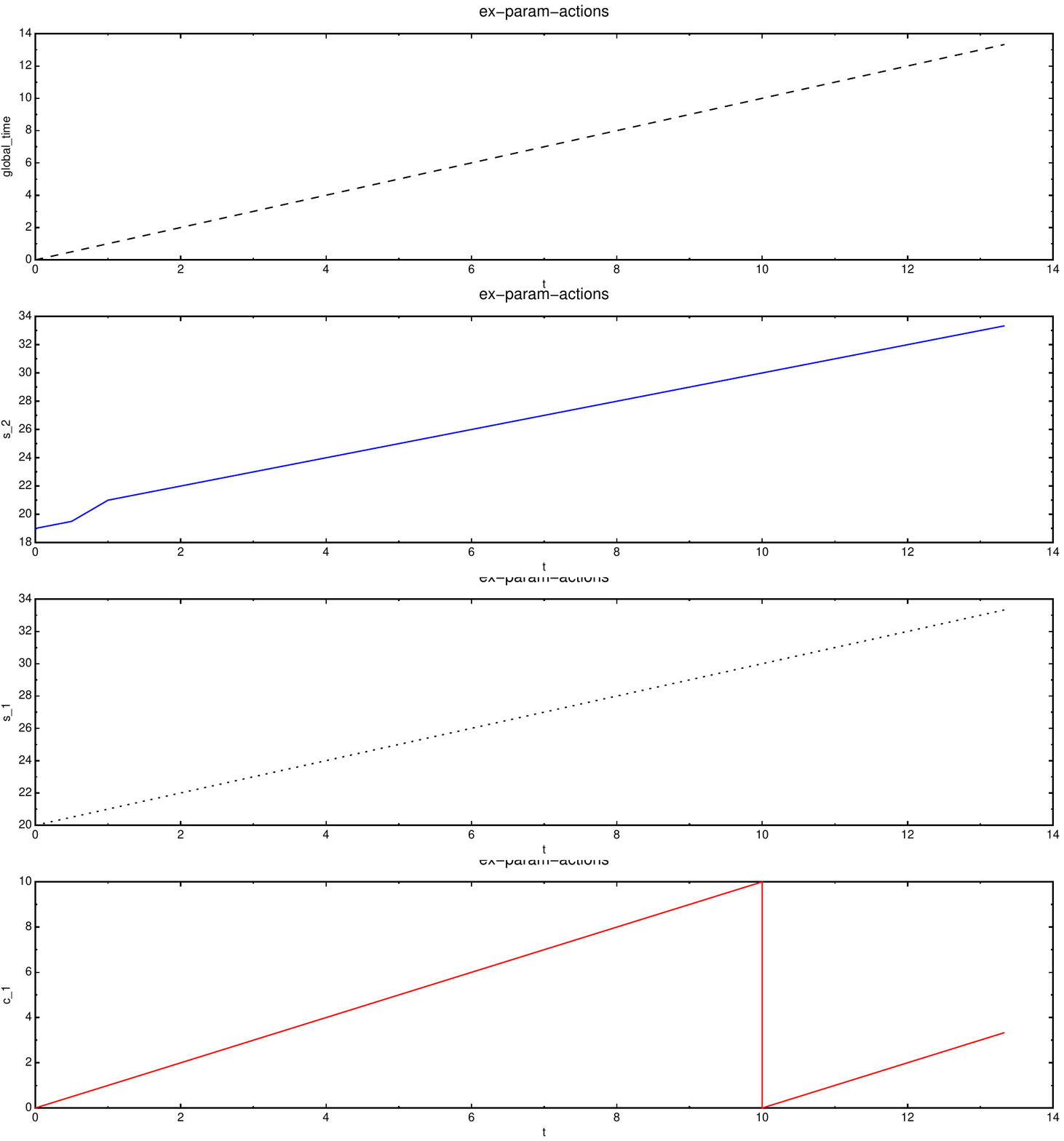}
		\caption{Example 2}
		\label{figex2signals2:neg2}
	\end{subfigure}
	
	\caption{Negative runs for \cref{figure:example-PTAS:param,figure:example-SBA-fastslow}}
	\label{fig:exparam:result:neg}
\end{figure}

\fi

\end{document}